\documentclass{aa}
\newcommand\arcdeg{\mbox{$^\circ$}\xspace} 
\usepackage{latexsym}		
\usepackage{graphicx}		
\usepackage{rotating}		
\usepackage{natbib}  
\usepackage{savesym}
\usepackage{pdflscape}
\usepackage{amssymb}
\savesymbol{doublespace}
\usepackage{wrapfig}
\usepackage{xspace}
\usepackage{color}
\usepackage{multicol}
\usepackage{mdframed}
\usepackage{hyperref}
\usepackage{subfigure}
\usepackage{lscape}
\usepackage{grffile}
\usepackage{standalone}
\standalonetrue
\usepackage{import}
\usepackage[utf8]{inputenc}
\usepackage{longtable}
\usepackage{booktabs}
\usepackage[yyyymmdd,hhmmss]{datetime}
\usepackage{fancyhdr}

\newcommand{\msun}{\ensuremath{M_{\odot}}\xspace}			

\newcommand{\hh}{\ensuremath{\textrm{H}_{2}}\xspace}			

\newcommand{\formaldehyde}{\ensuremath{\textrm{H}_2\textrm{CO}}\xspace}

\newcommand{\methanol}{\ensuremath{\textrm{CH}_3\textrm{OH}}\xspace}

\newcommand{\para}{\ensuremath{\textrm{p-H}_2\textrm{CO}}\xspace}

\newcommand{\threeohthree}{\ensuremath{3_{0,3}-2_{0,2}}\xspace}
\newcommand{\threetwotwo}{\ensuremath{3_{2,2}-2_{2,1}}\xspace}
\newcommand{\threetwoone}{\ensuremath{3_{2,1}-2_{2,0}}\xspace}
\newcommand{\fourtwotwo}{\ensuremath{4_{2,2}-3_{1,2}}\xspace} 
\newcommand{\methylcyanide}{\ensuremath{\textrm{CH}_{3}\textrm{CN}}\xspace}
\newcommand{\Rone}{\ensuremath{\para~S_{\nu}(\threetwoone) / S_{\nu}(\threeohthree)}\xspace}
\newcommand{\Rtwo}{\ensuremath{\para~S_{\nu}(\threetwotwo) / S_{\nu}(\threetwoone)}\xspace}

\newcommand{\hii}{H~{\sc ii}\xspace}

\newcommand{\kms}{\textrm{km~s}\ensuremath{^{-1}}\xspace}	

\newcommand{\pers}{\ensuremath{\mathrm{s}^{-1}}\xspace}

\newcommand{\percc}{\ensuremath{\textrm{cm}^{-3}}\xspace}
\newcommand{\perpc}{\ensuremath{\textrm{pc}^{-1}}\xspace}
\newcommand{\persc}{\ensuremath{\textrm{cm}^{-2}}\xspace}

\newcommand{\perkmspc}{\textrm{km~s}\ensuremath{^{-1}}\textrm{pc}\ensuremath{^{-1}}\xspace}	
\newcommand{\perkms}{\textrm{per~km~s}\ensuremath{^{-1}}\xspace}	
\newcommand{\um}{\ensuremath{\mu \textrm{m}}\xspace}    

\newcommand{\ammonia}{NH\ensuremath{_3}\xspace}
\newcommand{\twelveco}{\ensuremath{^{12}\textrm{CO}}\xspace}
\newcommand{\thirteenco}{\ensuremath{^{13}\textrm{CO}}\xspace}
\newcommand{\ceighteeno}{\ensuremath{\textrm{C}^{18}\textrm{O}}\xspace}
\def\ee#1{\ensuremath{\times10^{#1}}}

\def\eqref#1{Equation \ref{#1}}

\renewcommand\arcdeg{\mbox{$^\circ$}\xspace} 
\renewcommand\arcmin{\mbox{$^\prime$}\xspace} 
\renewcommand\arcsec{\mbox{$^{\prime\prime}$}\xspace}

\def\Figure#1#2#3#4#5{
\begin{figure*}[htp]
\includegraphics[scale=#4,width=#5]{#1}
\caption{#2}
\label{#3}
\end{figure*}
}

\def\FigureTwo#1#2#3#4#5#6{
\begin{figure*}[htp]
\subfigure[]{ \includegraphics[scale=#5,width=#6]{#1} }
\subfigure[]{ \includegraphics[scale=#5,width=#6]{#2} }
\caption{#3}
\label{#4}
\end{figure*}
}

\def\FigureTwoAA#1#2#3#4#5#6{
\begin{figure*}[htp]
\subfigure[]{ \includegraphics[scale=#5,width=#6]{#1} }
\subfigure[]{ \includegraphics[scale=#5,width=#6]{#2} }
\caption{#3}
\label{#4}
\end{figure*}
}

\newenvironment{rotatepage}%
{}{}

\def\RotFigureTwoAA#1#2#3#4#5#6{
\begin{rotatepage}
\begin{sidewaysfigure*}[htp]
\subfigure[]{ \includegraphics[scale=#5,width=#6]{#1} }
\\
\subfigure[]{ \includegraphics[scale=#5,width=#6]{#2} }
\caption{#3}
\label{#4}
\end{sidewaysfigure*}
\end{rotatepage}
}

\def\RotFigureThreeAA#1#2#3#4#5#6#7{
\begin{rotatepage}
\begin{sidewaysfigure*}[htp]
\subfigure[]{ \includegraphics[scale=#6,width=#7]{#1} }
\\
\subfigure[]{ \includegraphics[scale=#6,width=#7]{#2} }
\\
\subfigure[]{ \includegraphics[scale=#6,width=#7]{#3} }
\caption{#4}
\label{#5}
\end{sidewaysfigure*}
\end{rotatepage}
\clearpage
}

\begin{document}

\title{Dense gas in the Galactic central molecular zone is warm and heated by turbulence}
\titlerunning{APEX CMZ \formaldehyde}
\authorrunning{Ginsburg et al}
\newcommand{\eso}     {$^{1 }$}
\newcommand{\mpifr}   {$^{2 }$}
\newcommand{\saudi}   {$^{3 }$}
\newcommand{\naoj}    {$^{4 }$}
\newcommand{\pmo}     {$^{5 }$}
\newcommand{\nrao}    {$^{6 }$}
\newcommand{\casa}    {$^{7 }$}
\newcommand{\cfa}     {$^{8 }$}
\newcommand{\chalmers}{$^{9 }$}
\newcommand{\oxford}  {$^{10}$}
\newcommand{\mpa}     {$^{11}$}
\newcommand{\ljmu}    {$^{12}$}
\newcommand{\lmu}     {$^{13}$}

\author{Adam Ginsburg{\eso},
        Christian Henkel{\mpifr,\saudi},
        Yiping Ao{\naoj,\pmo},
        Denise Riquelme{\mpifr},
        Jens Kauffmann{\mpifr},
        Thushara Pillai{\mpifr},
        Elisabeth A.C. Mills{\nrao},
        Miguel A. Requena-Torres{\mpifr},
        Katharina Immer{\eso},
        Leonardo Testi{\eso},
        Juergen Ott{\nrao},
        John Bally{\casa},
        Cara Battersby{\cfa},
        Jeremy Darling{\casa},
        Susanne Aalto{\chalmers},
        Thomas Stanke{\eso},
        Sarah Kendrew{\oxford},
        J.M. Diederik Kruijssen{\mpa},
        Steven Longmore{\ljmu},
        James Dale{\lmu},
        Rolf Guesten{\mpifr},
        Karl M. Menten{\mpifr}
        }

\institute{
    {\eso}{\it{European Southern Observatory, Karl-Schwarzschild-Strasse 2, D-85748 Garching bei München, Germany\\
                      \email{Adam.Ginsburg@eso.org}}} \\ 
    {\mpifr}{\it{Max Planck Institute for Radio Astronomy, auf dem Hugel, Bonn}}
    {\saudi}{\it{Astron. Dept., King Abdulaziz University, P.O. Box 80203,
    Jeddah 21589, Saudi Arabia}}\\
    {\naoj}{National Astronomical Observatory of Japan, 2-21-1 Osawa, Mitaka, Tokyo 181-8588, Japan}
    {\pmo}{Purple Mountain Observatory, Chinese Academy of Sciences, Nanjing 210008, China}
    {\nrao}{\it{National Radio Astronomy Observatory, Socorro}}
    {\casa}{\it{CASA, University of Colorado, 389-UCB, Boulder, CO 80309}} \\ 
    {\cfa}{\it{Harvard-Smithsonian Center for Astrophysics, 60 Garden
    Street, Cambridge, MA 02138, USA}} \\ 
    {\chalmers}{\it{Department of Earth and Space Sciences,
                    Chalmers University of Technology}}
    {\oxford}{\it{Department of Astrophysics, The Denys Wilkinson Building, Keble Road, Oxford OX1 3RH}}
    {\mpa}{\it{Max-Planck Institut f\"{u}r Astrophysik, Karl-Schwarzschild-Stra\ss e 1, 85748 Garching, Germany}}
    {\ljmu}{\it{Astrophysics Research Institute, Liverpool John Moores
    University, IC2, Liverpool Science Park, 146 Brownlow Hill, Liverpool L3
    5RF, United Kingdom}}
    {\lmu}{\it{University Observatory Munich, Scheinerstr. 1, D-81679 München, Germany}}
    }

\date{Date: \today ~~ Time: \currenttime}

\abstract
{The Galactic center is the closest region in which we can study star formation
under extreme physical conditions like those in high-redshift galaxies.
}
{We measure the temperature of the dense gas in the central molecular zone
(CMZ) and examine what drives it.}
{We mapped the inner 300 pc of the CMZ in the temperature-sensitive
$J=3-2$ para-formaldehyde (\para) transitions.  We used the 
\threetwoone / \threeohthree line ratio to determine the gas temperature in
$n\sim10^4-10^5$ \percc gas.  We have produced temperature maps and cubes with
30\arcsec and 1 \kms resolution and published all data in FITS
form\thanks{The data can be accessed from \texttt{doi:10.7910/DVN/27601} and
are available from CDS via anonymous ftp to cdsarc.u-strasbg.fr (130.79.128.5)
or via \url{http://cdsweb.u-strasbg.fr/cgi-bin/qcat?J/A+A/}}.}
{Dense gas temperatures in the Galactic center range from $\sim60$ K to $>100$
K in selected regions.  The highest gas temperatures $T_G>100$ K are
observed around the Sgr B2 cores, in the extended Sgr B2 cloud, the 20 \kms and
50 \kms clouds, and in ``The Brick'' (G0.253+0.016).  We infer an upper limit on
the cosmic ray ionization rate $\zeta_{CR}<10^{-14}$ \pers.  
}
{
The dense molecular gas temperature of the region around our
Galactic center is similar to values found in the central regions of other
galaxies, in particular starburst systems.  The gas temperature is uniformly
higher than the dust temperature, confirming that dust is a coolant in the
dense gas.  Turbulent heating can readily explain the observed temperatures
given the observed line widths.  Cosmic rays cannot explain the observed
variation in gas temperatures, so CMZ dense gas temperatures are not dominated
by cosmic ray heating.  The gas temperatures previously observed to be high
in the inner $\sim75$ pc are confirmed to be high in the entire CMZ.}

\maketitle

\section{Introduction}
\label{sec:intro}
The central region of our Galaxy, the central molecular zone (CMZ), is the
nearest location in which we can study star formation in an extreme
environment with pressure, turbulent Mach number, and gas temperature much
higher than in local star-forming regions \citep{Morris1996a}.  While there
have been great leaps in our understanding of star formation in local
molecular clouds in the past decade, there remain many unanswered questions
about how star formation changes as gas becomes denser, more opaque, and more
turbulent, as it most likely was in galaxies at the peak of cosmic star
formation \citep{Kruijssen2013a}.

The Galactic CMZ has attracted a great deal of attention recently because it
has a much higher ratio of dense gas mass to star formation tracers than the
rest of the galaxy
\citep{Longmore2013a,Longmore2012b,Longmore2013b,Kruijssen2014c,Yusef-Zadeh2009a,Immer2012a}.
The star formation rate in this region therefore appears to be suppressed
relative to the expectations from nearby star-forming regions and nearby galaxy
disks, where the star formation timescale decreases with increasing gas
surface density
\citep{Kennicutt1998a, Kennicutt2012a, Leroy2013a, Heiderman2010a}.  Within the
CMZ, the central $\sim100$ pc\footnote{Assuming the IAU-recommended $d=8.5$ kpc
\citep{Ghez2008a, Gillessen2009b, Reid2009a, Gillessen2013b}.} ``ring'' including
the  dust ridge \citep{Lis1991a, Sofue1995a, Molinari2011a} contains most of
the dense gas.  \citet{Longmore2013a} recently proposed that the clouds along
the ``ring'' \citep[which is more accurately described as a
stream;][]{Kruijssen2015a} represent a time-ordered path from starless to
star-forming clouds, from ``The Brick'' (G0.253+0.016, cloud a) through the
lettered dust ridge clouds \citep[clouds b,c,d,e,f][]{Lis1999a}, ending at Sgr
B2.  In principle, this would allow us to study the time evolution of
protocluster clumps given an orbital model to describe their motion
\citep{Kruijssen2015a}.

The thermal properties of this gas are a crucial component for understanding
the conditions governing both star formation and interstellar chemistry.  The
gas temperature sets the sound speed in the gas and therefore the Mach number
within a turbulent medium.  It also governs the thermal Jeans mass.  Regardless
of whether turbulence or gravity controls the fragmentation scale, the gas
temperature is clearly important.  

Prior gas temperature measurements in the CMZ primarily used the popular
ammonia metastable inversion transition thermometer \citep[\ammonia (1,1) -
(7, 7);][]{Guesten1981a, Mauersberger1986a, Huettemeister1993a, Ott2014a},
which is
sensitive to moderate density gas \citep[$n(\hh) \sim 10^3-10^4$
\percc;][]{Shirley2015a}.  While this thermometer is generally reliable in cold
regions (e.g., the (1-1)/(2-2) ratio is accurate to $T_G\lesssim40$ K), the
population of the higher energy states \emph{may} be affected by formation
heating, a mechanism recently discovered to affect H$_3$O+ in the CMZ and
suggested to affect \ammonia, in which the excess energy from molecule
formation leaves a significant fraction of the molecules in a rotationally
excited state \citep[][]{Lis2014a}, so confirmation of the high temperatures
seen in \ammonia \citep{Mills2013a} can be used to assess the significance of
that mechanism.  \citet{Ao2013a} used the para-formaldehyde (\para) 218 GHz
thermometer \citep{Mangum1993a}, which is sensitive to denser ($n\sim10^4-10^5$
\percc) and hotter ($T_G>20$ K, usually $T_G\sim60$ K) gas, to map out the
inner $\sim75$ pc of the CMZ and found high temperatures comparable to those
found in previous studies.

There is a long-standing problem that the observed gas and dust temperatures do
not agree throughout the CMZ \citep{Guesten1981a, Molinari2011a, Ao2013a,
Clark2013a, Ott2014a}.  Photon-dominated region
(PDR) models predict that gas and dust temperatures should match at high column
densities
\citep{Hollenbach1999a}.  The observed discrepancy represents a significant
problem for understanding the importance of gas temperature and pressure in the
evolution of molecular clouds.  The dust temperature is readily estimated via
multiband continuum mapping and therefore is often used as a proxy for the
dense gas temperature.  While the discrepancy is not theoretically a problem,
as the gas and dust may remain collisionally uncoupled up to moderately high
densities given a high enough heating rate, it is an important empirical
difference between CMZ and Galactic disk clouds.  Dust and \ammonia derived
temperatures are usually assumed and often observed to agree in Galactic disk
clouds \citep{Pillai2006a, Dunham2010a, Juvela2012a, Battersby2014a},  but the
difference can no longer be ignored in the Galactic center.

Within both our own Galactic center and many nearby galactic nuclei, there is
controversy over which heating mechanisms dominate.  In the CMZ,
\citet{Ao2013a} were able to rule out X-ray heating as an energetically
important mechanism, leaving cosmic rays and turbulence (mechanical heating) as
viable candidates.  Studies of the CO spectral line energy distribution (SLED)
around Sgr A* and Sgr B2 argued for mechanical heating and UV heating,
respectively \citep{Goicoechea2013a,Etxaluze2013a}.  Nearby infrared-luminous
galaxies have properties that resemble these regions
\citep{Kamenetzky2012a,Kamenetzky2014a}, but the debate about the dominant
heating mechanism remains open in many galaxies, allowing for photon-dominated,
X-ray-dominated, mechanical/turbulent, or cosmic ray heating
\citep{Loenen2008a, van-der-Werf2010a, Papadopoulos2011a, Meijerink2011a,
Bayet2011a, Mangum2013a, Papadopoulos2013a}.

We have expanded the \formaldehyde mapping project of \citet{Ao2013a} to a
$\sim5\times$ larger area, including the dust ridge and the positive-longitude
turbulent clouds thought to be associated with the intersection point between
the x1 and x2 orbits \citep{Rodriguez-Fernandez2006b, Riquelme2013a}.  We
describe the new
observations in Section \ref{sec:observations}.  In Section \ref{sec:signal},
we describe the analysis process used to extract signal from the data cubes.
In Section \ref{sec:analysis}, we describe how we 
derive temperatures from the line ratios.
We discuss various implications of our data in Section
\ref{sec:discussion}.  In the appendices, we describe further details of the
data reduction process, include additional tables, and provide the source code
for all aspects of this project, from data reduction through figure generation
for the paper.

\section{Observations \& data reduction}
\label{sec:observations}

\subsection{Observations}
We observed the CMZ from $-0.4\arcdeg < \ell < 1.6\arcdeg$
with the APEX-1 (SHFI) instrument \citep{Vassilev2008a} on the Atacama Pathfinder
Experiment (APEX) telescope \citep{Gusten2006b} using the eXtended bandwidth
Fast Fourier Transform
Spectrometer (XFFTS) backend \citep{Klein2012a}.  The observations were
performed in service mode and were spread out over two years.  The time was
divided into 25 hours in June
2013, 75 hours in April-July 2014, and 50 hours in October 2014.  A final
set of observations was taken in April 2015; these data are not included in the
analysis in this paper but are provided in the delivered FITS files.  The
time was split between the ESO (E-093.C-0144A; 50h, E-095.C-0242A, 25h), MPIfR
(M-091.F-0019 and M-093.F-0009; 75h), and OSO (E-093.C-0144A; 25h) queues.

The 2013 observations were taken in $4\arcmin \times 4\arcmin$ patches, and the
frequency range covered was 217.5-220 GHz and 216-218.5 GHz in the two spectral
windows. Scans were performed along lines of constant RA and Dec.  These
observations used the same observing strategy and off position as
\citet{Ao2013a}.

The 2014 observing strategy was modified to use larger $8\arcmin \times
8\arcmin$ scans along lines of Galactic latitude and longitude.  The frequency
range was also shifted to cover windows over 217-219.5 and 218.4-220.9 GHz, thus
including the bright \thirteenco and \ceighteeno 2-1 lines.
Three off positions were used for these observations: (a) 17:52:06.854
-28:30:31.32, (b) 17:43:53.890 -28:07:04.68, and (c) 17:48:11.934 -29:44:41.83;
position (b) exhibits some \thirteenco emission from local clouds near 0 \kms
but no other emission lines, while both (a) and (c) appear to be entirely
clean.

Additionally, for the \para data, we incorporated the 41 hours of observations
from \citet{Ao2013a} using the older fast fourier transform spectrometer (FFTS)
backend.  These data covered 2 GHz of bandwidth, including all three of the
\para lines and SiO 5-4, but they did not cover the CO isotopologue lines.

The detected lines were \para \threeohthree 218.22219 GHz, \para \threetwotwo
218.47563 GHz, \para \threetwoone 218.76007 GHz, SiO 5-4 217.10498 GHz,
\methanol \fourtwotwo 218.44005 GHz, OCS 18-17 218.90336 GHz, HC$_3$N 24-23
218.32471 GHz, SO $6_5-5_4$ 219.94944 GHz, HNCO $10_{0,10}-9_{0,9}$ 219.79828
GHz, \ceighteeno 2-1 219.56036 GHz, and \thirteenco 2-1 220.39868 GHz.  All
listed frequencies are rest frequencies, but we do not include many, e.g.,
\methylcyanide 12-11, that are in the observed band but were detected only in
Sgr B2, since complete surveys of the rich molecular heimat in this band have
been compiled elsewhere \citep{Nummelin1998a, Belloche2013a}.  Of the detected
lines, the SO, HNCO, \ceighteeno, and \thirteenco were not covered by the
\citet{Ao2013a} data but were in the newer XFFTS data.  While we do
not analyze these lines in this paper, we provide data cubes that include
them.

The raw data were acquired with 32768 spectral channels in each window,
yielding 0.1 \kms resolution.  We expect to see no lines narrower than a few
\kms in the CMZ, particularly not with the relatively shallow observations we
have acquired.  We therefore downsampled the data by a factor of 8 to 0.8 \kms
resolution prior to resampling onto a 1 \kms grid to make the data more
manageable.  

The system temperature ranged from $120 < T_{sys} < 200$ K for the majority of
the observations, with a mean $T_{sys}=165$ K.  A small fraction (10\%) of
observations were in the range $200 < T_{sys} < 300$ K.  There were also a very
small number ($<1\%$) with much higher system temperatures ($300 < T_{sys} <
750$ K) that were flagged out (see Section \ref{sec:flagging}).

\FigureTwoAA
{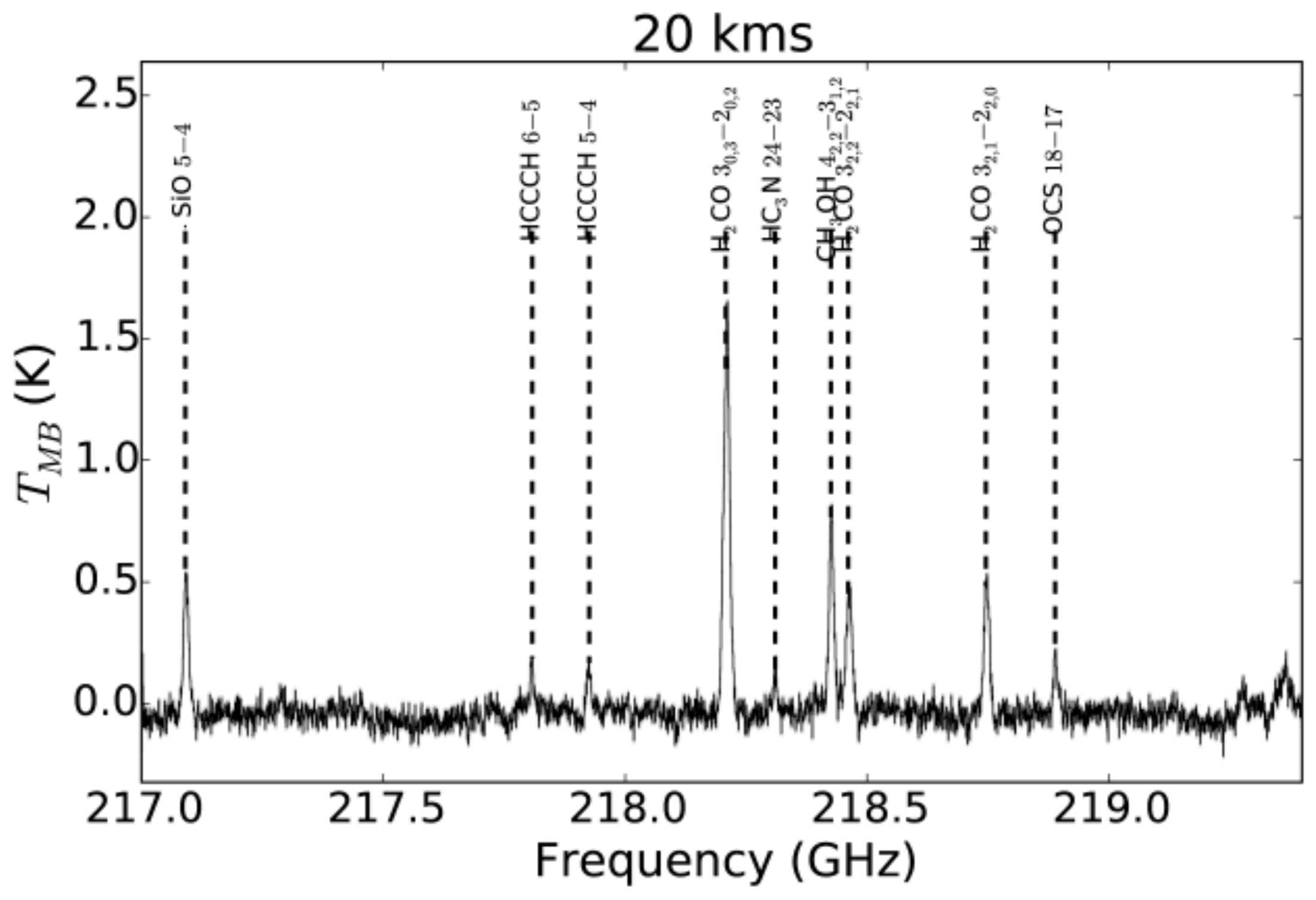}
{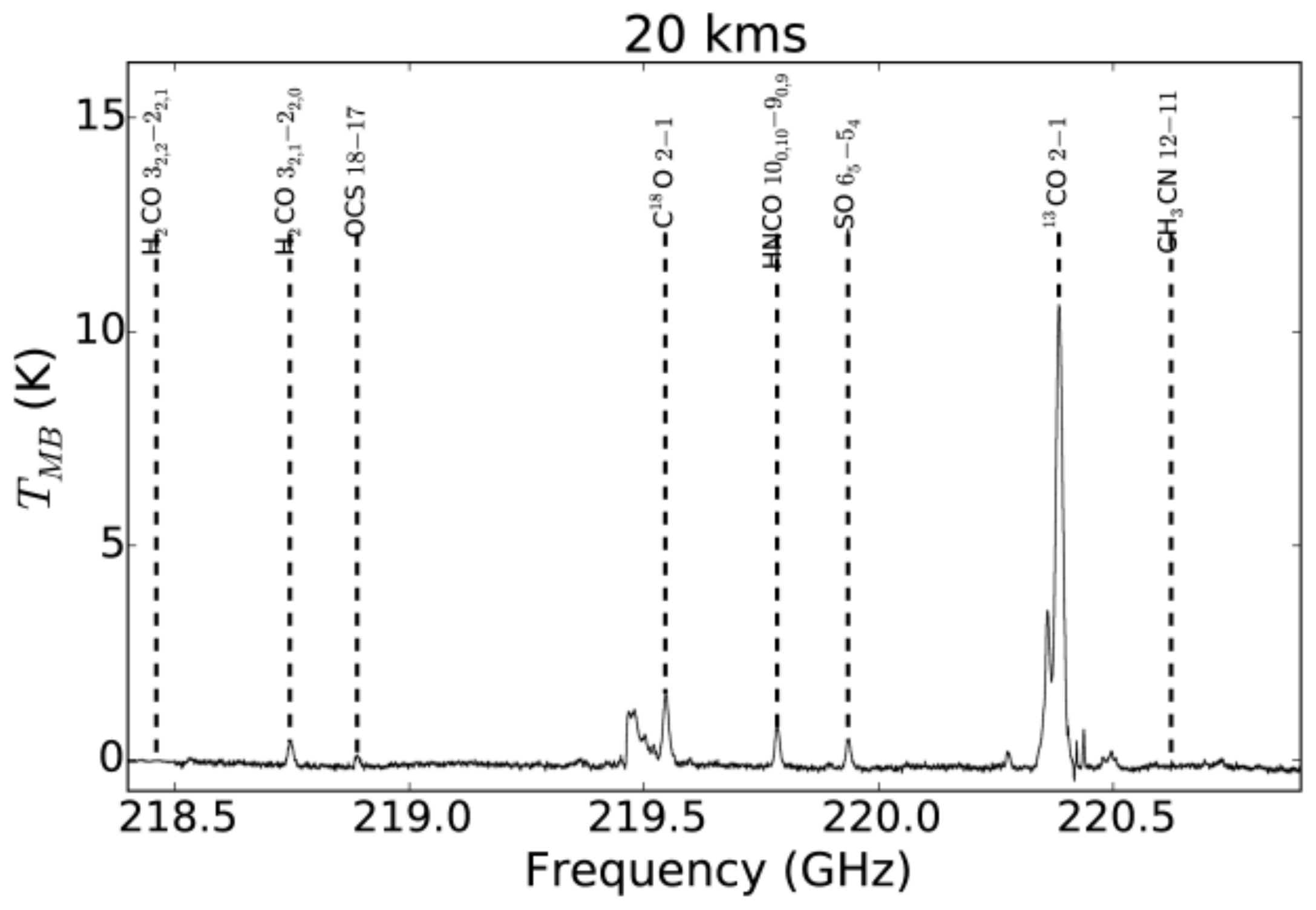}
{Spectra toward one \formaldehyde-bright cloud, the 20 \kms cloud
($\ell=359.892$, $b=-0.074$), over the full APEX band.  All detected lines are
identified.  The feature overlapping the \ceighteeno
2-1
line at about 219.5 GHz is the \twelveco line from the upper sideband,
suppressed by $\sim20$ dB.
There are artifacts at the band edges, e.g., below 217.0 GHz, because these
spectra average a few different tunings without accounting for missing data at
the band edges.}
{fig:fullspectra}{1}{6.5in}

\subsection{Reduction}
\subsubsection{Calibration}
Calibration was performed at the telescope using the standard APEX calibration
tools
\footnote{See the observing manuals:
\url{www.apex-telescope.org/documents/public/APEX-MPI-MAN-0012.png}
\url{www.apex-telescope.org/documents/public/APEX-MPI-MAN-0013.png} }.
These yield flux-calibrated spectra at each position with appropriate pointing
information.  Typical flux calibration uncertainties are $\sim10\%$ and
pointing errors $\sigma\lesssim2$\arcsec. 

There was a significant calibration error discovered at the APEX telescope during
a large segment of the 2014 observing campaign
(\url{http://www.apex-telescope.org/heterodyne/shfi/calibration/calfactor/}).
The calibration errors were of order $\sim15-25\%$.  We corrected the data using
the correction factors suggested on the calibration page, but the remaining
calibration uncertainty is higher in this data, $\sim15\%$ total rather than
the usual $\sim10\%$.

\subsubsection{Flagging bad spectra}
\label{sec:flagging}
Spectra were removed if they showed excessive noise compared to the
theoretical expectation given the measured system temperature.  As in
\citet{Ao2013a}, the threshold was set to $1.5\times$ the theoretical noise
from averaging two polarizations,
i.e. $3 T_{sys} (2 \Delta\nu t_{exp})^{-0.5}$, where $t_{exp}$ is the
exposure time (integration time) per spectrum in seconds and $\Delta\nu$ is the
channel width in Hz.  This approach resulted in $\sim0.2\%$ of the data being
removed.  

At the position of Sgr B2, the noise was significantly higher due to signal
from the continuum source.  We therefore disabled this flagging in a
$2.5\arcmin$ box around Sgr B2.

\subsubsection{Mapmaking}
The maps were made by computing an output grid in Galactic coordinates with
7.2\arcsec pixels and adding each spectrum to the cube at the appropriate
pixel\footnote{We use the term `pixel' to refer to a square data element
projected on the sky with axes in Galactic coordinates.  The term `voxel' is
used to indicate a cubic data element, with two axes in galactic coordinates
and a third in frequency or velocity} location.  In order
to avoid empty pixels and maximize the signal-to-noise, the spectra were added
to the grid with a weight set from a Gaussian with $FWHM=10\arcsec$, which
effectively smooths the output images from $FWHM\approx28\arcsec$ to
$\approx30\arcsec$.  See \citet{Mangum2007a} for more detail on the on-the-fly
mapping technique used here.  The spectra were averaged with inverse-variance
weighting.  The full weight equation is therefore:
\begin{equation}
    W_i = \exp\left(\frac{-\left((x-x_i)^2+(y-y_i)^2\right)}{2\sigma_{pix}^2}\right)
          \frac{1}{\sigma_{rms}^2}
\end{equation}
where $x_i$ and $y_i$ are the coordinate of the pointing in pixel space,
$\sigma_{pix}\approx10\arcsec/\sqrt{8\log{2}}/(7.2\arcsec \textrm{pix}^{-1})=1.38$,
and $\sigma_{rms}$ is the standard deviation measured along the spectrum in a
signal-free frequency region.

The position-position-velocity (PPV) cubes were created with units of
brightness temperature on the corrected antenna temperature scale ($T_A^*$).
The main beam efficiency is $\eta_{mb} = 0.75$ (gain $\sim39$
Jy/K)\footnote{\url{http://www.apex-telescope.org/telescope/efficiency/}}.
These values are noted in the FITS headers of the released data.

The maps achieved a depth of $\sigma=50-80$ mK in 1 \kms channels and
$FWHM=30\arcsec$ beams.  The noise is slightly lower (about 15\%) than expected
from the online APEX calculator because our noise measurements are made in
moderately smoothed maps.

\subsection{Baselining}
\label{sec:baseline}
The data showed significant baseline structure, leading to large-scale
correlated components in the resulting spectra.  The baselines were removed
over a velocity range $-150 < v_{LSR} < 150$ \kms by the
following procedure:
\begin{enumerate}
    \item identify bright PPV regions in the \para \threeohthree line using the
        technique described in Section \ref{sec:signal}
    \item mask the regions identified as \threeohthree-bright for each other 
        spectral line data cube
    \item fit a polynomial to the un-masked data
    \item subtract the fitted polynomial
\end{enumerate}
We used a 7th-order polynomial for this process because lower order baseline
removal left a significant and patterned residual.  A more detailed examination
of the baseline removal process is described in Appendix
\ref{sec:baselineappendix}.

\section{Signal extraction \& masking}
\label{sec:signal}
We use the method described partially in \citet{Ao2013a} and more thoroughly in
\citet{Dame2011b} to mask the data cubes at locations of significant signal in 
the brightest line. 
A noise map was created by computing the sample standard deviation over a
200 \kms range in which no signal was present.
We use the \para \threeohthree line, which is the brightest of the
\formaldehyde lines, to create the mask by
the following procedure:

\begin{enumerate}
    \item Smooth the data with a Gaussian of width two pixels in each direction
        (spatial, $\sigma=14.4\arcsec$, and spectral, $\sigma=2$ \kms)
    \item Create an inclusive mask of all pixels with brightness $T_A >
        2\sigma$ in the smoothed map
    \item Grow the mask from the previous step by one pixel in each direction
        (this is known in image processing as binary dilation)
\end{enumerate}

The \para \threeohthree mask was then applied to the \threetwoone and
\threetwotwo cubes created with the same PPV gridding.  There is some overlap
between the \methanol \fourtwotwo line and the \para \threetwotwo line in PPV
space, so we shifted the \para \threeohthree mask to the velocity of the
\methanol line in the \para \threetwotwo cube and excluded those pixels expected
to contain \methanol emission.
However, inspection of the \threetwotwo cube revealed that significant chunks
of the \threetwotwo signal were excised by this masking, so all further
analysis excludes the \threetwotwo line.

Integrated intensity maps are shown in Figure
\ref{fig:integ303}a for the \threeohthree map masked with the above method
and in Figure \ref{fig:integ303}b for the \threetwoone map
using the mask created from the \threeohthree data.

\RotFigureTwoAA
{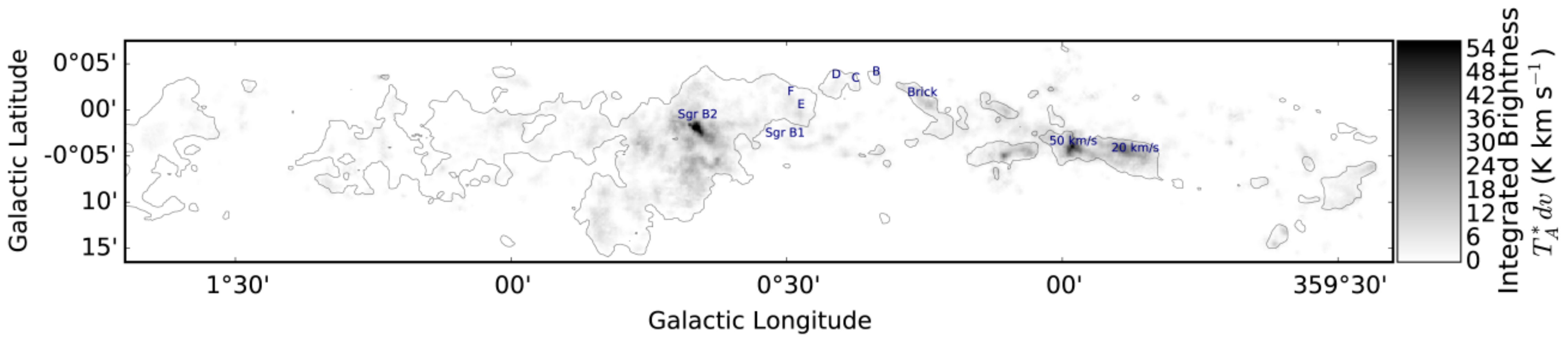}
{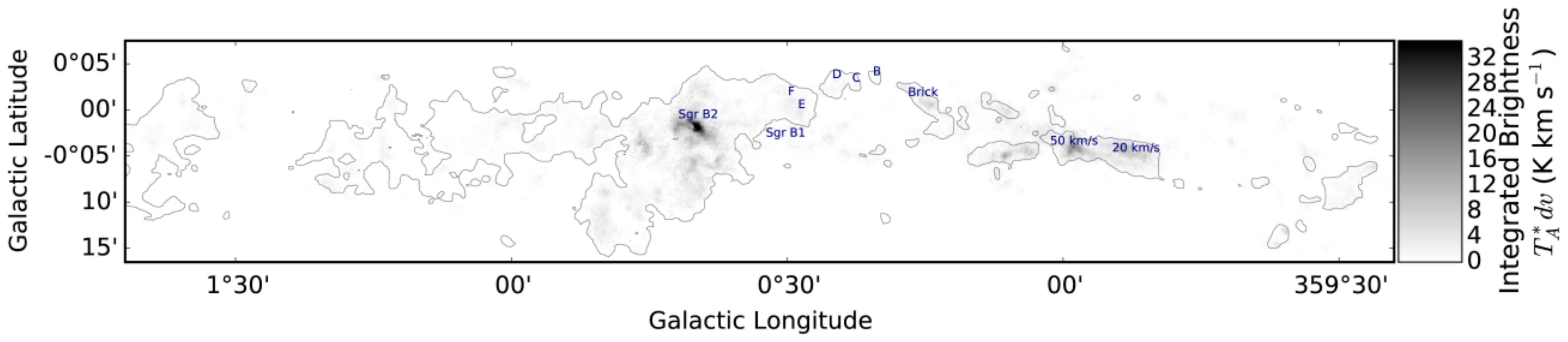}
{Integrated intensity (moment 0) maps of the (a) \para \threeohthree and (b)
\threetwoone lines.  The cubes were masked prior to integration using the
method described in Section \ref{sec:signal}.  Selected region names are
labeled.
 The thin black contours are at a column $N(\hh)=5\ee{22}$
\persc from the Herschel SED fit maps.
}
{fig:integ303}{1}{9.5in}

\subsection{Ratio maps}
\label{sec:ratio}
We have created \Rone ratio maps as a first step toward building temperature
maps.  In general, a higher \Rone indicates a higher temperature, and therefore
the ratio maps are a good proxy for \emph{relative} temperature.  We display
four different varieties of ratio map in Figures \ref{fig:ratiomaps} and
\ref{fig:ratiomapssm}.

Figure \ref{fig:ratiomaps} shows maps produced by creating ratio cubes, in
which each voxel has the value $R_1=\Rone$, then averaging across velocity with
constant weight for each voxel in Figure \ref{fig:ratiomaps}a and with the
weight set by the \threeohthree brightness value in Figure
\ref{fig:ratiomaps}b.  In both cases, the mask described in
Section \ref{sec:signal} was applied to the cubes before averaging over
velocity.

Ratio maps are subject to dramatic uncertainties in the low signal-to-noise
regime, which is inevitably reached in some parts of the map.  The error
distribution on the ratio becomes non-Gaussian as signal approaches zero,
approaching a pathological Cauchy distribution in which the mean becomes
undefined.
We therefore limited our ratio measurements to regions with significance
$>5\sigma$ in the \threeohthree line and used that line as the denominator
in the ratio.  By using the brighter line as the denominator, we avoid
divide-by-zero numerical errors.

\RotFigureTwoAA
{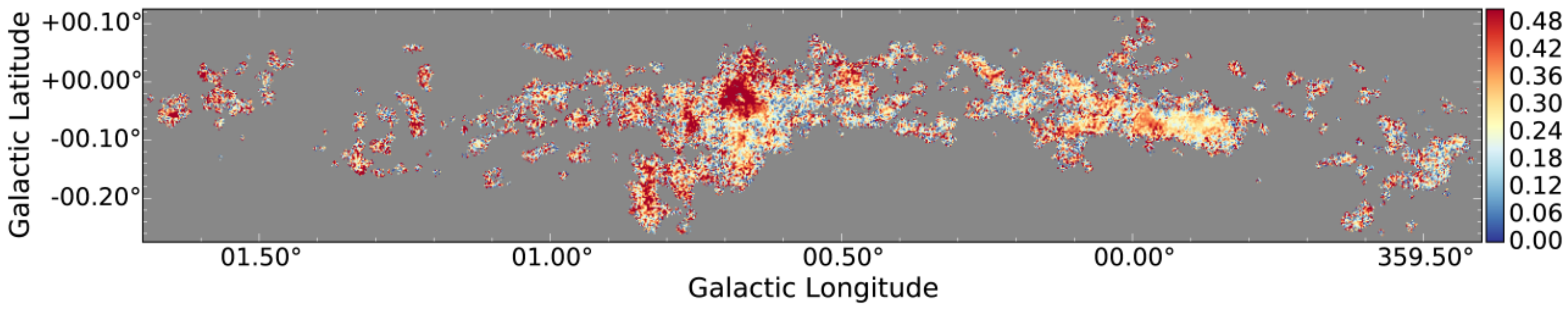}
{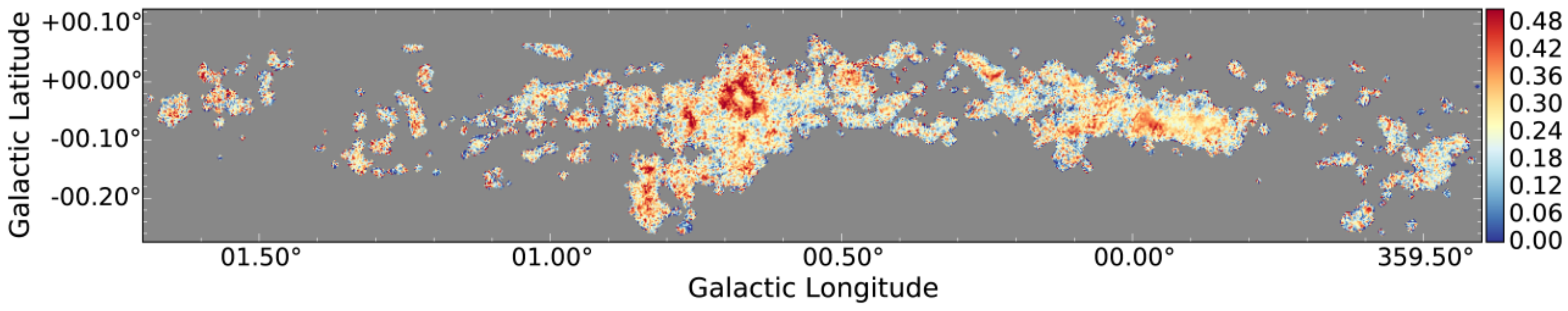}
{(a) Map of the ratio of the integrated emission of
\Rone.  The gray areas indicate pixels that were completely masked out in the
PPV cubes.  The Sgr B2 peaks exhibit \formaldehyde absorption and therefore do
not contain reliable measurements.  Higher ratios
(red) correspond
to higher temperatures.
\newline
(b) The same as (a), but the average along velocity has been
weighted by the \threeohthree brightness
\newline
}
{fig:ratiomaps}{1}{9.5in}

\subsection{Region extraction}
\label{sec:region}
In order to extract higher signal-to-noise measurements on selected regions, we
broke the data down into subsets using both a by-eye region selection drawing
regions using ds9 and a more systematic approach using a dendrogram-based clump
finding algorithm \citep[][\url{http://dendrograms.org/}]{Rosolowsky2008c}.
The dendrogram clump finding uses a watershed decomposition like other clump
finding methods, but it also extracts clumps at different brightness scales
(different heights in the watershed) and tracks how nested clumps are related
to each other.  Within a hierarchy of nested clumps, the brightest and smallest
is a `leaf' and all associated larger structures are `ancestors' (roots or
trunks and branches).  We refer to the clumps identified in this manner as
`dendrogram-extracted' clumps from here on.
Figures 1 and 3 in \citet{Rosolowsky2008c} give visual representations of this
hierarchy.

We ran the dendrogram extraction analysis on the \threeohthree PPV cube data.
We assumed a uniform noise per pixel $\sigma_{T}=\left<\sigma_T\right>=0.07$ K,
which corresponds to the mean noise per field.  We enforced thresholds of a
minimum number of voxels $N_{min}=100$, a peak threshold of $I = 3\sigma$, and
a splitting threshold $\Delta = 2\sigma$, which defines how large the
brightness difference between the peak pixel and an adjacent pixel must be in
order to divide the clump into two sources.  The exact values of these
parameters are not particularly important, as we are interested in general
trends with size-scale and galactocentric distance, but we caution against
overinterpretation of the resulting catalog as the number of sources and their
size and distribution can change dramatically with small changes in the
selected parameters.  The number and location of voxels included in the catalog
is, however, relatively robust against small parameter changes.

For each extracted clump, we measured the corresponding integrated and peak
emission in the \threeohthree, \threetwoone, and \thirteenco cubes.  We also
extracted the mean dust temperature and column density from SED fits to
Herschel HiGal 170-500\um maps \citep{Molinari2010a,Traficante2011a}.  The SED
fits were performed on background-subtracted data using an approach originally
described in \citet{Battersby2011a}, which includes details of the fitting
process and specification of assumed physical parameters.  We also extracted
dust column density and temperature using a more naive pixel-by-pixel approach
with no background subtraction
(\url{http://hi-gal-sed-fitter.readthedocs.org/en/latest/}) and found the two
to be consistent.  We assumed a dust to gas ratio of 100 and dust opacity
$\kappa_{505 GHz} = 4$ g \persc, the same as \citet{Battersby2011a} and
equivalent to that assumed in the Bolocam Galactic Plane Survey observations of
the CMZ with $\beta=2$ \citep[$\kappa_{271.1 GHz}=1.1$ g
\persc][]{Bally2010a, Aguirre2011a, Ginsburg2013a}. 

\subsubsection{Manual region extraction}
\label{sec:byeye}
The by-eye selected regions were extracted less systematically, instead
focusing on peaks in the \threeohthree or \threetwoone emission.  These
selections were made in two dimension rather than in the full PPV space.  We
included a series of large square $8\arcmin \times 8\arcmin$ apertures
corresponding to each observed field.  For each region extracted in this
fashion, we fit a 1-3 component model as described in Section
\ref{sec:spectralmodeling}, where each component consists of a 6-parameter
multi-Gaussian model.  An example spectrum extracted from a small region around
``cloud c'' is shown in Figure \ref{fig:cloudcspec}.  The extracted regions are
shown in Figure \ref{fig:regions}.

\section{Analysis: temperature determination}
\label{sec:analysis}
In this Section, we describe the techniques used to derive gas temperatures.
In Section \ref{sec:linemodeling} we model the line ratios with a radiative
transfer code.  We fit Gaussians to spectra extracted from the cubes and use
these fits to measure line ratios in Section \ref{sec:spectralmodeling}.  In
Section \ref{sec:dendromod}, we apply the same modeling approach to the
dendrogram-extracted clumps from Section \ref{sec:region}.  We use the modeling
results from Section \ref{sec:spectralmodeling} and \ref{sec:dendromod} to
justify a simplified modeling approach that we apply to the entire map of the
CMZ in Section \ref{sec:formaldehydetemmap}.

\subsection{Radiative transfer modeling}
\label{sec:linemodeling}
We use RADEX \citep{van-Der-Tak2007a} and a related solver (Fujun Du's
\texttt{myRadex}; \url{https://github.com/fjdu/myRadex}) to create model grids
for the p-\formaldehyde lines over a grid of 20 densities ($n=10^{2.5}$-$10^{7}$
\percc), a fixed assumed\footnote{The line gradient is selected because
observed line widths have FWHM larger than 5 \kms in 30\arcsec (1.2 pc) beams
throughout the CMZ.  The temperature measurements are
robust against this assumption, with a typical change of $dT/T < 25\%$ for a
gradient $4\times$ larger or smaller.} line gradient of 5
\perkmspc, 30 \formaldehyde column densities $N(\formaldehyde) =
1\ee{11}-1\ee{15.1}$ \persc, and 50
temperatures from 10 to 350 K.  The grid was created using the \texttt{pyradex}
wrapper of RADEX\footnote{\url{https://github.com/adamginsburg/pyradex}} and
collision rates retrieved from LAMDA \citep{Wiesenfeld2013a}.  We then
upsampled these grids by spline interpolating between grid points to acquire a
high-resolution ($T_G\times n \times N = 250\times100\times101$) grid covering
$10<T_G<350$ K, $10^{2.5} < n < 10^7$ \percc, and $10^{11} < N < 10^{15.1}$
\persc\perkmspc.  For our analysis, we used the \texttt{myRadex} grid; it is
consistent with the \texttt{RADEX} grid over most of parameter space but
suffers fewer numerical instabilities over the region we are interested in.

\subsection{Spectral modeling}
\label{sec:spectralmodeling}
We extracted spectra averaged over selected regions and fitted
a 6-parameter model to the full 218-219 GHz spectral range using
\texttt{pyspeckit} \citep{Ginsburg2011c}.  The fitted parameters are the
amplitude of the \formaldehyde \threeohthree line, the centroid
velocity\footnote{We assume the \methanol and \formaldehyde lines have the same
centroid velocity, and we find no counterexamples to that assumption within our
data.}, the line width ($\sigma$), the ratio $R_1 = \Rone$, the ratio $R_2 =
\Rtwo$, and the
amplitude of the \methanol \fourtwotwo line\footnote{The parameters $R_2$ and
the \methanol brightness are essential for fully fitting the spectrum, but they
are not used for further analysis.}, where $S_\nu$ is the surface
brightness.  For these spectra, we performed a different baseline subtraction
to that described in Section
\ref{sec:baseline}.  We first fit the described 6-parameter model, then fit a
spline curve to the residual smoothed to a 50-channel (50 \kms) scale.  We then
re-fit the model to the data.  In all but the lowest signal-to-noise cases,
this approach resulted in consistent ratio measurements between the two fits
and with the other methods.  Figure \ref{fig:twospectra_hotcold} shows two
example spectra that exhibit significantly different ratios with the spline
baseline fit indicated.  We use the baseline-subtracted measurements for
further analysis.

\FigureTwo
{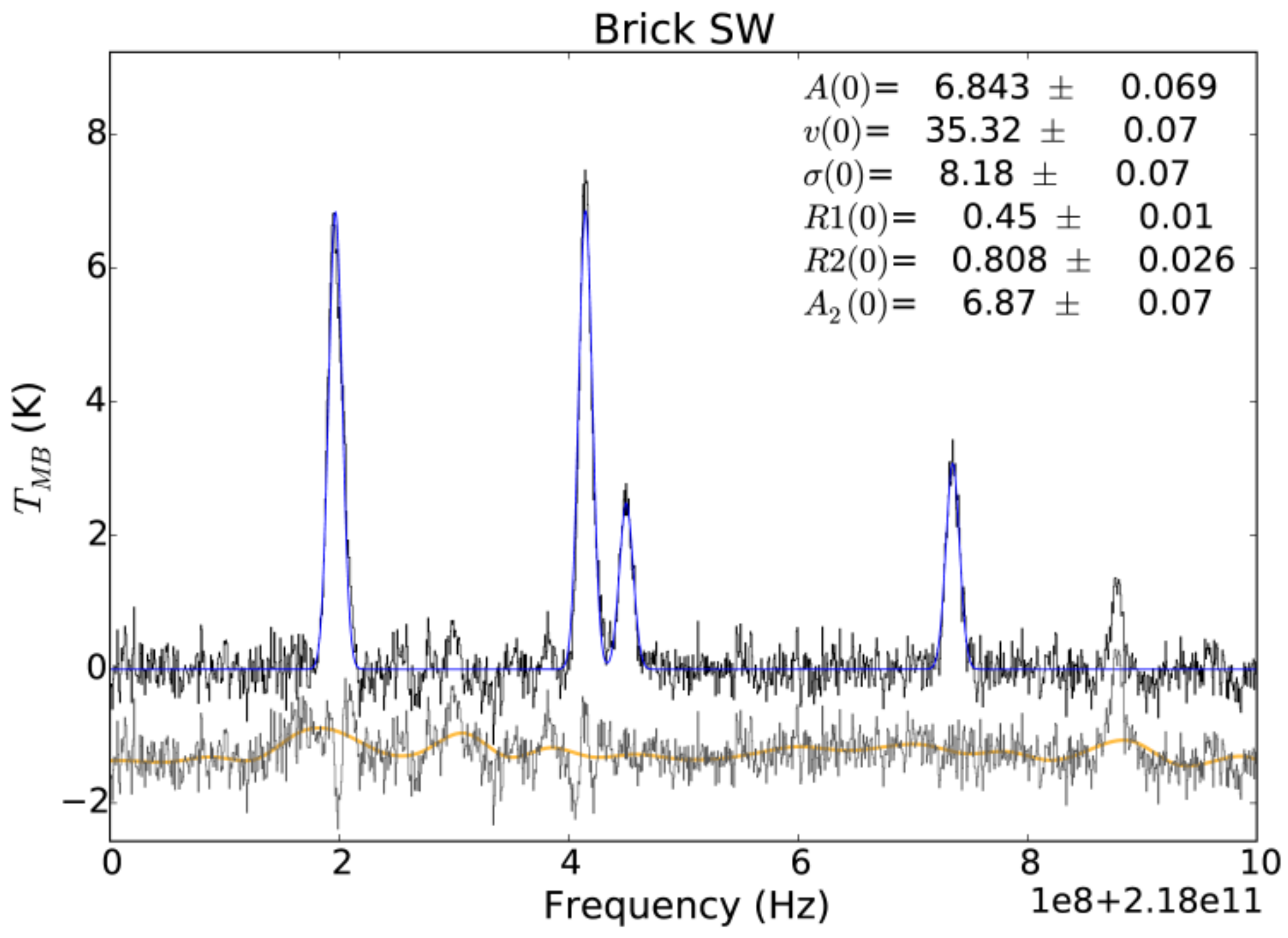}
{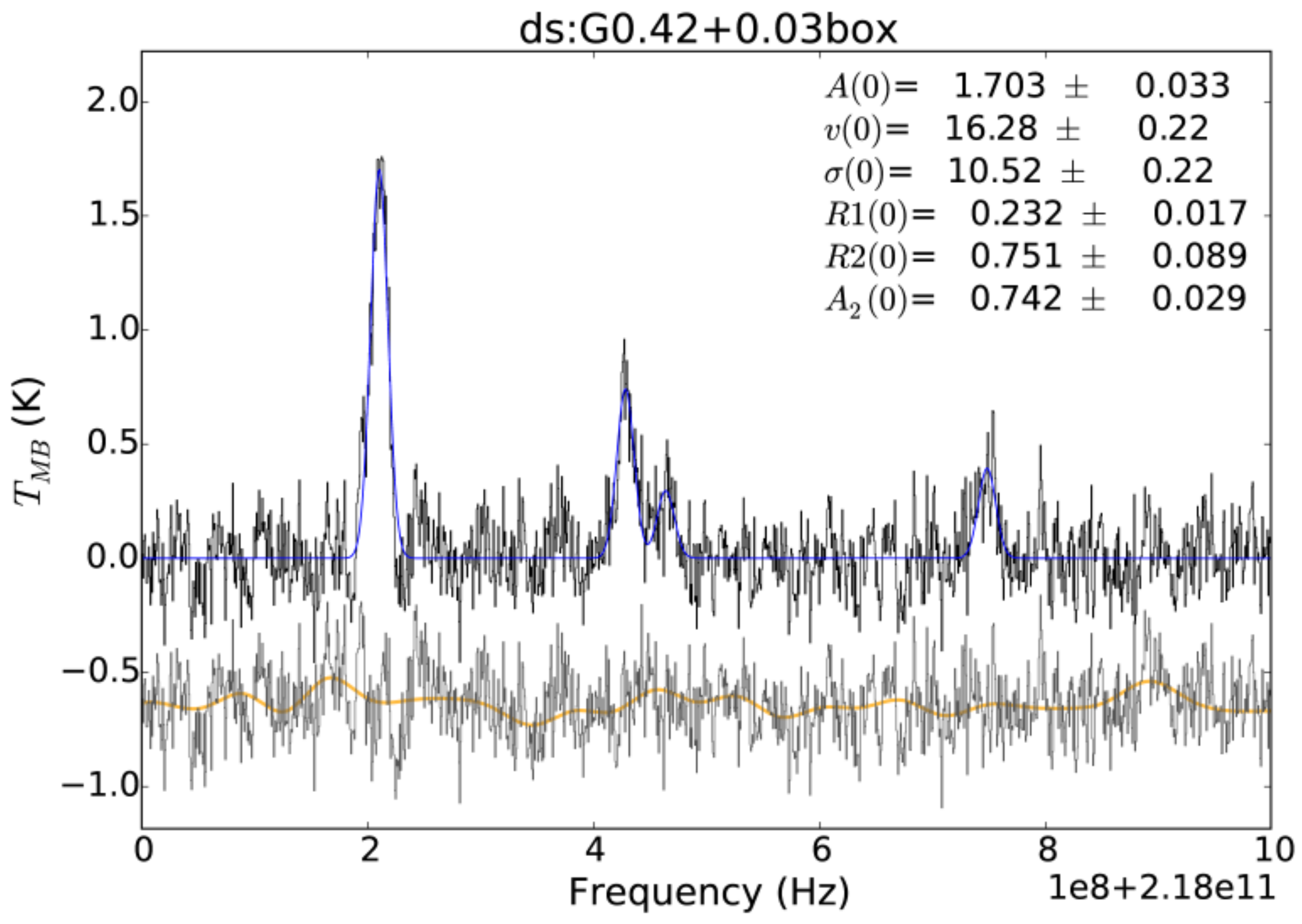}
{
Simultaneous fits to four lines, from low to high frequency, \para
\threeohthree, \methanol \fourtwotwo, \para \threetwotwo, and \para \threetwoone
for (a) The Brick ($\ell=0.241$, $b=0.006$, $r=25\arcsec$)and (b) a box
centered on G0.42+0.03 ($\ell=0.415$, $b=0.032$,
$\approx80\arcsec\times80\arcsec$).  The fitted parameters
shown in the legend are the amplitude of the \formaldehyde \threeohthree line,
the centroid velocity, the line width ($\sigma$, not FWHM), the ratio $R_1 =
\Rone$, the ratio $R_2 = \Rtwo$, and the amplitude of the \methanol \fourtwotwo
line.  These spectra have significantly different ratios and therefore derived
temperatures: The Brick has $T_G\gtrsim100$ K while G0.42+0.03 has
$T_G\approx40-50$ K.  The bottom black spectrum is the residual, and the orange
wiggly spectrum shows the spline fit used to remove baseline ripples.
The OCS 18-17 218.90336 GHz and HC$_3$N 24-23 218.32471 GHz lines are also
detected, but not fitted, in The Brick.
In these examples, we are showing cases where the \methanol \fourtwotwo and
\para \threetwotwo lines are not too badly blended, but along many lines of
sight, especially those within $\sim20\arcmin$ of Sgr B2, they are
irretrievably blended.
}
{fig:twospectra_hotcold}{1}{3in}

For each extracted region (see Section \ref{sec:region}), we created $\chi^2$
grids over the modeled temperature-density-column density parameter space.  We
measured $\chi^2$ with five independent constraints: the line ratio $R_1$, the
\formaldehyde abundance, the total column density of \hh, a lower limit on the
\hh volume density, and an upper limit on the beam filling
factor\footnote{Where the model predicts a brightness temperature higher than
observed, we can invoke a low filling factor to explain the discrepancy.  If
the model predicts a brightness temperature too low,
we can rule it out.}.  We use the
line \emph{ratio} as the primary constraint rather than line brightnesses to
avoid uncertainties due to the filling factor of the emitting gas: even for
diffuse clouds, the filling factor of the emitting regions may be $ff<<1$ if
the emission is isolated to compact shocked regions, as expected if highly
supersonic $\mathcal{M}>10$ turbulence is energetically dominant in the clouds.
We use an abundance $N(\para)/N(\hh) = X(\para) =
10^{-9.08\pm1}$, allowing for more than two orders of magnitude variation in
the \formaldehyde abundance
\citep{Ginsburg2013a, Carey1998a, Wootten1978a, Mundy1987a}.  To constrain the
total column density, we use the Herschel dust maps (see Section
\ref{sec:region}) to derive an \hh column
density, which has a nominal $\sim2-3\times$ uncertainty\footnote{The
statistical uncertainty is much smaller than the systematic unknowns, including
the dust opacity per unit mass, the presence of multiple line-of-sight cloud
components, and the presence of multiple temperature components within any
given cloud.}.  We treat the error as $10\times$ (instead of $\sim2-3\times$)
to account for the additional uncertainty in the abundance of \para.
The effective range of $N(\para)$ is therefore $10^{11.5} < N < 10^{15.5}$
\persc, allowing for $10^{21.5} < N(\hh) < 10^{23.5}$ \persc and $\pm1\sigma$
in log abundance.
To
constrain the dust density, we assume the selected area has a mass given by the
Herschel dust-derived mass and a spherical volume, which sets a conservative
(but nonetheless highly constraining) lower limit on the volume density.

These constraints are shown projected onto the three planes of our fitted
parameter space in the multi-paneled Figures
\ref{fig:coltemconstraints}-\ref{fig:parsonbrightness} for one example region.
The fitted parameters are displayed in Figure \ref{fig:parsonbrightness}.

While $R_2$ can provide significant constraints on various parameters, we do
not use it for our general analysis because of the ambiguity imposed by the
overlap with the \methanol \fourtwotwo line.  Blending between the \methanol
\fourtwotwo and \formaldehyde \threetwotwo line occurs frequently, in $\sim1/4$
of all sightlines.

\Figure{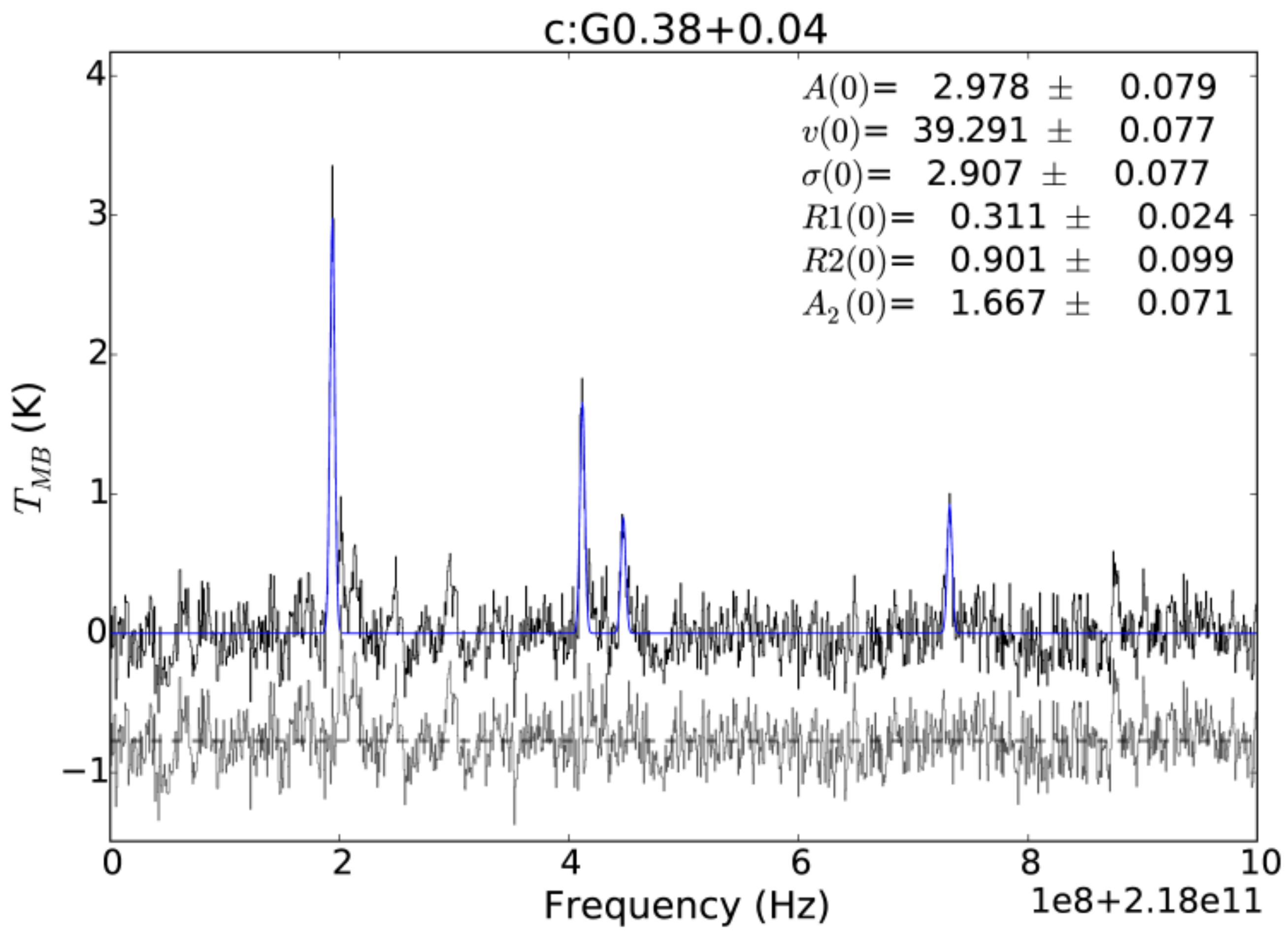}
{Fitted spectrum of ``cloud c".  The fitted parameters and their corresponding
errors are shown in the legend as in Figure \ref{fig:twospectra_hotcold}.}
{fig:cloudcspec}{0.5}{3.5in}

\subsection{Clump modeling}
\label{sec:dendromod}
We derived temperatures for each dendrogram catalog clump using the same
measured quantities as in Section \ref{sec:spectralmodeling} as constraints.
The $\chi^2$ grid described in Section \ref{sec:linemodeling} was used to
extract the temperature and associated error.

We plot the derived temperature for all clumps against the measured line ratio
in Figure \ref{fig:ratiovstem}.  This figure includes both `leaves' (the most
compact of the nested clumps) and `ancestors' (any clump that contains a
smaller clump within it)
from the dendrogram extraction.  The figure illustrates the effects of assuming
higher or lower density on the temperature; the derived temperature is only weakly dependent
on the assumed density. The lower density models are excluded by
the observed volume density lower limit (assuming spherical geometry) from the
dust.  We use an assumed density $n(\hh)=10^4$ \percc when a uniform assumed
density is required in further analysis.

\Figure{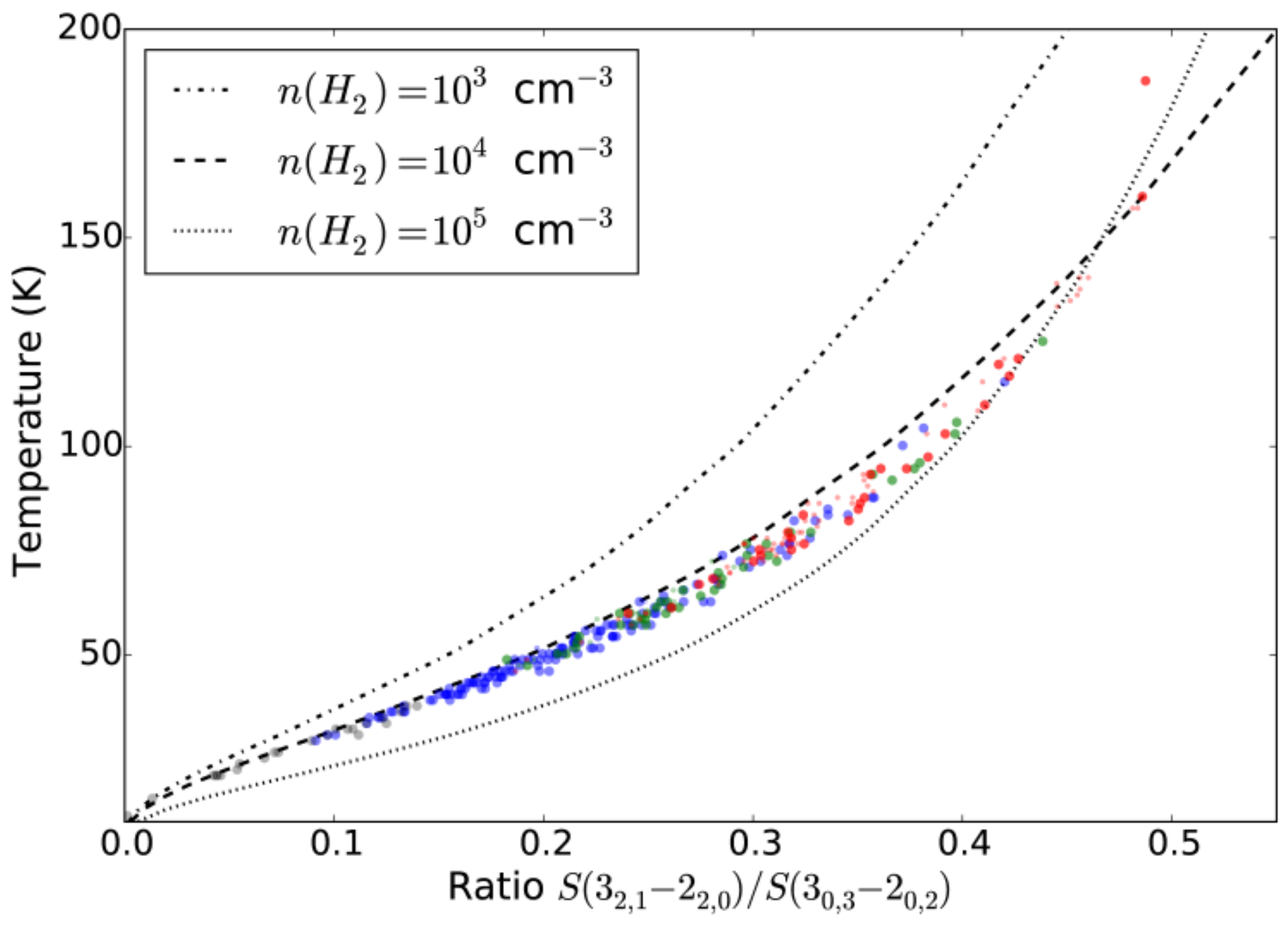}
{The derived temperature vs the measured ratio \Rone for the dendrogram clumps.
The big symbols represent leaves (compact clumps) and the small symbols
represent larger and therefore lower-density ancestor structures.
The temperature includes constraints from the assumed constant \formaldehyde
abundance and the 
varying column density.  The points are color coded by signal-to-noise in the
ratio $R_1$, with gray $S/N < 5$, blue $5 < S/N < 25$, green $25 < S/N < 50$,
and red $S/N > 50$.  The black lines show the modeled temperature as a function of
$R_1$ for three different assumed densities with a constant assumed abundance
$X_{\formaldehyde} = 1.2\ee{-9}$.
While the measured values appear consistent with the $n=10^4$ \percc curve, the
systematic errors permit any density in the range $10^4~\percc < n < 10^5
~\percc$; this plot most importantly shows that $n\sim10^3$ \percc is not
plausible and that the temperature does not differ very much for a given value
of $R_1$ over the plausible range of densities.
}
{fig:ratiovstem}{0.5}{3.5in}

\subsection{Temperature maps}
\label{sec:formaldehydetemmap}
Using the relations shown in Figure \ref{fig:ratiovstem}, we converted the
ratio map shown in Figure \ref{fig:ratiomaps} to a temperature map in Figure
\ref{fig:temmapnosmooth} with a constant abundance $X(\formaldehyde)=1.2\ee{-9}$ 
and two constant densities $n=10^4$ (Figure \ref{fig:temmapnosmooth}a/b) and
$n=10^5$ \percc (Figure \ref{fig:temmapnosmooth}c).  Because this map includes
the full line-of-sight integrated emission, there are many regions where
multiple independent components are being mixed.

These temperature maps do not have uniform reliability.  There is some spatial
variation in the noise, but the variation in the signal is much more important.
In regions where the peak signal is low, the reliability of these maps is
substantially reduced.  We have also applied a mask to the lower panels of
Figure \ref{fig:temmapnosmooth} to make the low-reliability regions blend into
the background by adding a foreground gray layer with opacity set by the
inverse S/N (see Appendix \ref{appendix:smoothed}).

The effective range of detected temperatures is $\approx40-150$ K. While lower
and higher temperatures appear in the map, those measurements should not be
trusted.  Below $T_G\lesssim40$ K, all temperature measurements in the map are
effectively $1\sigma$ upper limits\footnote{When averaged into regions in later
sections, we measure some temperatures below this limit, albeit still at low
significance.}.  This
limit is due to a combination of sensitivity and excitation: below 40 K, the
\para lines are observed to be faint; the \threetwoone line is not detected.
In regions of high column density, we detect both lines, and in all of these
regions, we have measured high temperatures ($T_G>40$ K).  
Since $T_G\sim40$ K corresponds to $R_1\sim0.15$, a $5\sigma$ temperature
measurement would require $T_B(\threetwoone)=(5\times0.07\mathrm{K}) = 0.35$ K
and
$T_B(\threeohthree) = (5 \times 0.07 \mathrm{K}) /
0.15 = 2.33$ K.  This $T_B(\threeohthree)$ is slightly brighter than the
brightest detection in our map,
and all regions detected with such high brightness
temperatures also turned out to have higher $T_B(\threetwoone)$, and therefore
higher $R_1$ and temperature.
The higher measured temperatures
become lower limits above $\sim150$ K because the \para line ratio is
intrinsically insensitive to higher temperatures, as can be seen in Figure
\ref{fig:parsonbrightness} (rightmost column), in which the modeled line
brightnesses are both nearly constant above $T_G\gtrsim150$ K.

\RotFigureThreeAA
{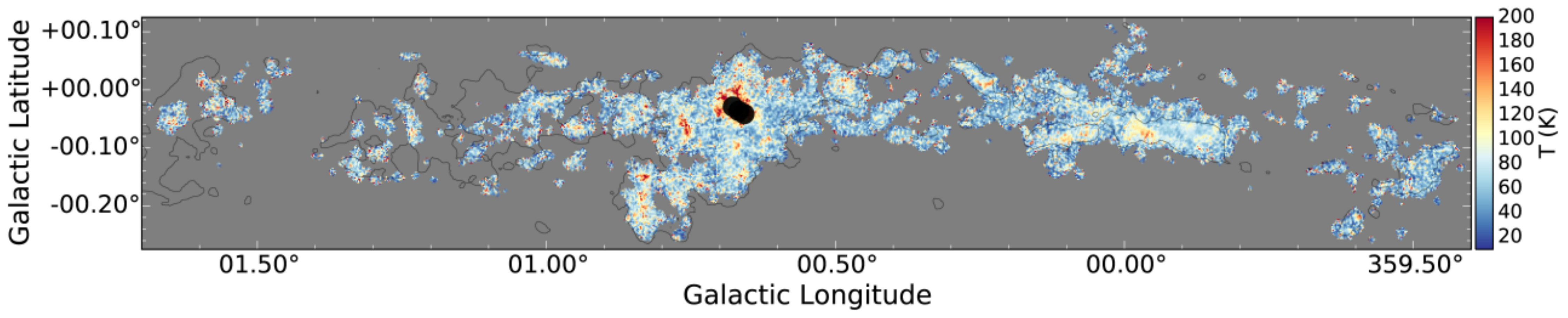}
{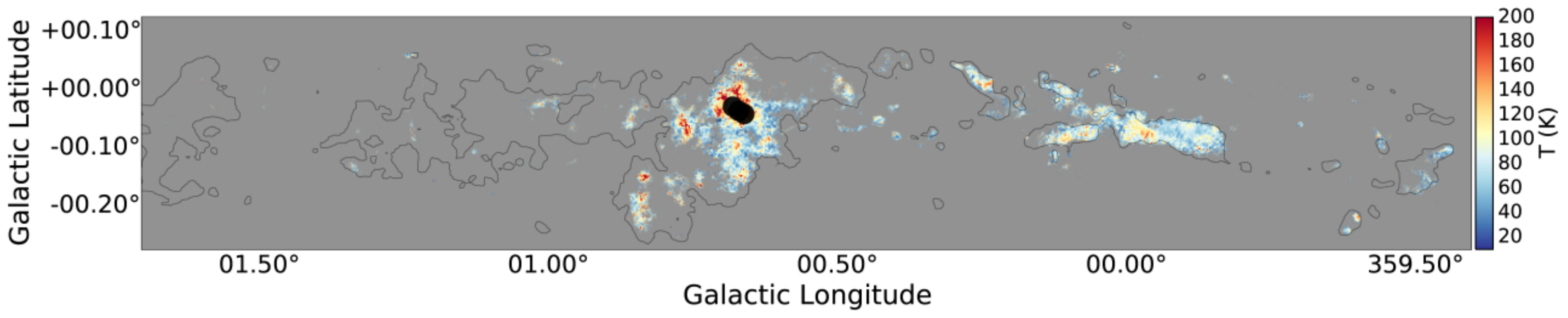}
{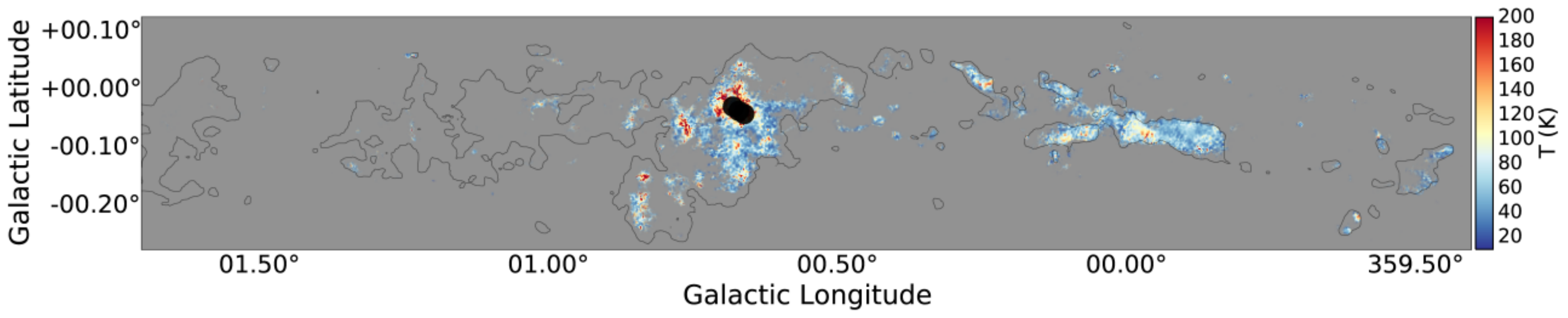}
{Temperature maps assuming fixed $X_{\para}=1.2\ee{-9}$
and $n(\hh)$.  The maps in (b) and (c) are masked by signal-to-noise such that
regions that have lower signal-to-noise are more gray.  The thin black contours
are at a column $N(\hh)=5\ee{22}$ \persc
from the Herschel SED fit maps.
(a) $n(\hh) = 10^4$ \percc, no mask;
(b) $n(\hh) = 10^4$ \percc;
(c) $n(\hh) = 10^5$ \percc
}
{fig:temmapnosmooth}{1}{9.5in}

\RotFigureTwoAA
{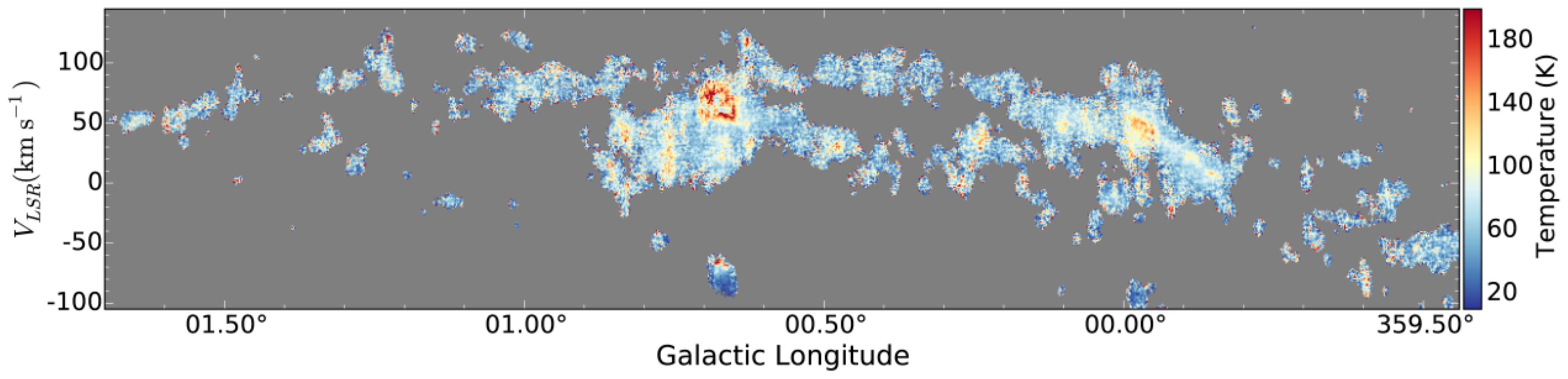}
{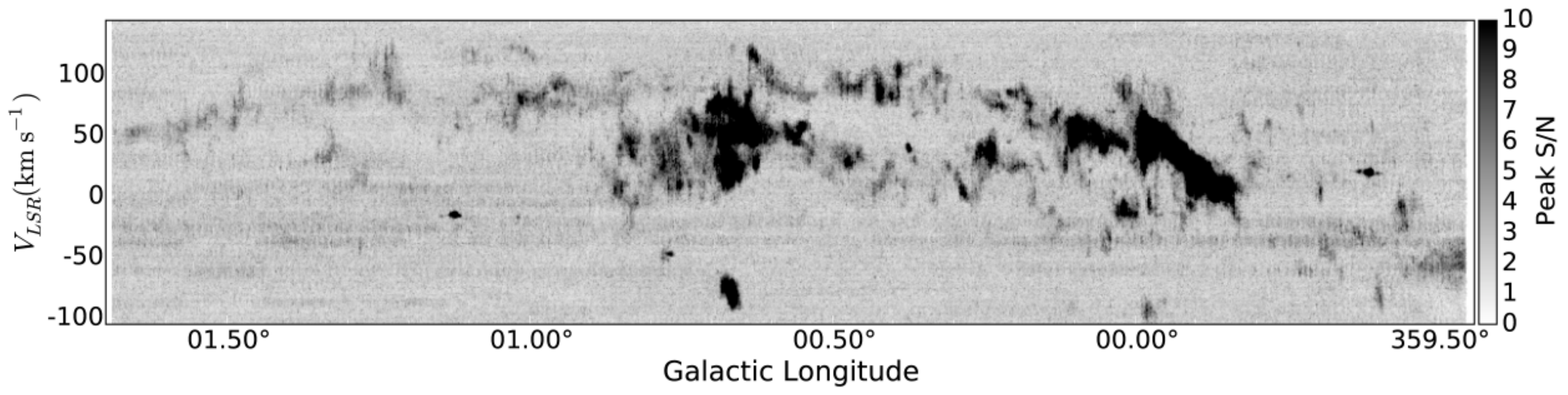}
{(a) A Galactic longitude-velocity diagram made by taking the
\threeohthree-weighted average of the $R_1$ PPV cube along Galactic latitude
and using the $n=10^4$ \percc curve
from Figure \ref{fig:ratiovstem} to derive temperature.  (b) A Galactic
longitude-velocity map of peak signal-to-noise along Galactic latitude, which
is used as a mask in (a).  The apparently cold clumps at -80 \kms in Sgr B2 and
-90 \kms in Sgr A are from a different line, HC$_3$N 24-23: the gas
temperatures from these positions are meaningless.  The apparent high
temperatures around the cloud edges are artifacts caused by lower
signal-to-noise ratios at the edges and should be ignored.
}
{fig:pvtem}
{1}{9.5in}

\clearpage
\section{Discussion}
\label{sec:discussion}
\subsection{Gas heating and thermal balance}
\label{sec:heating}
Following \citet{Ao2013a} and \citet{Papadopoulos2010a}, we examine the energy
balance in the CMZ gas to understand the temperatures we measure.  The two key
processes they identified are cosmic ray and turbulent dissipation (shock)
heating.  We ignore all UV-driven heating processes in this section as we are
interested in gas at high extinction ($A_V > 10$)\footnote{While some emission
likely comes from PDRs, we expect these regions to have small filling factors
and typically lower densities than cloud centers.  At densities $n<10^4$
\percc, \para 3-2 emission is very weak and would contribute little to the
observed line emission.  Volumes within the actively star-forming regions
Sgr B2, Sgr C, and near the circumnuclear disk may be locally affected
by strong UV heating from young and forming stars, but the bulk of the dense
gas is not.  Finally, PDR models \citep[e.g.,][]{Hollenbach1999a} predict that
$T_{D} \approx T_{G}$ at high densities, while we show that $T_{G} >
T_{D}$; see Section\ref{sec:td_ne_tg}.}.
Other processes, e.g.,  direct (proto)stellar heating
or supernova bubble interaction may be relevant locally, but cannot heat the
gas globally.  \citet{Ao2013a} ruled out diffuse X-rays as a significant
heating source, with an energy density $\sim3$ orders of magnitude too low to
explain observed temperatures.

We determine the expected gas kinetic temperature by assuming that the heating
and cooling processes are in equilibrium.  This is generally a good assumption
on the scales we observe, as typical gas cooling rates via line or dust cooling
around $T_G\sim100$ K and $n\sim10^5\percc$ are $dT/dt \sim 10^{-20}
\mathrm{erg~s}^{-1}\percc \sim 23 (n/10^5 \percc) \mathrm{K~kyr}^{-1}$
\citep[][and see below]{Goldsmith2001a}.  The cooling timescale is shorter than
the timescales for any energy injection mechanisms that may plausibly heat the
gas.  The supernova rate is $<1$ kyr$^{-1}$ over the whole CMZ, and supernova
remnants stop producing cosmic rays rapidly, so the supernovae are
expected to contribute to a diffuse cosmic ray population rather than local
stochastic heating incidents.  Turbulence in these clouds is dynamically driven
on scales of 10-30 pc \citep{Kruijssen2015a}.  Assuming the turbulent
dissipation scale is close to the sonic scale with a sound speed $c_s = 0.76
(T_G/100\mathrm{K})^{1/2}$ \kms, the gas temperature will be well mixed on milliparsec
spatial scales on kiloyear timescales.

We expand upon the \citet{Ao2013a} analysis by solving for the temperature
assuming that both cosmic ray ionization and turbulent heating are present and
that the dust temperature is nonzero.  We assume the heating rate for cosmic
rays is constant per cubic centimeter $\Gamma_{CR} = n \zeta_{CR}$, and the
heating rate from turbulence $\Gamma_{turb} = n \sigma^3 / L$, where $\sigma$
is the 3D velocity dispersion and $L$ is the length scale.  We used
\texttt{DESPOTIC} \citep{Krumholz2014c} to determine the equilibrium
temperature.  We included CO, O, C+, C, o-\hh, p-\hh, and HD\footnote{The
collision rate files for \hh and HD were provided by Wing-fai Thi from work by
the ProDiMo team from \citet{Le-Bourlot1999a, Wrathmall2007a, Flower2000a,
Wolniewicz1998a}.
Input abundances were $X(CO)=10^{-4}$, $X(\thirteenco)=5\ee{-7}$,
$X(\ceighteeno)=5\ee{-8}$, $X(C)=5\ee{-7}$, $X(C+)=5\ee{-10}$, $X(O)=5\ee{-6}$,
$X(HD)=10^{-5}$, and the ortho/para ratio of \hh $OPR=3$.} as line
coolants\footnote{We assume a very low abundance for C+, $X_{C+}=5\ee{-10}$,
assuming that cloud shielding will prevent photoionization and collisions with
\hh will destroy CR-generated C+ rapidly, which is confirmed by the
\citet{Nelson1999a} chemical model.  \citet{Lis1998a} and \citet{Lis1999a}
observed bright C+ emission from The Brick; we interpret this emission as
coming from a heated outer skin of the cloud at lower densities not coincident
with the observed \formaldehyde emission.}.  We used the large velocity
gradient (LVG) approximation for optical depth calculation.  The resulting
curves for $T_G(\sigma_v)$ are shown in Figure \ref{fig:temvsfwhm_regions} (see
Appendix \ref{sec:despoticplots} for further details).  We
have assumed a size scale $L=5$ pc, a gradient $dv/dr = 5$ \kms \perpc, and a
cosmic ray ionization rate (CRIR) $\zeta_{CR}=10^{-17}$ \pers, with exceptions
noted in the legend.  We assume a dust temperature $T_D=25$ K, but we assume
lower dust radiation temperatures because the dust is optically thin at long
wavelengths in most regions.  We have assumed a constant interstellar radiation
field (ISRF) of $G_0$, which should only weakly affect dust temperatures and
has no effect on the gas at the high column densities being modeled.  A higher
ISRF may be appropriate but does not change our results, since a 1000 $G_0$
field provides the same gas heating rate as a $\zeta_{CR}=2.5\ee{-17}$ \pers,
a rate much lower than the measured CRIRs in the CMZ.
\citet{Clark2013a} similarly find that the ISRF has little effect on the gas
temperature.

A first key point is that we measure gas temperatures $T_G\sim60$ K in many
clouds everywhere throughout the CMZ.  Since cosmic rays have very high
penetration depths, we expect that cosmic ray heating is uniform
everywhere within the CMZ, independent of column density\footnote{However,
low-energy cosmic rays will be affected by the strong magnetic fields in the
CMZ.  Our naive assumptions gloss over much of the physics of cosmic ray
interaction with molecular cloud gas, but they are consistent with typical
assumptions adopted in the literature and are necessary for a constant cosmic
ray ionization rate (a $\zeta_{CR}$) to be meaningful.}.  If this uniformity
holds, then we can set an upper limit on the cosmic ray ionization rate,
$\zeta_{CR} \lesssim 10^{-14}$ \pers, since a higher rate would result in a
minimum gas temperature above what we have measured.  The limit is consistent
with many previous measurements of the CRIR, though
it rules out some of the more extreme values \citep{Yusef-Zadeh2013b,Goto2013a}.
For a fiducial upper limit measurement, we use $n=10^{4}$ \percc, $dv/dr=5$
\kms \perpc, $T_{rad,D}=10$ K, and $\zeta_{CR} = 10^{-14}$ \pers, which
yields $T_{G}=72$ K in the absence of any other heating mechanisms.  This
upper limit could be violated if the gas in all observed clumps is about an
order of magnitude denser than the dust-derived average density, which is not
likely.  This limit leads to the conclusion that cosmic ray heating is either
not dominant in the CMZ or not uniform, as uniform cosmic ray heating cannot
satisfy the requirement for both high ($>100$ K) and low ($\lesssim40$ K)
temperatures to be present in the CMZ.

We have information on the linewidth, size scale, and density for each
individual region, albeit with great uncertainties on each. To assess the role
of heating by turbulent energy dissipation,  we use these measurements to
predict the gas temperature.  We use the effective radius,
which is the geometric mean between the moment-derived major and minor
axes\footnote{The effective radius is often computed as the square root of the
total area (in pixels), but \citet{Rosolowsky2006a} show that this is usually a
very bad approximation, so we use the Gaussian-corrected (see next footnote)
moment-derived radius.}, of the selected sources as the length scale, the
measured linewidth\footnote{We correct the linewidth and the radius of
dendrogram-extracted clumps for clipping assuming the underlying source is
Gaussian following \citet[][Appendix B]{Rosolowsky2005b}.}
$\sigma_v$, the
measured dust temperature $T_{D}$ scaled down by an optical depth correction
$1-\exp\left[-N(\hh)/10^{24}~\persc\right]$, and we assume the velocity gradient
$dv/dr = 5$ \kms \perpc.  Of these measurements, the linewidth $\sigma_v$ is
dominant in determining the predicted temperature and is the least uncertain.
This analysis is performed both for the hand-selected apertures, which have
arbitrary radii but well-measured linewidths, and the dendrogram leaf sources.

\FigureTwo
{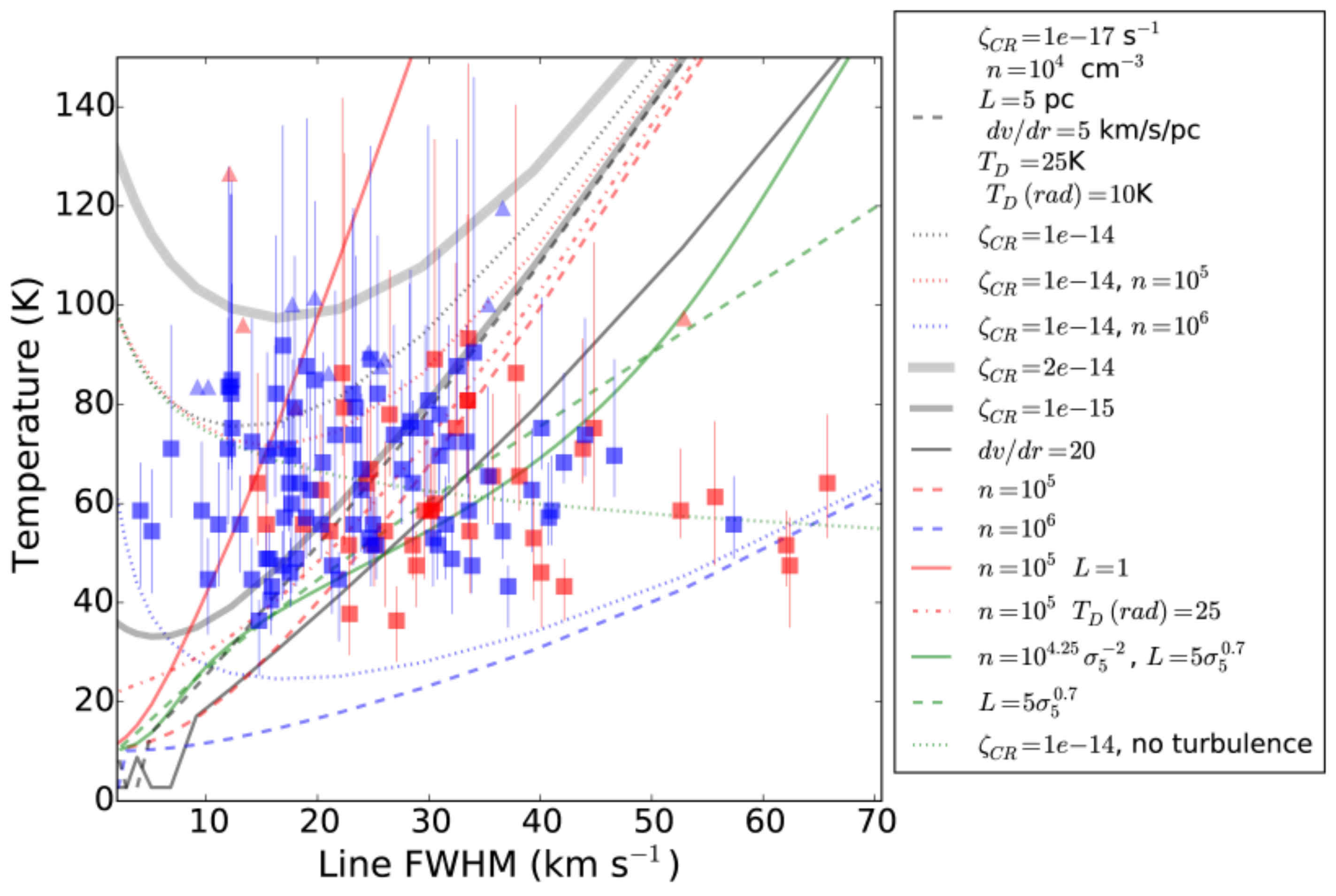} 
{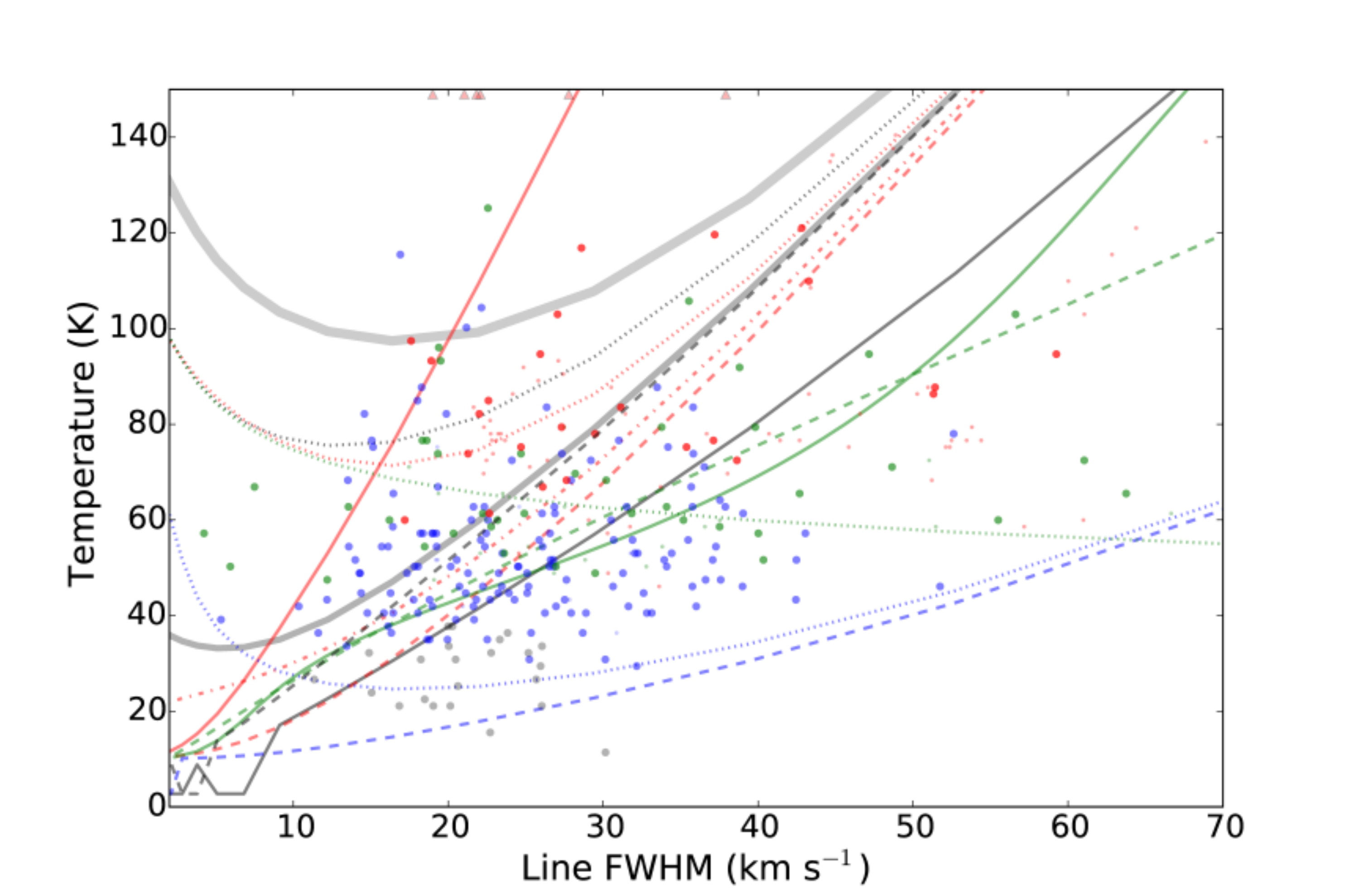} 
{
Plots of gas kinetic temperature vs linewidth.  The lines in both plots show
modeled temperatures as a function of line width given different assumptions as
computed by \texttt{DESPOTIC}.  
A version of this plot with no data is shown in Figure
\ref{fig:tvssigma_despotic}.
The legend
shows the various assumptions, with the fiducial assumptions indicated at the
top, associated with the dashed gray line; these fiducial values are kept fixed
except where indicated.  The colors of the curves each correspond to a
different density, while different line styles indicate changes in other
physical constants.  The green solid curve shows a pressure equilibrium case
($n\propto\sigma^{-2}$) with a Larson relation ($L\propto\sigma^{0.7}$) based
on \citet{Shetty2012a}.
(a) 
The fitted temperature as described in Section \ref{sec:spectralmodeling} plotted
against the fitted line width for the by-eye selected regions.  The blue
symbols are compact `clump' sources and the red symbols are large-area
($8\arcmin$) square regions.  The very broad line regions, FWHM $\gtrsim$ 40
\kms, are sometimes affected by significant baseline issues, and it is possible
that the temperature has been underestimated as a result (underestimated
\threetwoone lines, and therefore underestimated temperatures, are possible
because of the lower signal-to-noise of the \threetwoone line; the
\threeohthree line is less likely to be affected).
(b) The same as (a) but for the dendrogram-extracted sources. The points are
color coded by signal-to-noise in the ratio $R_1$, with black $S/N < 5$, blue
$5 < S/N < 25$, green $25 < S/N < 50$, and red $S/N > 50$.  Small, faint points
indicate ancestors in the dendrogram, while bright, large points indicate leaves.
Lower limits on linewidth and temperature are indicated with triangles on the
edge of the plot.  
Note that in both panels, any correlation is even weaker than that observed in
Figure 6 of \citet{Huettemeister1993a}, who make the same plot for \ammonia (4,4)
and (5,5).  \citet{Riquelme2013a} also found no correlation using \ammonia over a similar
range of velocity dispersions.
While there is no clear correlation between velocity dispersion and
temperature, most data points fall within a range that is consistent with
turbulent heating \emph{if} the appropriate size scale is selected; see Figure
\ref{fig:turbtem} for plots that account for the differing source sizes.
No single cosmic ray ionization rate is
capable of reproducing our observations, and the higher CRIRs ($\zeta_{CR} >
10^{-14}$ \pers) are inconsistent with most of the data.
}
{fig:temvsfwhm_regions}
{1}{3.5in}

The results of the turbulence-predicted temperature analysis are shown in 
Figure \ref{fig:turbtem}, which plots our measured temperature against
the predicted temperature assuming turbulent heating is active and the
cosmic ray heating is negligible ($\zeta_{CR}=1\ee{-17}$ \pers).  For the
hand-extracted regions, there is no strong trend in the data, but most data
points can be explained by turbulent heating alone, perhaps with cosmic rays
setting a high floor temperature.  There are a few points with measured
temperatures below the turbulence prediction.  Some of these regions
include at least one velocity component that seems
to be genuinely cool and broad.  A possible explanation for these regions is
that their linewidths are an indication of bulk motion rather than turbulent
motion, and therefore they may be good candidates for expanding bubble edges,
either supernova or \hii-region driven.  Alternatively, they may contain denser
gas and therefore be able to cool more efficiently; these regions would then be
good regions to search for star formation.

The dendrogram-extracted points plotted include only those with a Gaussian
correction factor $<3$, as larger corrections are regarded as unreliable
extrapolations.  Since dendrogram-extracted sources are necessarily coherent in
position-velocity space, they are less likely than the hand-selected features
(which were selected in position-position space) to represent unassociated
line-of-sight features.

The predicted temperatures for dendrogram clumps frequently exceed the measured
temperatures by a large margin, often a factor of 2 or more.  There are two
classes of explanation: First, that the cooling rate has been underestimated.
This is entirely plausible, as we have assumed solar neighborhood chemistry
(specifically, a CO abundance $X_{CO}=10^{-4}$ \hh, and negligible abundances
for other
coolants), which does not necessarily hold under CMZ conditions.  The other
possibility is that we have overestimated the line width or underestimated the
size scale of the clump via our extraction method.  While we have tried to
account for this source of error by correcting for clipping, we are nonetheless
extrapolating our measurements.

The predicted temperatures for the hand-extracted regions are more uncertain.
Most of the hand-extracted regions agree with the prediction at a 1$\sigma$ and
all but a few at a 2$\sigma$ level.  Given that the sizes used for these
regions are uncertain by a large factor (for square apertures especially), there
is no significant disagreement between $T_{G,turb}$ and $T_{G,measured}$.

\FigureTwo
{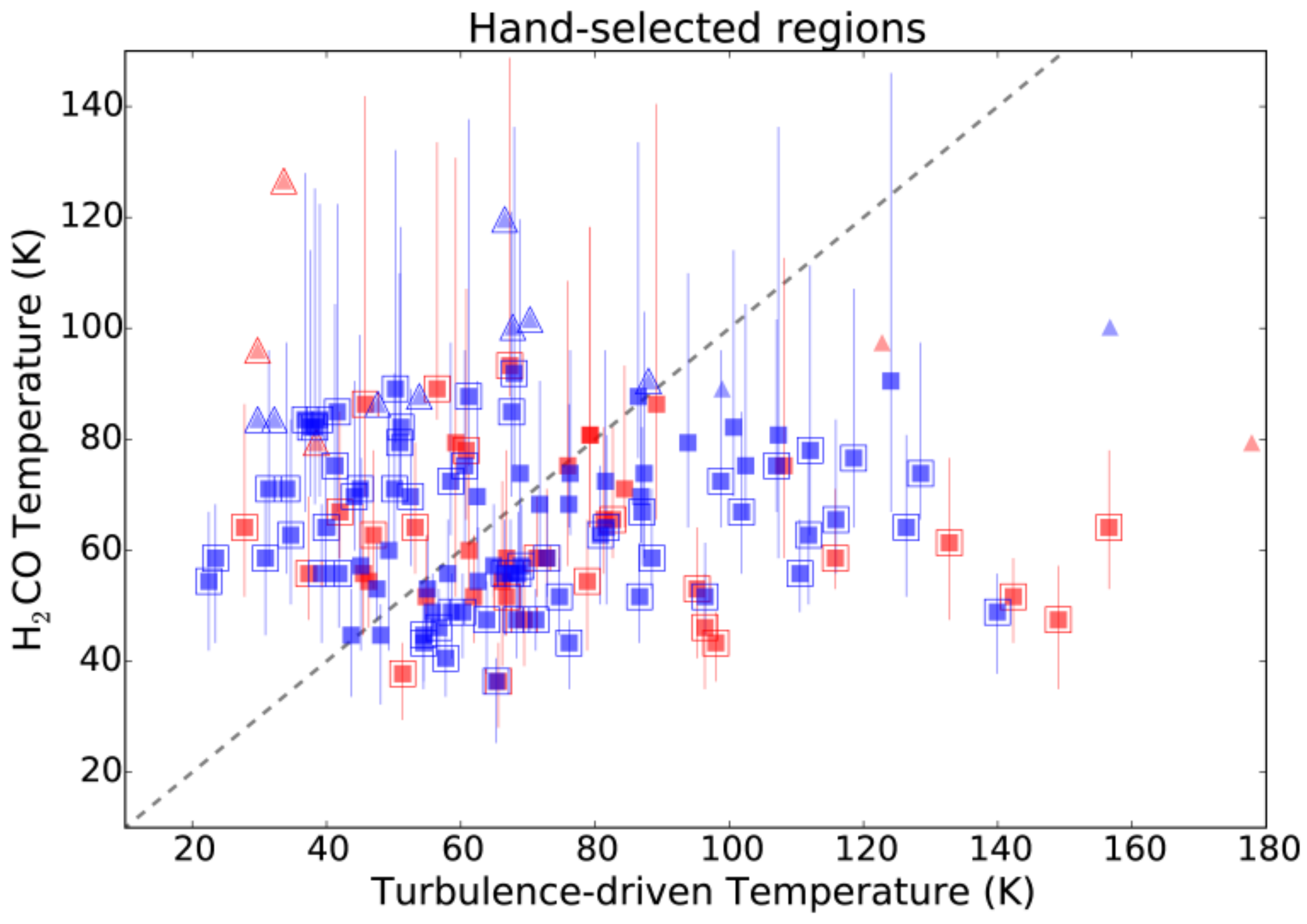} 
{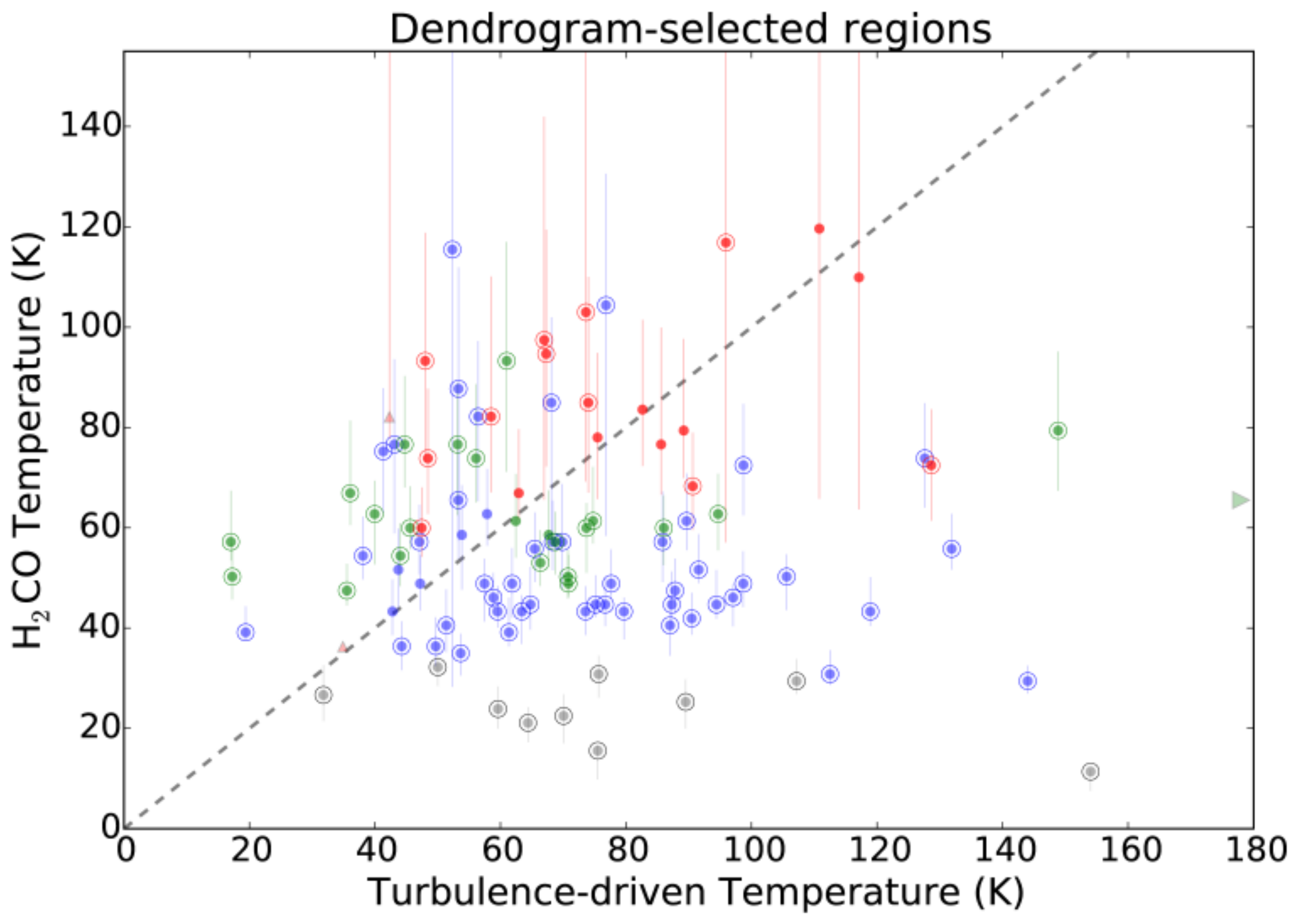} 
{Plots of the \para-measured temperature against the predicted
turbulent-driven temperature as discussed in Section \ref{sec:heating}.
While no trend is obvious in the data, most of the high signal-to-noise
data points are fairly close to the  $T_{G,turb} = T_{G,measured}$ line.
There are few points with $T_{G,measured} > T_{G,turb}$, so turbulence generally provides
enough - and sometimes too much - energy to heat the gas to the observed temperatures.
The regions with predicted temperatures higher than observed require additional
explanation: either their heating rate has been overestimated or their cooling
rate underestimated.
(a) The hand-selected regions as described in Figure
\ref{fig:temvstem_regions}; the blue symbols are compact `clump' sources and
the red symbols are large-area square regions. 
Triangles indicate regions with lower limits on
the measured temperature, i.e. any region with
a measured $1-\sigma$ lower limit on the temperature $T_G>150$ K.  The gray
dashed line shows the $T_{G,turb} = T_{G,measured}$ relation.  Regions that are inconsistent
with the relation at the $1-\sigma$ level are outlined.
(b) The same as (a) for the dendrogram-extracted regions.  Regions with a large
Gaussian correction factor $f_g>3$ are excluded.  Regions with $T_{G,turb}>180$ K
are indicated with triangles.
The points are color coded by signal-to-noise in the ratio $R_1$, with gray
$S/N < 5$, blue $5 < S/N < 25$, green $25 < S/N < 50$, and red $S/N > 50$. 
If the lower signal-to-noise points in blue and gray are ignored, the agreement
between the predicted temperature and observed temperature is fairly good.
}
{fig:turbtem}{1}{3.5in}

To explore the possibility that our cooling rates are underestimated, we
examine a particular clump from the catalog, with $n=10^{4.2}$ \percc,
$\sigma=8.6 f_g = 14.3$ \kms, $L = 2r = 2(0.588)f_g = 1.65$ pc
(beam-deconvolved), $T_D=21$ K, and $T_{D,rad} = 1.2$ K, where $f_g$ is the
Gaussian correction factor.  This clump has a
predicted temperature $T_{G}=156$ K, but a measured temperature $T_{G}=91$
K.  With local chemistry, the dominant coolant is CO by 2 orders of magnitude.
In \texttt{DESPOTIC}, we set chemical equilibrium using the \citet{Nelson1999a}
chemical network code and find that the predicted O abundance is increased by
an order of magnitude, which in turn makes oxygen a significant coolant with
$\Lambda_{O} \approx 0.5 \Lambda_{CO}$, where $\Lambda$ is the cooling rate in
erg \pers \percc (other coolants remain negligible).   The predicted
equilibrium temperature is then $T_{G}=114$ K.  While this temperature is
still a bit high, this example shows that the modified chemistry at these high
temperatures should be accounted for when predicting the gas temperature.

We may be able to further constrain the heating mechanisms using this same data
set if more precise velocity structure information is obtained for the clouds.
Better velocity structure measurements can be achieved by combining line
fitting analysis with systematic clump extraction in order to avoid using the
extrapolated correction factors.  This analysis is left for a future work.

\subsection{Temperature variation with spatial scale or radius}
Photon-driven heating mechanisms are subject to attenuation, so clouds with
significant X-ray or UV heating may have heated shells and cool interiors.
Turbulent and cosmic ray heating experience no such effects, but they exhibit a
different dependence on gas density than the corresponding cooling mechanisms.
We would also expect cooler interiors if the clouds are centrally condensed, since
more efficient dust cooling can dominate heating processes at very high
densities.  If clouds are internally heated by gravitational collapse and star
formation, the cloud interior could be warmer than its exterior.  

We have examined the radial profiles of temperature and velocity
dispersion for selected clouds.  We measured the ratio $R_1$ and the line width
as a function of aperture size by extracting progressively larger circular
apertures centered on the same point.  For each spectrum extracted, we have
fitted the 6-parameter model described in Section \ref{sec:spectralmodeling} to
derive the ratio and linewidth.

Different variations with scale are observed.  For example, around `The Brick',
the ratio decreases from $\sim0.55$ at the position $\ell=0.237\arcdeg$,
$b=0.008$, which is coincident with the highest measured temperature in The
Brick (within a single spectrum), to $\sim0.33$ when averaged over a radius
$\sim70\arcsec$ (Figure \ref{fig:brickradial}a).  By contrast, Figure
\ref{fig:brickradial}b shows a highly turbulent example where the temperature
is lowest at the center and higher on larger scales.  
These observations may imply that two different heating mechanisms, one
internally driven (The Brick) and one external (G1.23-0.08), are active within
the CMZ.

These two objects are not meant to be representative of the whole CMZ, and the
radial properties of clumps need to be studied in more detail to draw firm
conclusions about photon-driven heating processes and thermal balance on
different size scales.  

\FigureTwo
{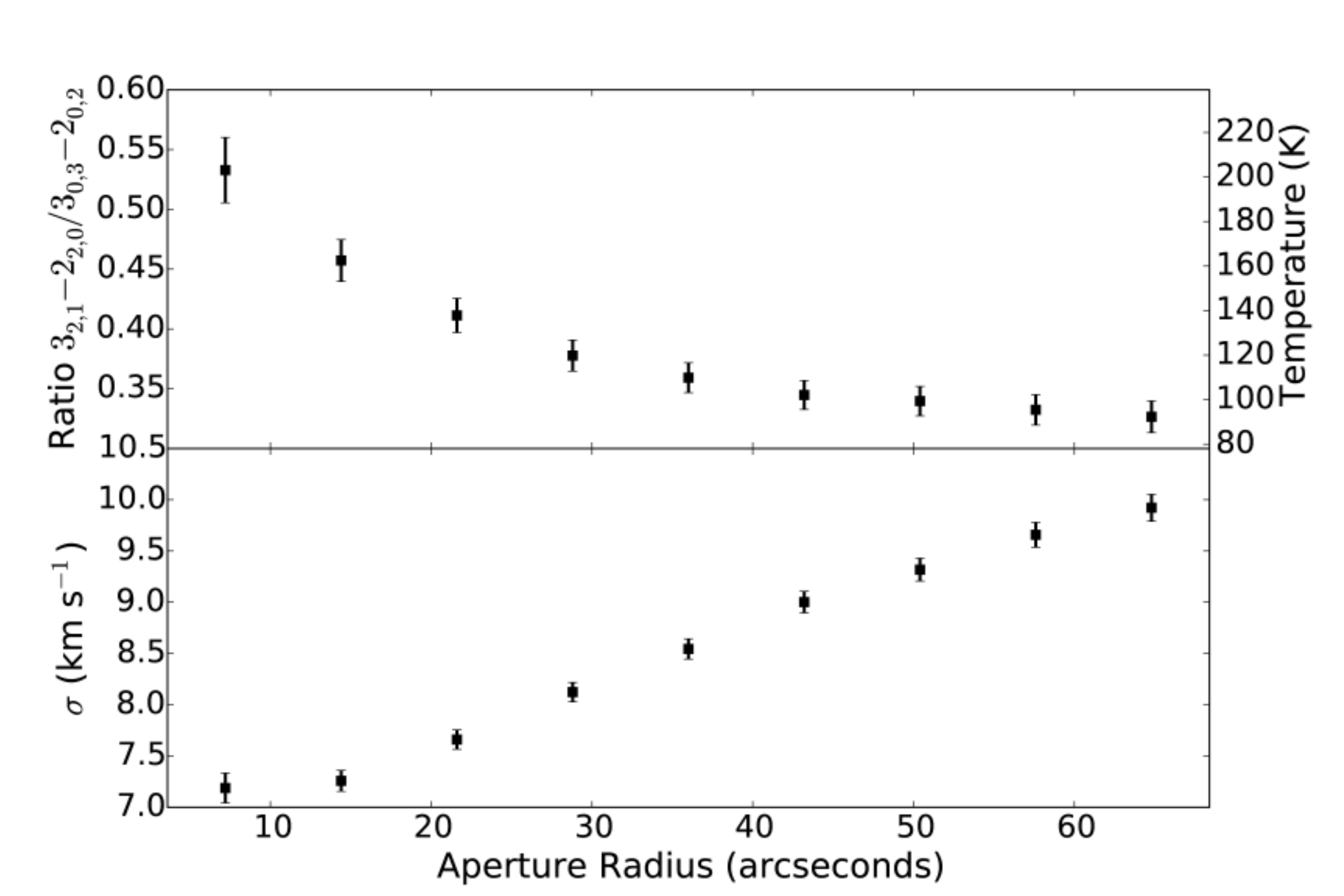}
{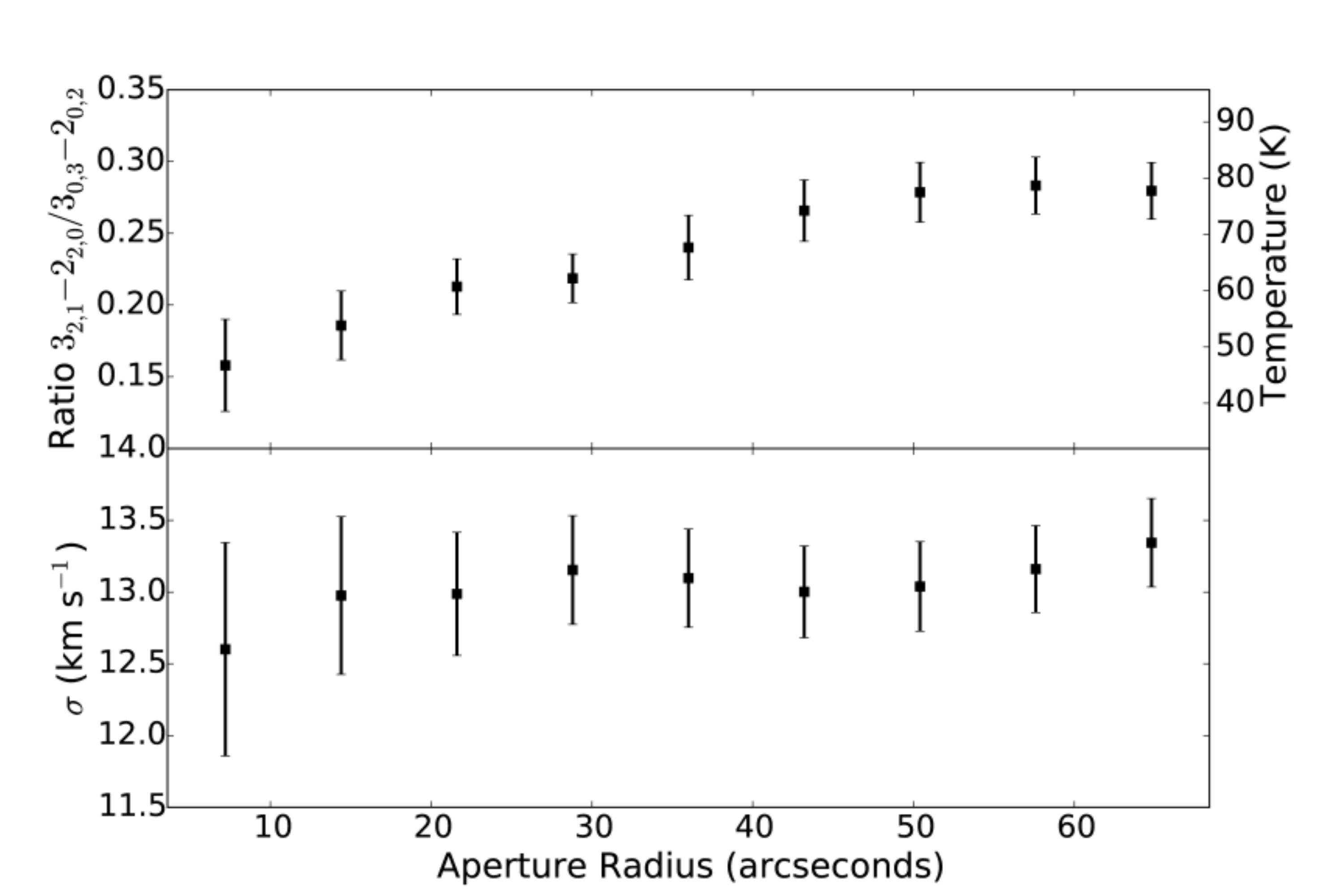}
{ (a) Radial plots centered on the southwest portion of `The Brick'.
The top panel shows the ratio \Rone as a function of aperture size, starting
from a single pixel.
The derived temperature using Figure \ref{fig:ratiovstem} is shown on the right
axis; temperatures $\gtrsim150$ K should be regarded as lower limits.
The bottom panel shows the line width $\sigma = FWHM/2.35$
as a function of aperture size.  
(b) The same as (a) but for a region centered on G1.23-0.08, a local
peak in the \para \threeohthree emission.  In this case,
the ratio \Rone is lower at the center and increases toward larger radii.
}
{fig:brickradial}
{1}{3.5in}

\subsection{Difference between gas and dust temperatures}
\label{sec:td_ne_tg}
Similar to the results of \citet{Ao2013a} for the inner $\sim75$ pc, we find
that the \para and dust temperatures are not the same anywhere within the
inner $R\lesssim200$ pc.  Figure \ref{fig:temvstem_regions} shows the fitted \para
temperature vs the fitted dust temperature from HiGal.  The majority of the
\formaldehyde data points are well above the $T_{D}=T_{G}$ line
independent of the method used to extract the temperature.  The few sources
that are consistent with $T_{D}=T_{G}$ represent an interesting subset
of potentially very cool clouds, though they all have low signal-to-noise
ratios in the present data set.  Overall, though, our data clearly indicate
that, on $\sim$pc scales in the CMZ, dust is a coolant and the gas and dust are
not in equilibrium.  This disagreement means that gas temperatures are
inconsistent with PDR models \citep{Hollenbach1999a} and confirms that UV
heating is not a significant driver of the gas temperatures.

\FigureTwo
{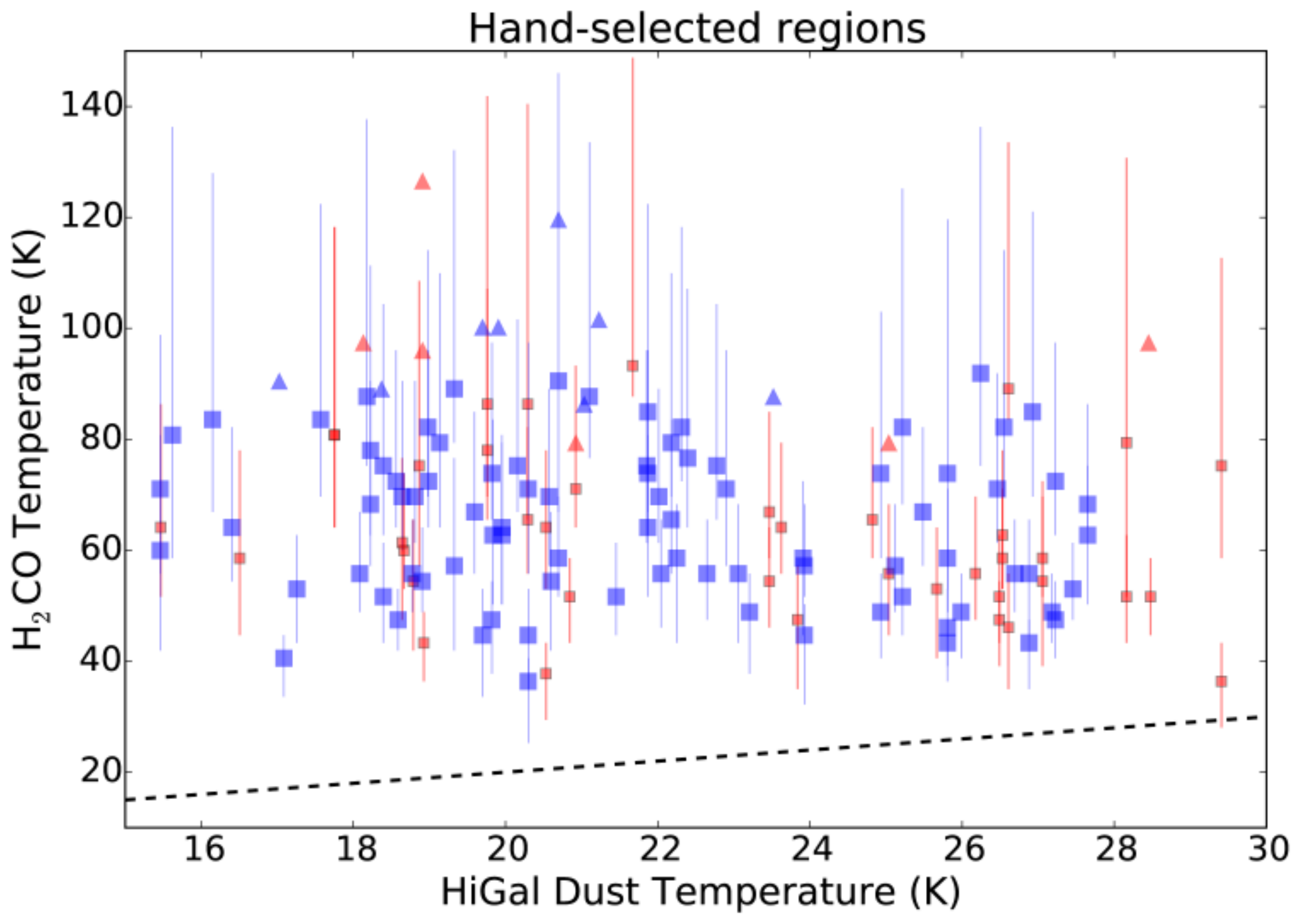} 
{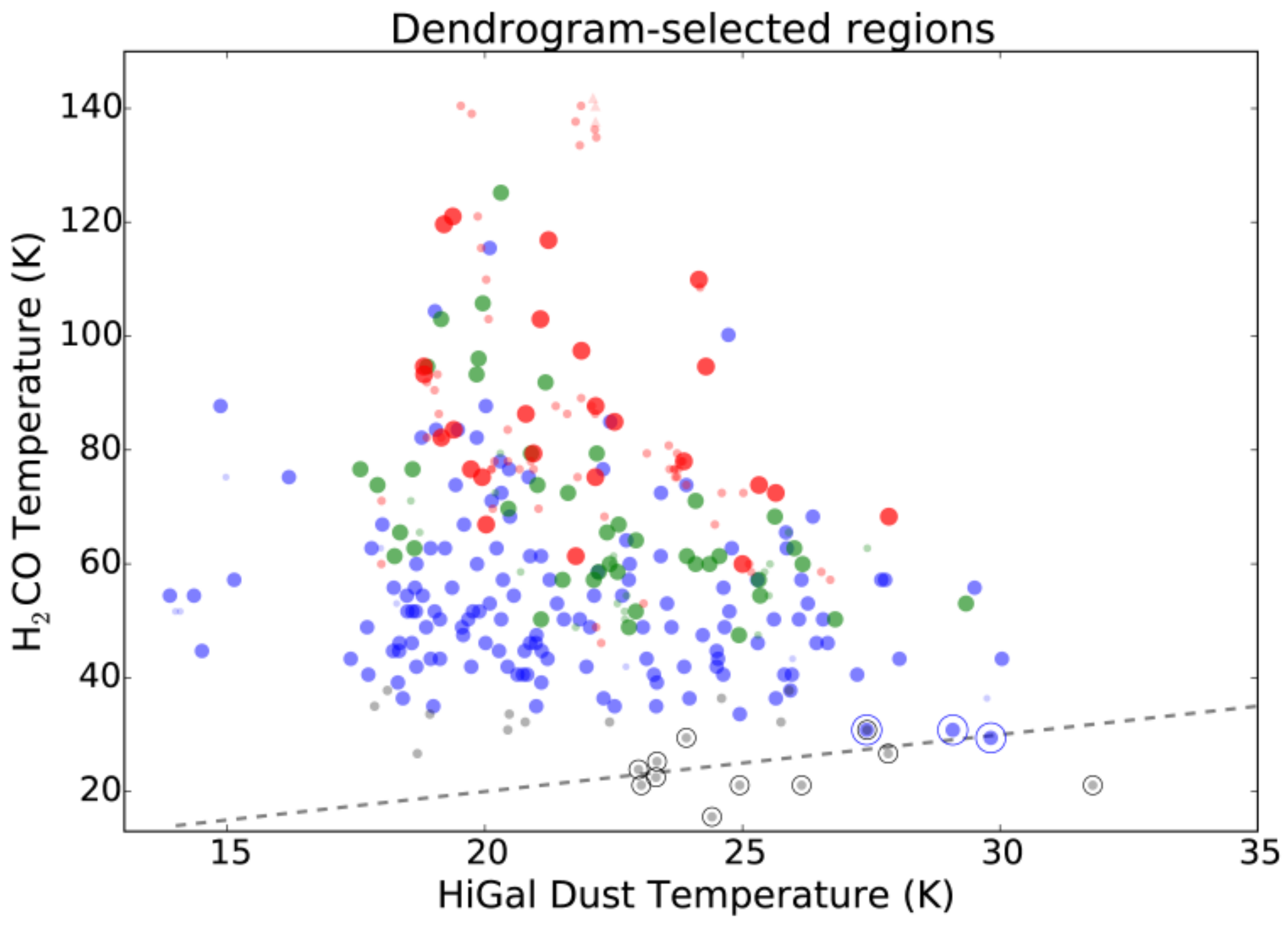} 
{(a) The fitted temperature of region-averaged spectra as described in
Section \ref{sec:spectralmodeling} plotted against the HiGal fitted dust
temperature.  As in Figure \ref{fig:temvsfwhm_regions}, the blue symbols are
compact `clump' sources and the red symbols are large-area square regions.  The
black dashed line shows the $T_G = T_D$ relation.  Nearly all of the data
points fall above this relation, and very few are consistent with it at the
3$\sigma$ level.
(b) The same as (a) but for the dendrogram-extracted sources, showing a similar
discrepancy between dust and gas temperature.  The points are color coded by
signal-to-noise in the
ratio $R_1$, with gray $S/N < 5$, blue $5 < S/N < 25$, green $25 < S/N < 50$,
and red $S/N > 50$.  }
{fig:temvstem_regions}
{1}{3.5in}

\subsection{Spatial variations in gas temperature and orbit comparison}
There are significant spatial variations in the gas temperature.  The most
obvious is in and around the Sgr B2 complex, where very high gas temperatures
are measured.  High gas temperatures, $T_G\sim100-120$ K, were previously
measured in a ridge to the Galactic northeast of Sgr B2 N using \methylcyanide
\citep[][Figure 4b]{de-Vicente1997a}, which we now confirm, though our data
show $T_G\gtrsim150$ K.  \citet{Ott2014a} also showed relatively high \ammonia
temperatures in this hot ridge, and \citet{Etxaluze2013a} found a
high-temperature component ($\sim50-84$ K) in their high-J CO spectral line
energy distribution analysis.
The Brick, Sgr A, and the 50 \kms and 20 \kms clouds exhibit the
highest temperatures outside of Sgr B2, with $T_G>100$ K, in agreement with measurements
in high-J transitions of \ammonia by \citet{Mills2013a}.  The western portion of the
CMZ, around G359.5, shows cooler temperatures $40 $K$ < T_G < 60$ K.  
Along the `dust ridge' (see Section \ref{sec:intro}), the high-latitude portion is slightly warmer than the
low-latitude region; this difference is most evident when comparing The Brick
to the region just south of it and at $\ell=0.5\arcdeg$.
Figure \ref{fig:temvslonplot} shows the temperature of dendrogram clumps as a
function of longitude.  The high temperatures ($T_G\gtrsim60$ K) persist to high
longitudes, including the enigmatic turbulent yet starless G1.6 region
\citep{Menten2009a}.

\Figure
{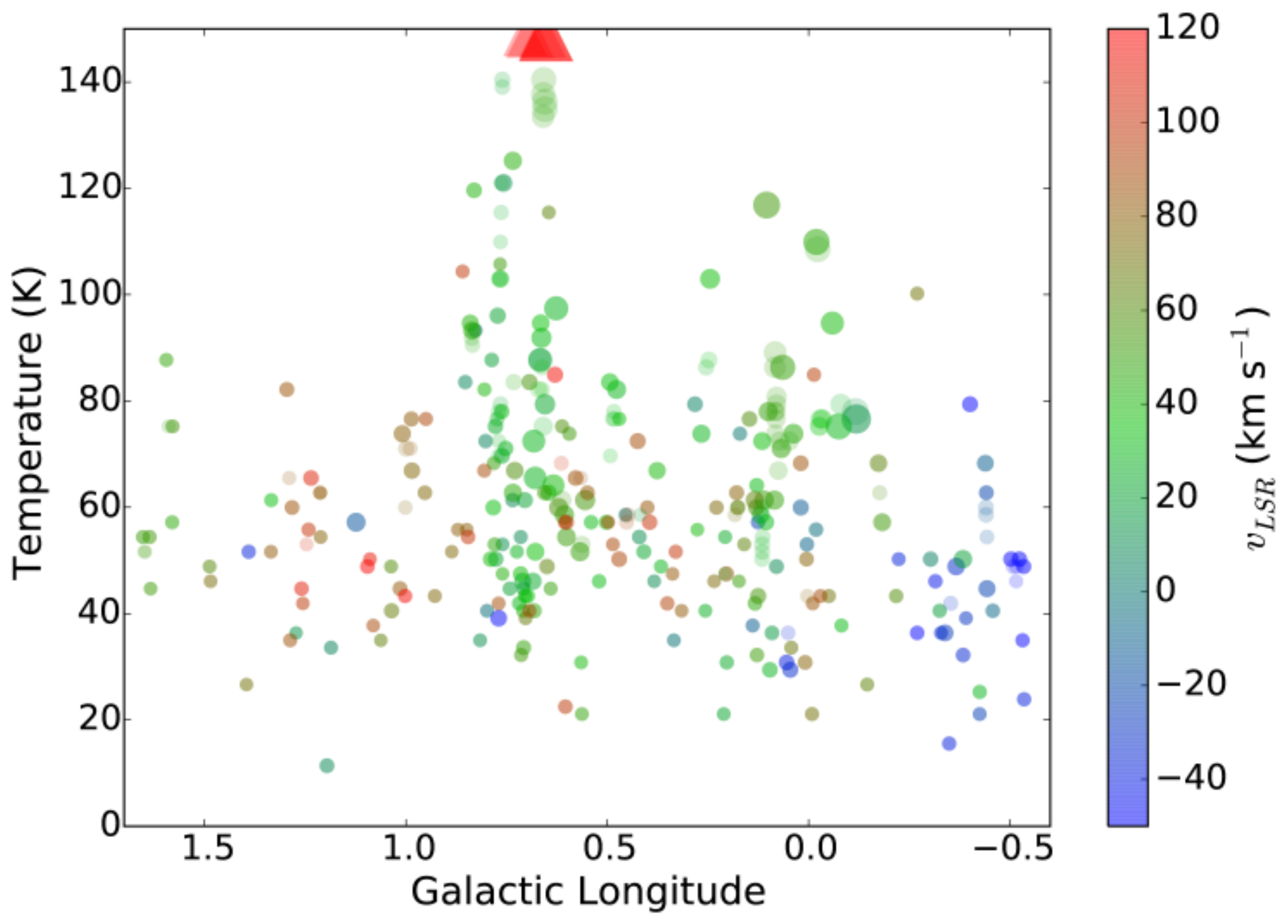}
{Derived temperature as a function of longitude colored by LSR velocity for
dendrogram-extracted data.  The points have a size proportional to the
\threeohthree line brightness.  The more transparent points reflect ancestors
in the dendrogram tree, while the opaque points show leaves.  Lower limits
for some of the Sgr B2 regions are shown as red upward-pointing triangles near
the top of the plot.}
{fig:temvslonplot}{0.4}{3.5in}

In order to search for explanations for the temperature variations we observe,
we compared our data to predicted cloud orbits.  Figure \ref{fig:kdlorbit} shows
the orbital fit from \citet{Kruijssen2015a} and a position-velocity slice
through the fitted temperature cube following that orbit.  The orbit does not
align with all cloud components, especially along the green segment, but is still
very useful for
tracking the properties of clouds in common environments.  Assuming we can use
the orbit PPV predictions to locate the clouds in 3D, we can begin to
understand the geometry of the CMZ.  It appears that the
far-side clouds (green segment, 1 Myr < $t$ < 2 Myr), are slightly cooler
than many of those on the near side, such as ``The Brick'' (red segment).

We assigned a time to each dendrogram-extracted clump within 35 \kms and 6 pc
of the \citet{Kruijssen2015a} orbit and plotted their temperatures against that
time in Figure \ref{fig:kdltemvstime}.  We observed a hint of a linear increase
in temperature with time at the first crossing.  The possible linear increase
in temperature over the first $\sim0.5$ Myr would support the
\citet{Kruijssen2015a} hypothesis that the ``ring'' represents a single gas
stream following an eccentric orbit with regular pericenter passages by Sgr A*.
In this model, the pericenter passage drives additional turbulence and induces
collapse, which implies increased turbulent energy dissipation (hence heating)
after the pericenter passage.  However, the possible trend is also reasonably
consistent with a constant temperature over the same range, so we caution
against overinterpretation. 

\RotFigureTwoAA
{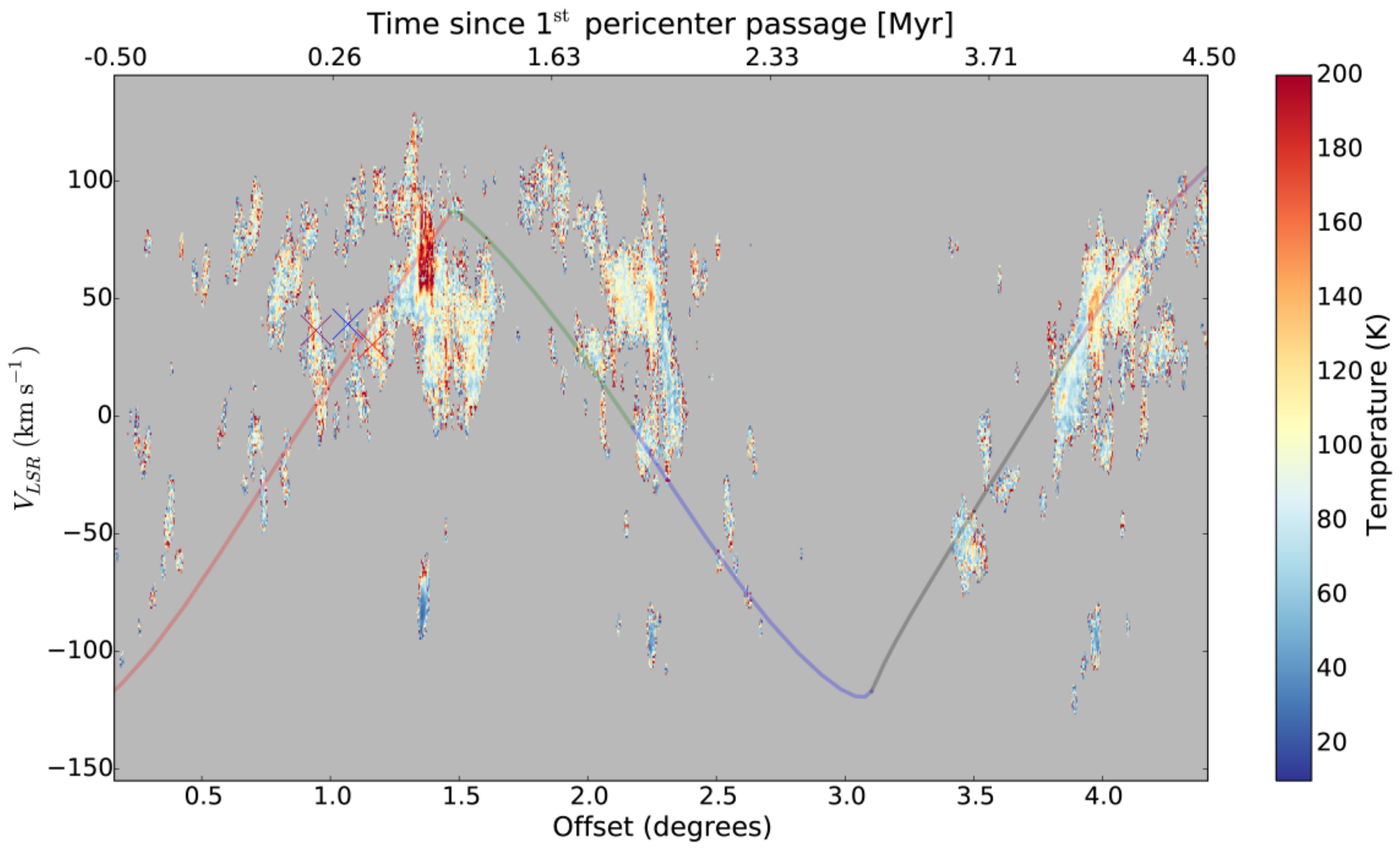}
{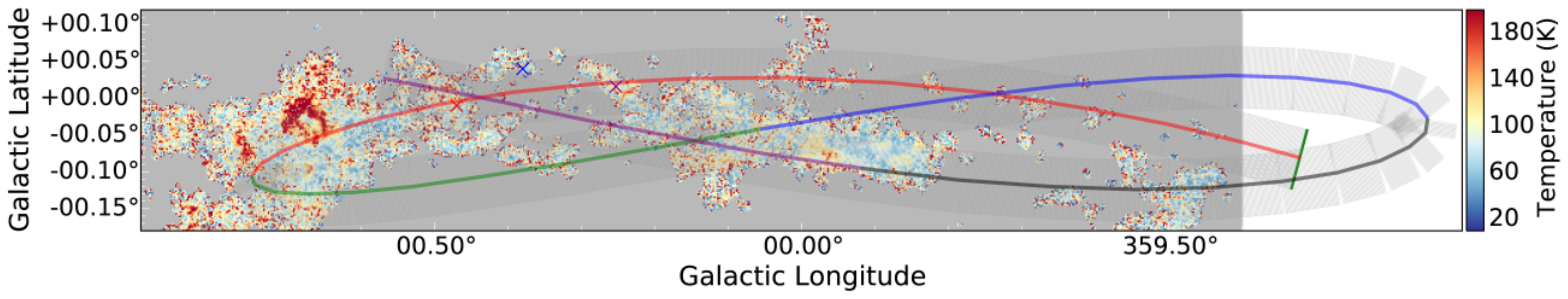}
{(a) A position-velocity slice through the temperature cube along the
\citet{Kruijssen2015a} orbit.  
Gray regions have no gas detected or inadequate signal to measure a
temperature.
(b) The path used to create the slice shown in (a).  The path consists of a series
of transparent $7.2\arcsec\times300\arcsec$
rectangles over which the temperature has been averaged.  The green bar at
$\ell=359.3\arcdeg$ indicates the starting point (offset$=0$) of the position-velocity
slice.  The orbit is outside of our observed field to the west of Sgr C.  
The X's mark The Brick (purple), cloud d (blue), and cloud e (red) to provide
landmarks for comparison between the two figures.
}
{fig:kdlorbit}{0.5}{7in}

\FigureTwo
{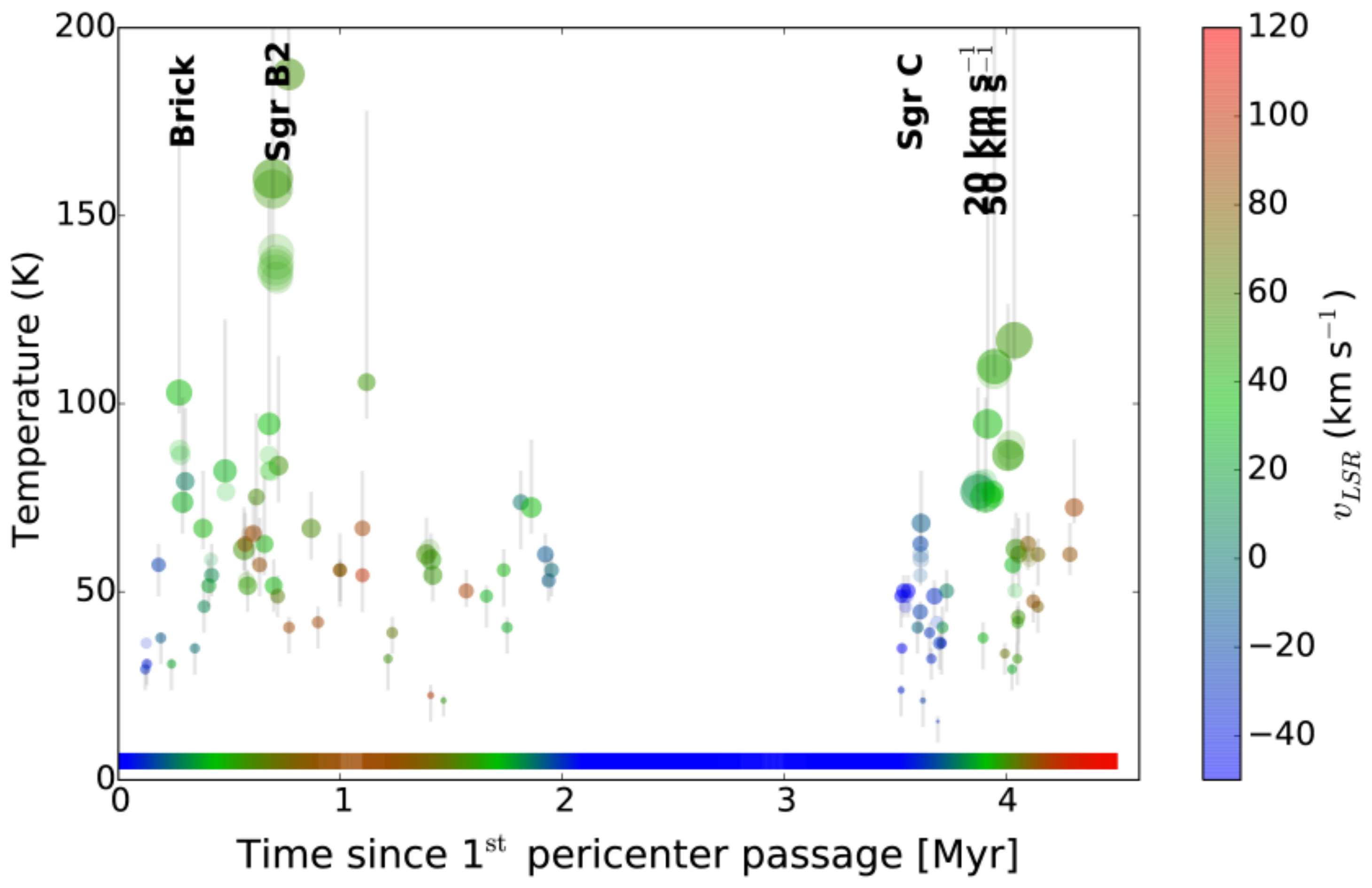}
{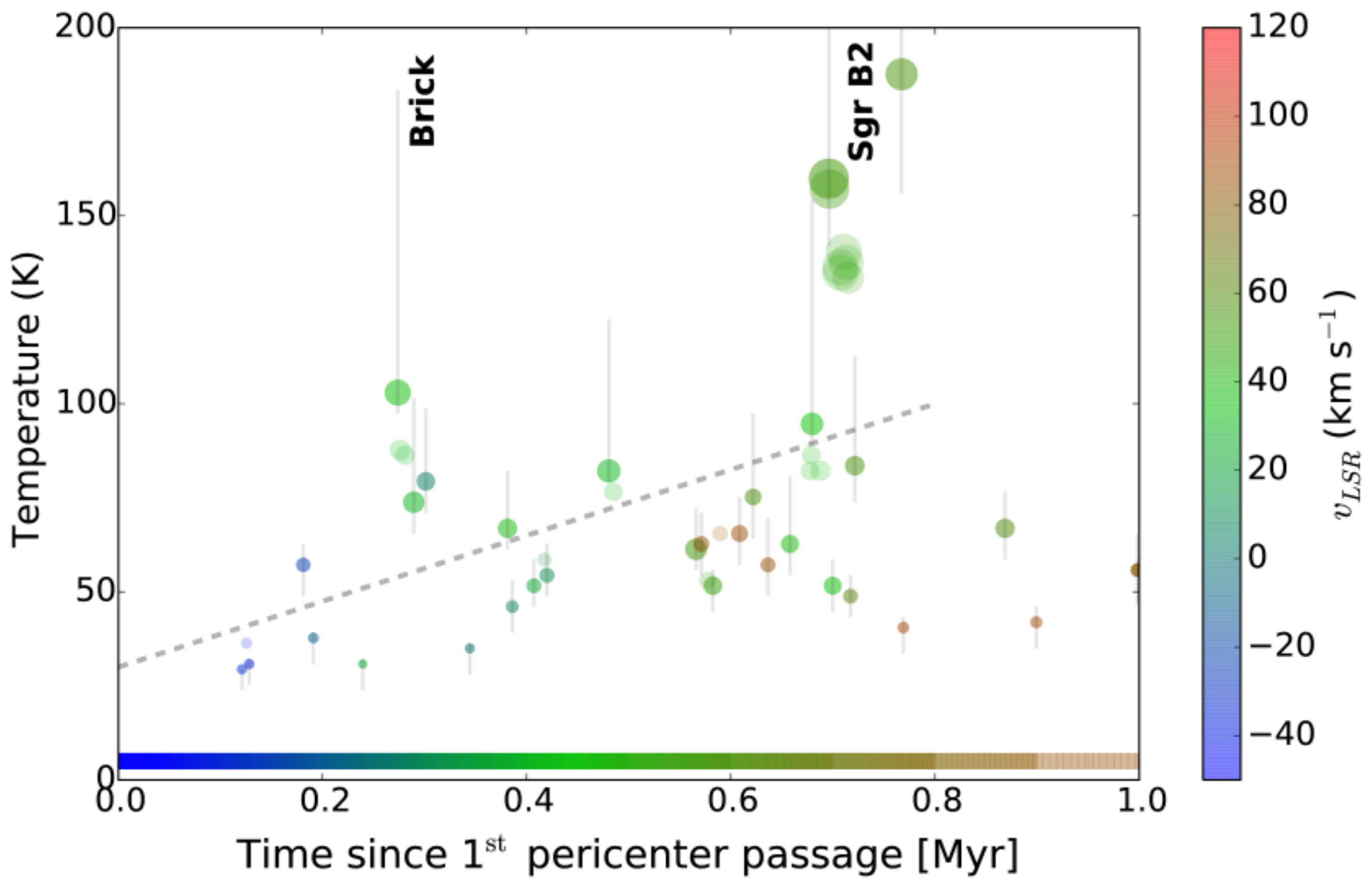}
{The observed temperature vs time since the most recent pericenter passage of
The Brick and the dust ridge clouds on the \citet{Kruijssen2015a} orbit for
dendrogram-extracted clumps within 35 \kms and 12 pc of the orbital path.  
(a) shows the entire orbit
while (b) shows an expanded
view of the first Myr.  The colored bar along the bottom of the figure shows the
modeled orbit velocity at each position.
In both panels,
the darker symbols represent `leaves'
in the dendrogram structure, i.e. compact clumps, while the fainter symbols
represent their parent structures. There is a \emph{hint} of an 
increase in temperature with time, from an initial temperature $T_G\lesssim50$ K
to $T_G\gtrsim100$ K over a period $\tau\sim0.8$ Myr, as illustrated by the
dashed line in panel (b); note that the cool features from 0.7-0.9 Myr are
at a significantly different velocity and therefore cannot be part of the same
stream.  The Brick is an outlier (it is hotter) whether or not there is a
trend.  The symbol sizes are proportional to the \para \threetwoone brightness
and are meant to give
an indication of how reliable the temperature measurements are.  The gray
vertical lines through the circles indicate the formal $1-\sigma$ errors on the
temperature.
}
{fig:kdltemvstime}{0.4}{3.5in}

\subsection{Comparison to \ammonia measurements from \citet{Ott2014a}}
\label{sec:ammoniacompare}

\citet{Ott2014a} measured the \ammonia (1,1) and (2,2) lines over a subset of
the region we mapped, from $-0.2\arcdeg$ to $0.8\arcdeg$.  These two lines are
a frequently used thermometer sensitive to gas temperatures $5 \lesssim T_G
\lesssim 40$ K, with sensitivity up to $\sim80$ K but requiring a larger
$T_{rot}$ to $T_{kin}$ correction, which means the uncertainties at high
temperatures are larger \citep[][their Figure 1]{Mangum2013a}.  The \ammonia lines
have lower critical densities than the \formaldehyde 3-2 transitions by a few
orders of magnitude, $n_{crit}(\textrm{\ammonia}~(1,1))\sim1\ee{3}$ \percc
while $n_{crit}(\para~ \threeohthree)\sim8\ee{5}$ \percc and
$n_{eff}(\textrm{\ammonia}~(1,1))\sim1e3$ while
$n_{eff}(\para~\threeohthree)\sim1\ee{5}$
\citep[][]{Shirley2015a}\footnote{$n_{crit}$ is defined to be the ratio between
the spontaneous decay coefficient, the Einstein $A_{ul}$ value, to the sum of
the collision rates into and out of the upper level.  $n_{eff}$ is the density
required to provide a 1 K \kms flux density given a temperature and a reference
column density.  For either density measure, the species with the higher $n$
will trace denser gas.}.  The energy levels of the
\threeohthree and \threetwoone lines are 21 and 68 K respectively, while
p-\ammonia (1-1), (2-2), and (4-4) have $E_U = 1, 42, $ and $178$ K; the (4-4)
line is therefore required to accurately measure temperatures over the range
covered by the \para J=3-2 lines.

The temperatures derived from the \ammonia line ratio may be lower for a
variety of reasons.  The lower critical density means that \ammonia is
measuring more diffuse gas (which is apparently colder) that is not detected at
all in \formaldehyde.  Since \formaldehyde only provides measurable
temperatures above $\sim50$ K in most locations, there may be low-column gas
that was too faint to be measured with \formaldehyde.  However, since
\citet{Guesten1981a} found temperatures comparable to the \formaldehyde
temperatures using higher transitions of \ammonia, there are either multiple
temperature components in the diffuse gas or peculiar excitation effects
affecting one or both of the thermometers (\citet{Guesten1981a} and
\citet{Morris1973a} suggest that collisional excitation of the (2,1) level may
de-populate the (2,2) level of \ammonia, for example).  The presence of more
diffuse gas traced by H$_3^+$ at $T_{gas}\sim350$ K \citep{Goto2011a,Goto2014a}
suggests that a cold and diffuse component traced by low-J \ammonia is
unlikely.

Figure \ref{fig:nh3tmap} shows the \citet{Ott2014a} temperature map in the same
color scheme that we have used in Figure \ref{fig:temmapnosmooth}.  There are some
trends that are common between the two maps: the Sgr A complex and Sgr B2
complex are the warmest regions, the diffuse gas surrounding these is
somewhat cooler, and the positive latitude component of the
\citet{Molinari2011a} ``ring'' is slightly warmer than the negative latitude
component.

The most significant difference occurs in the southwest portion of the Sgr B2
cloud.  In this region, the \ammonia temperature ranges from the high ($T_G>80K$)
temperatures of the Sgr B2 N hot ridge into a cooler background $T_G\sim30-40$ K.
By contrast, the \formaldehyde temperatures stay high, indicating gas
temperatures $T_G>150$ K extending to the southwest of Sgr B2 S.  This difference
probably indicates that there is a cool, diffuse component being detected in
\ammonia, while \formaldehyde is detecting a hotter, denser component.  This
region will provide a good comparison target for other thermometric
observations.  

Further comparison of the \ammonia and \formaldehyde thermometers is necessary,
and will be possible with future ATCA and GBT surveys of higher ammonia
transitions.  These will be aided by observations of both higher and lower
\formaldehyde transitions \citep{Mangum1993a} to verify the presence of a cool,
diffuse component.

\RotFigureTwoAA
{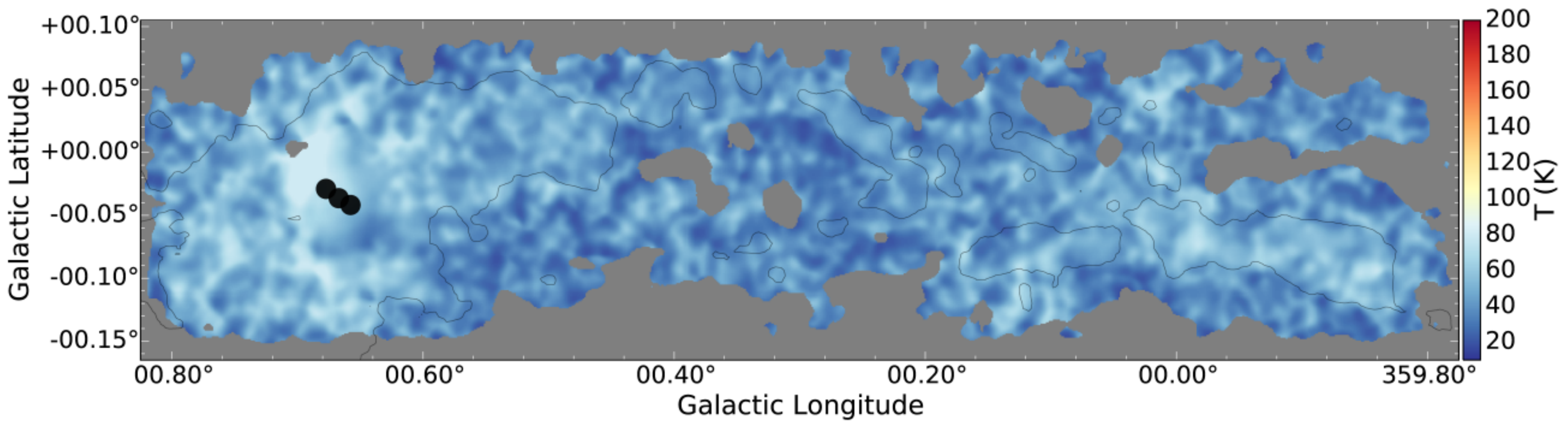}
{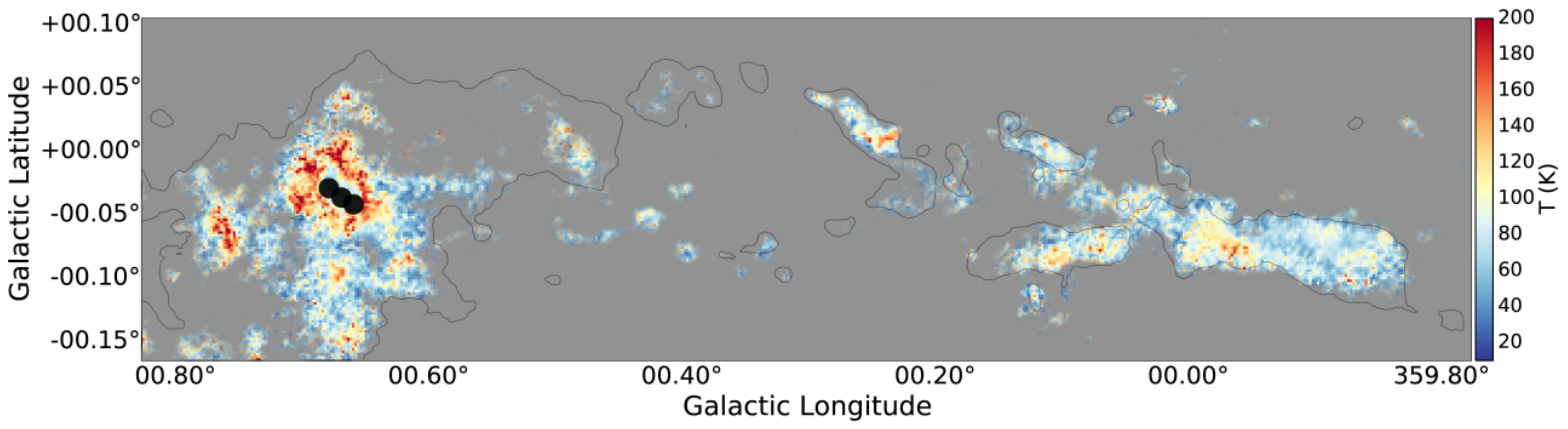}
{(a) A temperature map derived from the \ammonia (1,1)/(2,2) line ratio from
\citet{Ott2014a}.  The \ammonia thermometer has a more limited temperature
range when using only these two lines, so the map is cut off at $T_G=80$ K.
(b) The \para-derived temperature map over the same region from Figure
\ref{fig:temmapnosmooth}a
}
{fig:nh3tmap}
{1}{9.5in}

\subsection{Comparison to other measurements with \para}
\label{sec:h2cocompare}
The largest survey of \para temperatures prior to this work was by
\citet{Ao2013a}, who observed the inner 75 pc.  They performed a similar
analysis to ours but over a smaller area.  They reported similar temperatures,
with some small differences due to a different set of collision rates adopted
(we used \citet{Wiesenfeld2013a}, they used \citet{Green1991a}).  The \citet{Ao2013a}
survey found uniformly high temperatures within the bright Sgr A cloud complex.
Our survey finds a greater variety in gas temperatures due to the larger area
covered.

There have also been observations of \para temperatures with interferometers.
\citet{Johnston2014a} noted a very high temperature peak, $T_G>320$ K, using these
lines of \para toward The Brick, G0.253+0.015.  The ratio of the
SMA to APEX measurements of both the \threeohthree and \threetwoone line in the
southwest portion of The Brick is $\approx0.25$, but the ratio $R_1\approx0.45$
is approximately the same within the $r=25\arcsec$ aperture we used.  The high
temperature in \citet{Johnston2014a} comes from smaller regions in which a
higher $R_1\approx0.71$ is observed.

Above 300K, the collision rates provided by LAMDA
\citep{Green1991a,Schoier2005a,Wiesenfeld2013a} must be extrapolated, which
means the models do not produce reliable results.  Very high $R_1$ values above
about 0.7 cannot be reproduced by high temperatures with the current suite of
collision rates if the temperature is bounded to $T_G<300$K, but instead require
that the lines be optically thick and thermalized and therefore that the gas be
at very high densities.   Given the low brightness temperatures observed
($\sim2$ K even in the SMA $\sim4$\arcsec beams), if the emission is optically
thick, it must be coming from an extremely small area (it must have a small
filling factor).  A small filling factor for the cloud-scale APEX observations
is inconsistent with the observed large extent of the emission and the low
interferometer to single-dish ratio.  We discuss this topic further in Section
\ref{sec:thickorwarm}.  We conclude, therefore, that a lower limit on the gas
temperature in The Brick, $T_G\gtrsim120$ K, is consistent with our
observations, and that the \citet{Johnston2014a} high $R_1$ measurement should
be regarded as either a lower limit $T_G\gtrsim300$ K or a suggestion that
there is sub-arcsecond optically thick \para \threeohthree emission at the
brightness peaks.

\subsection{Is the gas warm or just optically thick?}
\label{sec:thickorwarm}
As noted in Section \ref{sec:h2cocompare}, there is degeneracy in the modeling
that allows the observed ratios to be produced by very dense gas in addition to
hot gas.  Indeed, ratios approaching $R_1=1$ occur for relatively low
temperatures ($T_G\sim30$ K) at very high densities ($n>10^{6}$ \percc)
and very high column densities ($N(\para)>10^{16}$ \perkms \persc) because
the lines become optically thick and have LTE excitation temperatures.

Since the mean densities at 30\arcsec resolution are always significantly lower
($n\lesssim\mathrm{few}\ee{5}$ \percc), such high-density,
optically thick gas would have to have a very low filling factor.  The peak
\para \threeohthree brightness ranges from $0.2 \lesssim T_{mb} \lesssim 2$ K
in the detected regions, implying a filling factor upper limit $ff\lesssim 1-7\%$ if
we assume the resolved brightness temperature equals the dust temperature,
$T_B=T_D$.
In practice, the filling factor would have to be lower still, since any
optically thin \para in surrounding lower-density gas would favor the
\threeohthree line and contribute to a lower $R_1$.

The strongest argument against an optically thick, cold gas explanation for our
observations is the extremely high column
density required.  At typical \para abundances $X\sim10^{-9}$, or even more
extreme abundances $X\sim10^{-8}$, the implied local column density is
$N(\hh)\gtrsim10^{25(24)}$ \persc in a single velocity bin.  Column densities
up to $N(\hh)\sim10^{24}$ \persc are observed, but with much lower filling
fractions $ff\sim10^{-4}$, in The Brick \citep[][their Figure
4]{Rathborne2014b}.  Even in The Brick, one of the densest clouds in the
dust ridge besides Sgr B2, higher column densities are not observed.  So, while
it remains possible that optically thick
\para is responsible for the observed line ratios, it is very unlikely,
especially outside the densest clouds.

One remaining possibility is that the \para lines become optically thick
because of dramatically reduced turbulent line spreading and therefore
increased line trapping.  Such an effect would require a transition from
supersonically turbulent to coherent (subsonically turbulent) gas motion
\citep[e.g.,][]{Pineda2010a} and could reduce the required column densities by
$\sim100\times$.   \citet{Kauffmann2013a} observed a few clumps within The
Brick with narrow velocity dispersions that could represent subsonic regions,
and such clumps could in principle mimic the observed signal since their
relative line-of-sight velocities are high.  However, there are not enough of
these to account for the observed line brightness, and there were also equally
many high-velocity-dispersion clumps detected. 

A related argument is that the high temperatures come from a very small
fraction of the gas heated in high velocity shocks, i.e. that we are seeing
intermittent high temperatures \citep[e.g.,][]{Falgarone1995a}.  Given the high
observed velocity dispersions, high velocity shocks capable of heating the gas
far above the observed temperatures are possible.  Very high velocity shocks,
with $v_s \gtrsim 50 \kms$ will dissociate molecules, so in this scenario the
emission must come from intermediate velocity shocks \citep{Neufeld1989a}.  The
primary distinction between this intermittent model and the more generic
turbulent dissipation heating model is timescale: in the intermittent model,
the shock-heated gas is not well-mixed with the rest of the gas and therefore
represents a distinct phase.  

\subsection{The CMZ average}
For comparison to extragalactic observations, we have included a spectrum
averaged over the whole $\sim300$ pc extent of our survey in Figure
\ref{fig:wholecmzspec}.  The peak amplitude is 50-60 mK and the measured
line ratio $\Rone = 0.25$, corresponding to a temperature $T_G=65$ K at $n=10^4$
\percc, $T_G=63$ K at $n=10^{4.5}$ \percc, or $T_G=48$ K at $n=10^5$ \percc.  
Over the same area, the measured mean dust temperature from HiGal is 23 K,
though on such large scales it would be more appropriate to use Planck or WMAP
data to perform this measurement.
While the spectral fit does not capture all components of the line, the
extracted line ratio is representative of the CMZ-wide average.  This
temperature is very close
to the \ammonia-measured temperature in M83 \citep[56 K;][]{Mangum2013a}, which
has a similarly low dust temperature $T_{D,M83} = 31$ K over a 600 pc
region\footnote{We also examine the HOPS \citep{Walsh2011a} \ammonia
observations of the CMZ over the same region as our \formaldehyde survey, and
find that the measured (1-1)/(2-2) ratio is 1.48, identical to that measured
for M83 to within measurement errors, so the ammonia temperature on these large
scales is also consistent.}.  Maffei 2, NGC 1365, and NGC 6946 also exhibit
similar gas and
dust
temperatures to our Galactic center in the \citet{Mangum2013a} sample,
suggesting that these galaxies all have analogous gas thermal structures in
their inner regions.  The other galaxies in that sample, especially the
starburst galaxies, have significantly warmer gas and dust, with temperatures
closer to those measured for Sgr B2 than for the whole CMZ.

M82's inner regions present an interesting contrast to the CMZ.
\citet{Muhle2007a} report high \para-derived temperature $T\sim200$ K, while
\citet{Weis2001a} found \ammonia-based temperature $T\sim60$ K.
Taken at face value, these measurements imply that both thermometers show
increasing temperatures in starbursts, but they each trace different gas.

\Figure
{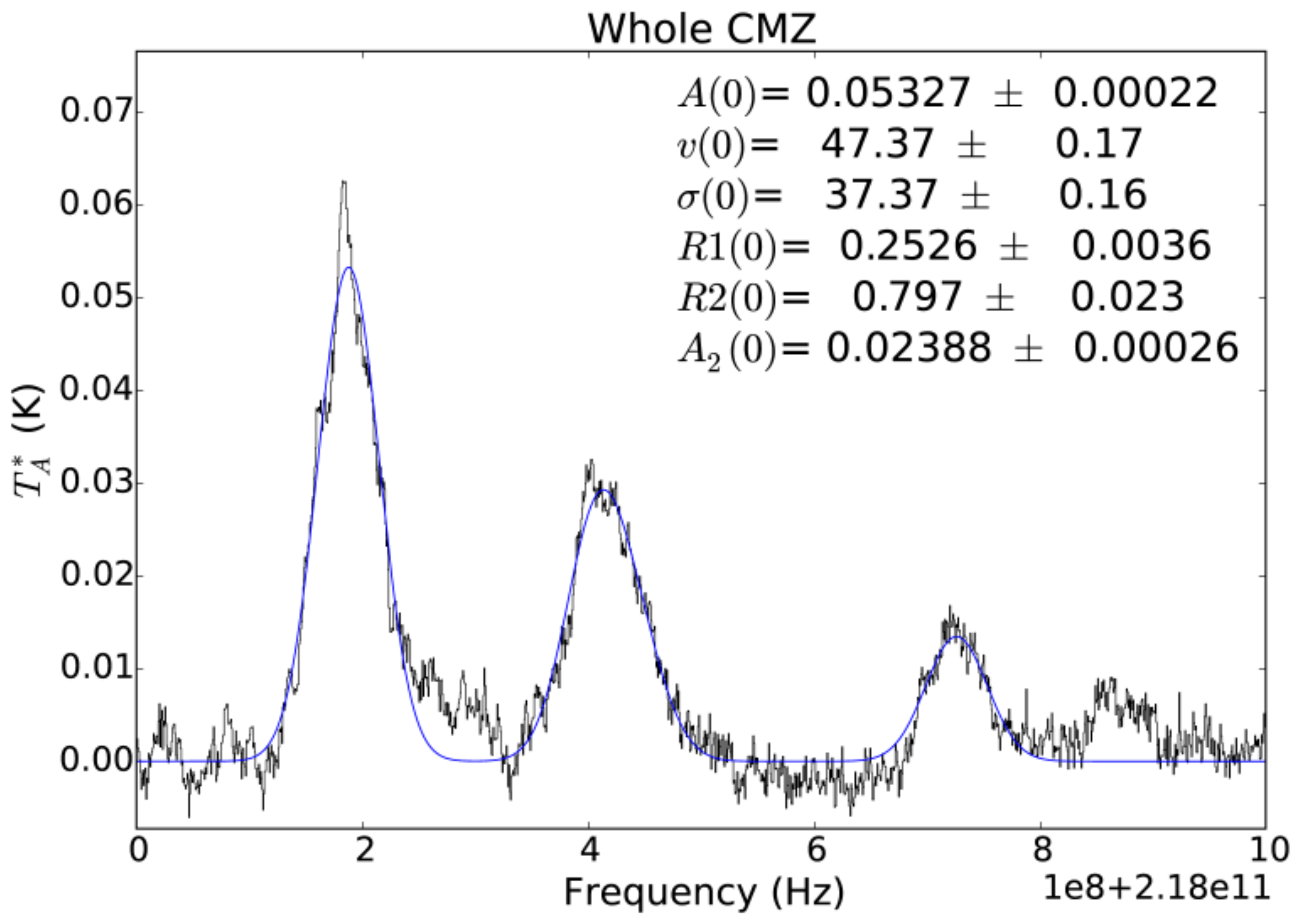}
{The spectrum of the whole survey averaged over all pixels.  The fit
parameters, along with the nominal errors on the parameters, are shown in the
legend.  A single-component fit was used, though many subtler individual
components are evident.  Assuming a constant gas density $n=10^{4.5\pm0.5}$
\percc, the measured ratio implies $T_{G,CMZ} = 63_{-16}^{+2}$ K.}
{fig:wholecmzspec}{0.5}{7in}

\subsection{Implications for star formation}

The star formation properties of the CMZ appear to be different than in the
rest of the Galaxy \citep{Yusef-Zadeh2009a, Yusef-Zadeh2010a, Immer2012a,
Longmore2013b}.  It generally appears deficient in star formation tracers
relative to its dense gas mass.  \citet{Kruijssen2014c} and
\citet{Krumholz2015b} suggested that one of
the main driving factors in this low star formation rate is an increased
threshold for gravitational collapse caused by turbulent pressure.
\citet{Rathborne2014b} support this claim with observations that the gas column
probability distribution function in The Brick is consistent with a purely
turbulent origin.

Our observations further confirm that turbulence is an energetically dominant
process in the CMZ.  We provide an important constraint on the isothermal sound
speed, $c_s = 0.5-0.9$ \kms, which is important for setting the Mach number
that is an essential ingredient in turbulence calculations.  While the high
$c_s$ implies a lower Mach number, the observed line widths more than make up
for it; CMZ gas is still more turbulent than Galactic disk gas.

The high temperature also means that the thermal Jeans mass is about a factor
of 5 higher than in local clouds (assuming $T_{G,CMZ}=60$ K, $T_{G,local}=20$
K, $M_{J,CMZ}(n=10^5 \percc) \approx 9$ \msun, $M_{J,local}(n=10^5 \percc)
\approx 2$ \msun); naively this implies a larger typical fragmentation scale
and perhaps a preference toward forming high mass cores \citep{Larson2005a},
though the higher density in CMZ clouds should balance this.  However, many
modern theories of star formation now invoke turbulence to provide the core
fragmentation spectrum, so the thermal Jeans scale is less relevant for setting
core masses \citep{Krumholz2005c, Hennebelle2011a, Padoan2011b, Federrath2012a,
Hennebelle2013a, Hopkins2013a}.  The high temperatures and high Jeans mass
provide an interesting opportunity to demonstrate that turbulent fragmentation,
which occurs at the sonic scale, rather than Jeans fragmentation is responsible
for setting the peak position of the core mass function.  If turbulence is
responsible for setting the core fragmentation scale
\citep{Offner2013b,Hopkins2013a}, it should be possible to find sub-Jeans-mass
fragments, and in the warm CMZ it should be easier than in cold, less turbulent
local clouds to distinguish the Jeans scale from the sonic scale.

The Equation of State (EOS) in many simulations of star-forming regions is
generally assumed to follow a fixed form that matches observations of gas in
the solar neighborhood \citep{Jappsen2005a,Bonnell2006b,Dale2012a}.  They
assume that $T_G = T_D$ at $n\sim10^5$ \percc.  Our measurements suggest that
gas at this density is still uncoupled from the dust, such that the first
`inflection point' in the EOS should be at a higher characteristic density,
$n_c\gtrsim10^6$ \percc.  Future simulations of CMZ-like environments should
account for this difference.

\section{Conclusion}
We present the largest gas temperature map of the CMZ to date.  We release the
data at \url{http://thedata.harvard.edu/dvn/dv/APEX-CMZ-1mm} and the source
code for the entire project at
\url{https://github.com/adamginsburg/APEX_CMZ_H2CO}.  

The main old result from this paper is that the gas temperature in the galactic
center molecular clouds is relatively high (50-120 K) and apparently fairly
uniform.  The temperatures which we derive are higher than the dust temperature
\citep[][see their conclusion]{Guesten1981a}.  We have added new information
on the large scale of this uniformity and on the association of these
temperatures with particular physical mechanisms, but overall we confirm those
early results.  There is warm ($T_G\gtrsim60$ K) dense gas pervading the CMZ out
to a radius $\sim200$ pc.

We have examined the thermal balance in CMZ clouds.
The high but variable gas temperatures throughout the CMZ suggest that
turbulent heating is the dominant heating mechanism in Galactic Center dense
gas.  Cosmic rays are not dominant, but may be important in less turbulent
sub-regions within the CMZ.  We are able to place an upper limit $\zeta_{CR}
< 10^{-14}$ \pers in the dense gas because a higher CRIR would result in a
higher floor temperature than we observe.  Dust is a coolant in CMZ dense gas,
and no other heating mechanisms should be able to affect the centers of dense
clouds.

There are many regions with expected temperatures based on heating from
turbulence substantially higher than observed.  These regions may have
chemistry significantly different from that in local clouds, with more line
coolants (e.g., atomic oxygen) available due to the high gas temperature.
While we did not investigate the chemistry of CMZ clouds in detail, we suggest
that exploring chemistry in $\sim60-100$ K, $n\gtrsim10^4$ \percc material will
be essential for understanding observations of molecular lines in the CMZ and
in other galaxies with similar conditions.  

The \formaldehyde temperatures are uniformly higher than \ammonia (1-1)/(2-2)
temperatures but similar to temperatures derived from higher \ammonia
transitions.  The lower temperatures observed in the lower critical density
tracer imply that there is a cool, low-density component of the molecular gas.
This perplexing result is contrary to expectations that the dense gas should be
the coldest and deserves further study.

Our new observations are broadly consistent with the orbital model of the CMZ
dense gas described by \citet{Kruijssen2015a}.  While there are some individual
discrepancies, it at least appears that the near-side clouds (The Brick and the
lettered clouds on the path to Sgr B2) have consistent thermal properties.
There is a hint that the dust ridge clouds fall along a stream of progressively
increasing temperature as a function of age along the predicted orbits.
However, the observational uncertainties are also consistent with a constant
temperature along the ridge, so these data cannot be taken as direct proof of
the orbital model.

\textbf{Acknowledgements}:
We thank the staff and observers at APEX for carrying out the service-mode
observations.  We are grateful to Arnaud Belloche, Axel Wei{\ss}, and Carlos de
Breuck for assistance in developing the observing strategy, and Per Bergman for
his assistance in understanding the SHFI-1 baseline issues.  We thank Katharine
Johnston for providing the SMA \para \threetwoone data cube for The Brick.  We
thank Erik Rosolowsky for advice concerning treatment of the biases inherent in
clump extraction, Padelis Papadopoulos for providing commentary on a draft of
the paper, Neale Gibson for discussions about parameter constraints, and the
referee for a helpful review.
T.P. acknowledges support from the Deutsche
Forschungsgemeinschaft (DFG) via
the SPP (priority program) 1573 `Physics of the ISM'.
This research was supported by the DFG cluster of excellence `Origin and
Structure of the Universe' (JED).

\textbf{Code Bibliography}:

\begin{itemize}
    \item sdpy \url{https://github.com/adamginsburg/sdpy}
    \item \texttt{FITS\_tools} \url{https://github.com/keflavich/FITS_tools}
    \item aplpy \url{http://aplpy.github.io}
    \item pyradex \url{https://github.com/adamginsburg/pyradex}
    \item myRadex \url{https://github.com/fjdu/myRadex}
    \item pyspeckit \url{http://pyspeckit.bitbucket.org} \citep{Ginsburg2011c}
    \item astropy \url{http://www.astropy.org} \citep{Astropy-Collaboration2013a}
    \item wcsaxes \url{http://wcsaxes.readthedocs.org}
    \item spectral cube \url{http://spectral-cube.readthedocs.org}
    \item pvextractor \url{http://pvextractor.readthedocs.org/}
    \item ipython \url{http://ipython.org/} \citep{Perez2007a}
    \item DESPOTIC \url{https://sites.google.com/a/ucsc.edu/krumholz/codes/despotic} \citep{Krumholz2014c}
\end{itemize}

\appendix
\section{Baseline removal}
\label{sec:baselineappendix}
The baselines in our APEX-1 (SHFI) data were particularly problematic, more than is
usual in modern heterodyne obervations.  The 218 GHz window we have observed is
particularly sensitive to resonances within the APEX-1 receiver that vary on
$<1$ minute timescales; our off-position calibrations were performed about once
per minute and therefore were not rapid enough to mitigate this problem
completely.  The baselines can broadly be described as smoothly varying ripples
on the scale of $\sim1/10$ the spectral window, plus more rapidly varying
ripples on 20-40 \kms scales.  In principle, this is a straightforward problem
of identifying the fourier components associated with each of these scales and
subtracting them.

In practice, we discovered that it was not possible to remove the dominant
baseline structure on either scale without significantly affecting the
underlying spectral data.  We simulated a variety of fourier-space suppression
approaches by adding synthetic signal to baseline spectra extracted from the
first few principle component analysis (PCA) components of the real spectra.
The PCA extraction approach is able to pull out the dominant baseline
components very effectively, but it inevitably includes significant signal in
the top few most correlated components, especially for the strong \formaldehyde
and \thirteenco lines.  We therefore abandoned it for the final data reduction.

For reference, we show an example spectrum that we believe to consist entirely
of baseline ripples in Figure \ref{fig:badbaselines}.

\Figure{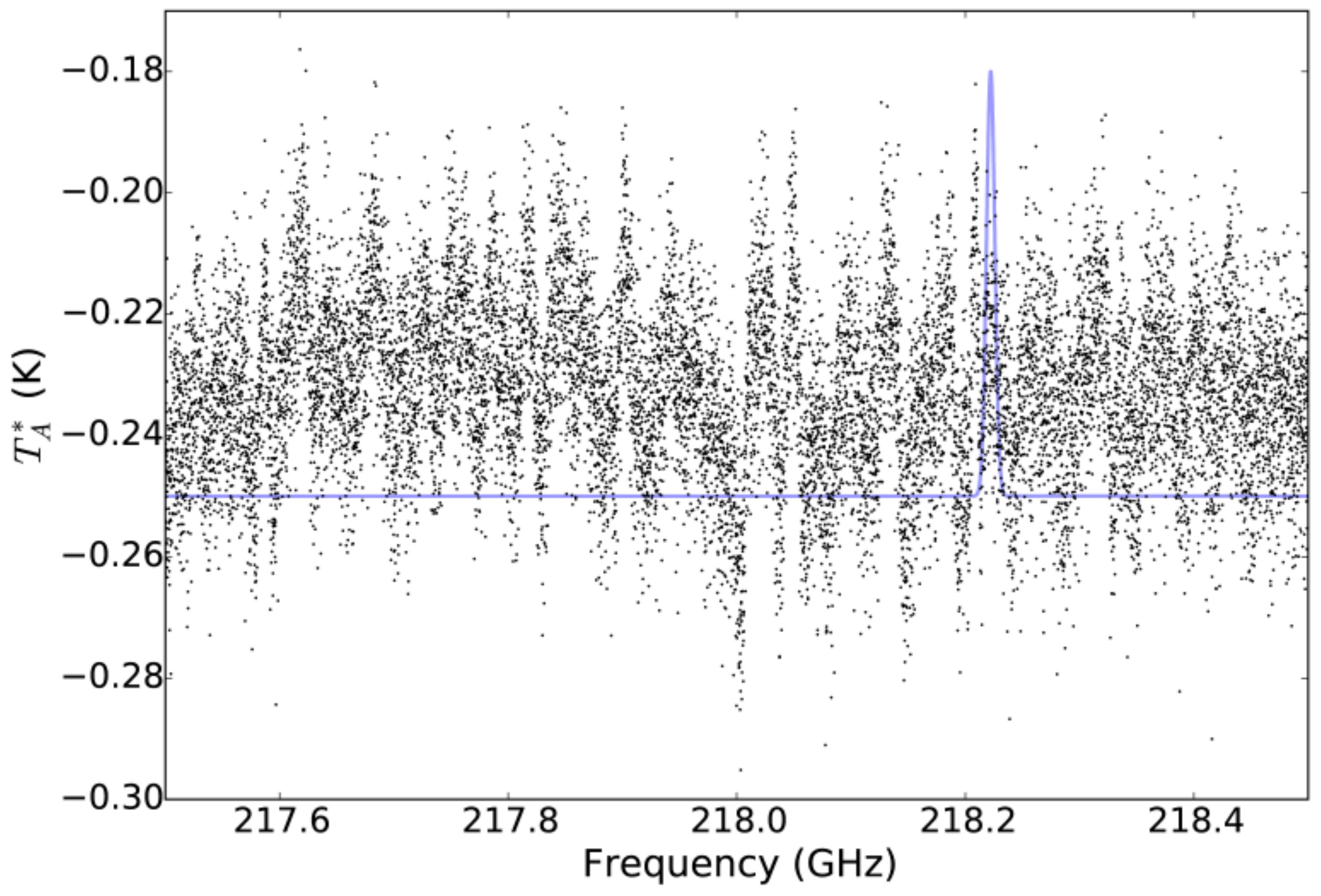}
{An example showing some of the worst baselines observed.  The plotted spectrum
is from an observation on April 2, 2014, showing the average of the 5\% worst
spectra.  The blue curve shows a $\sigma=5$ \kms (FWHM$=11.75$ \kms) line centered
at the 0 \kms position of \para \threeohthree,
illustrating that the baseline `ripples' have widths comparable to the observed
lines.  While we selected the worst 5\% in this case, nearly all spectra are
affected by these sorts of baselines, and the shape and amplitude varies
dramatically and unpredictably.  The variation, unpredictable though it is,
works in our favor as it averages out over multiple independent observations.
The 218 GHz region shown here is also the worst-affected; the 220 GHz range
that includes the \thirteenco lines generally exhibits smoother and
lower-amplitude baseline spectra.}
{fig:badbaselines}{0.6}{3.5in}

\section{Spectral fits for regions and apertures}
We provide a catalog of line ratios measured from aperture-extracted spectra
along with figures showing the best fit model for each spectrum.  We have
extracted spectra in regions with significant \para emission and in commonly
studied individual regions.  This catalog is intended to provide archival value
from our data set and enable future studies of individual regions.

In the hand-extracted source table, three types of region are included: 
\begin{enumerate}
    \item Circular regions-of-interest (Figure \ref{fig:regions}a)
    \item Rectangular regions-of-interest (Figure \ref{fig:regions}b)
    \item $8\arcmin \times 8\arcmin$ square observation fields (Figure \ref{fig:regions}c)
\end{enumerate}
The fit tables include multicomponent spectral fits with and without
spline-based baseline removal (see Section \ref{sec:spectralmodeling}).
Spatial parameters for the box (\texttt{GLON}, \texttt{GLAT},
\texttt{boxheight}, and \texttt{boxwidth}) or circle (\texttt{GLON},
\texttt{GLAT}, and \texttt{radius}) regions are included.  The average dust
column and dust temperature are reported (\texttt{higaldusttem} and
\texttt{higalcolumndens}).  The best fit (maximum likelihood) parameters for
gas temperature, \para column, and \hh density are included along with their
$\pm1-\sigma$ marginalized limits.  The full column description is in the
\texttt{tables/README.rst} file in the source code repository, along with the
tables in IPAC format.

The dendrogram property table is also included in the repository and described
in the \texttt{README}.

\RotFigureThreeAA
{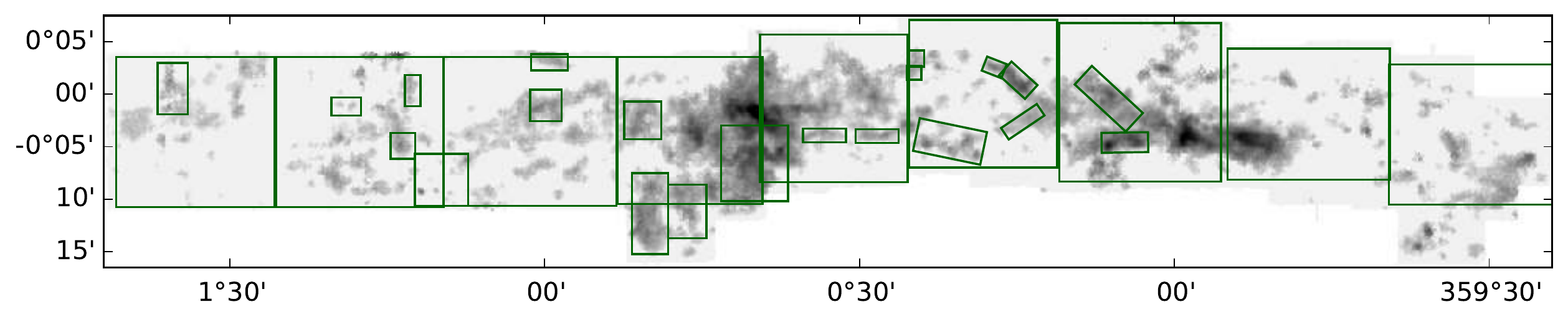}
{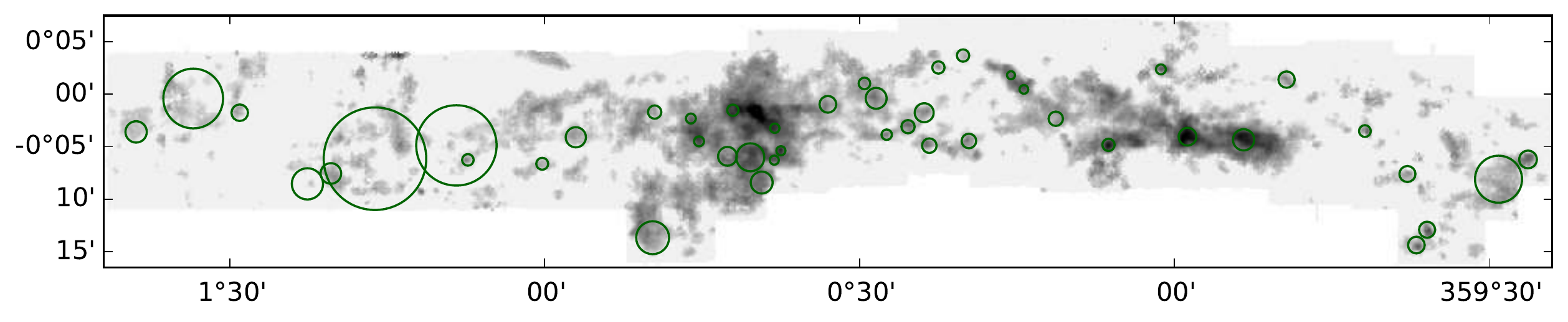}
{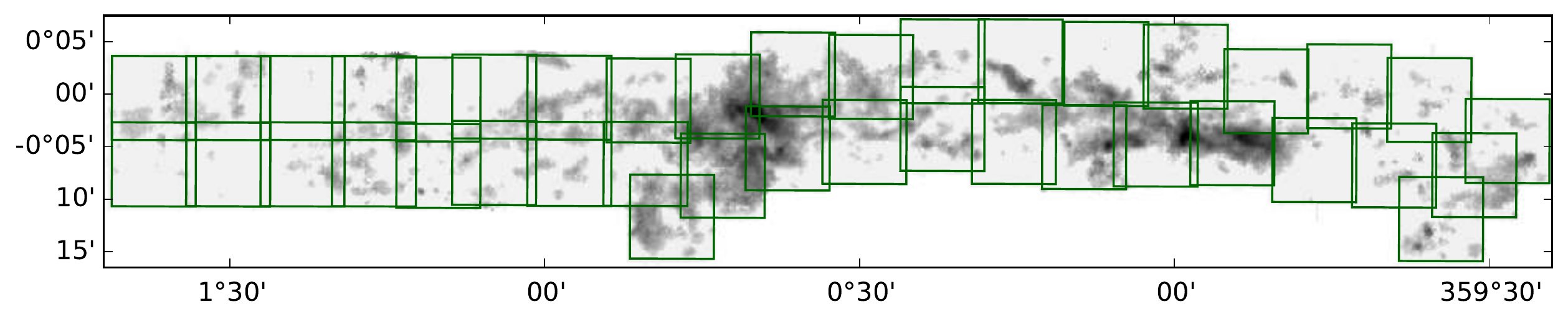}
{Figures showing the extracted regions.  The background is the integrated
masked \para \threeohthree image. (a) Box regions created for large-scale averaging
and to track the morphology of individual clumps  (b) Circular regions selected
to highlight points of interest (c) $8\arcmin \times 8\arcmin$ Box regions
matching the observational setup in 2014.}
{fig:regions}{1}{9.5in}

\section{Source code and data release}
The reduced and raw data are made available via the CfA dataverse
(doi:10.7910/DVN/27601).  The reduced data, including cubes, integrated images,
and catalogs, are also made available at the CDS via anonymous ftp to
cdsarc.u-strasbg.fr (130.79.128.5) or via
http://cdsweb.u-strasbg.fr/cgi-bin/qcat?J/A+A/ (specific URL to be filled in by
the Journal).

The source code for this project in its entirety is available in the attached
tar file and on the internet at
\url{https://github.com/adamginsburg/APEX_CMZ_H2CO}.  Because the archives are
public, we include scripts to download and process the raw data so that all
steps of the analysis performed here can be performed by any individual with
access to a computer.  We provide a script \texttt{download\_raw\_files.py}
that will retrieve the raw \texttt{.apex} data from the ESO archive and from
the dataverse; it requires a valid ESO archive account to use. The total data
reduction process takes $\sim30$ hours on a 48-core, 2012-era linux machine,
though most steps are not parallelized so the process may be faster on more
recent machines with fewer cores.  About 300 GB of free space are required to
store the raw, intermediate, and reduced data products, though the final
products are $<30$ GB, with an actual size depending on whether the
baseline-subtracted versions are kept separate from the original files.  The
download process may take substantially longer than the data reduction.

To fully reproduce the data products in this paper, some effort is required to
set up the appropriate directory structure.  The file
\texttt{reduction/run\_pipeline\_cleese.py} provides a useful template for
informing the code about the appropriate directory location, and
\texttt{reduction/README.rst} includes more detailed instructions.  The raw
\texttt{.apex} data files need to be stored in a common location, and output
directories need to be specified for the merged data files, each epoch's data
(one for the \citet{Ao2013a} data, one for the 2013 data, and one for the
2014/2015 data), diagnostic plots, and the subcubes for each molecule.  The
function \texttt{do\_everything} in \texttt{reduction/make\_apex\_cubes.py}
will then run the pipeline end-to-end.

\section{A detailed example showing parameter constraints}
In Section \ref{sec:spectralmodeling}, we discussed the fitting approach used
to determine a best-fit temperature and uncertainties on that temperature.
Here, we go through a detailed example showing projections of the full 3D
temperature-density-column parameter space we explored and the constraints
each measurement places in each phase space.

Figures \ref{fig:coltemconstraints}, \ref{fig:denscolconstraints}, and
\ref{fig:denstemconstraints} show the 3D parameter space of the LVG models
viewed from its three unique faces.  Each figure has 4 panels, each of which
shows the effect of one of our measurements in constraining the available
parameter space.  All panels include the marginalized likelihood contours.
Figure \ref{fig:parsonbrightness} shows slices through the LVG models at the
maximum likelihood position.  The panels show the predicted line brightness of
the \threeohthree and \threetwoone lines in each direction through the LVG
parameter cube.

Figure \ref{fig:oneddistributions} shows the marginalized one-dimensional
distributions for each parameter along with the constraints provided by each
measurement.  Note that the posterior is not simply the product of the other
displayed distributions; for example, very high temperatures require low
volume and high column densities that occupy a small fraction of the total
integrated likelihood.

In the script file \texttt{example\_parplot\_constraints.py} in the associated
repository, we examine the systematic uncertainties in the fitting process.
Using a known density, temperature, and column, we then use the fitting process
applied to the dendrogram objects and the hand-extracted regions to derive our
measured temperatures.  Over a range $10^4 < n < 10^7$ \percc, the maximum
likelihood estimator recovers the true temperature to within 25\% for all
temperatures, and to 10\% over the range $10^{4.25} < n < 10^{5.5}$ \percc.  At
densities lower than $n<10^4$ \percc, the recovered temperature is
significantly lower than the true temperature, which is expected since we
impose a prior that $n>10^4$ \percc.  Curiously, the expectation value is more
biased than the maximum likelihood estimator, with errors reaching 30\% and
with a much stronger dependence on the input temperature.

\Figure{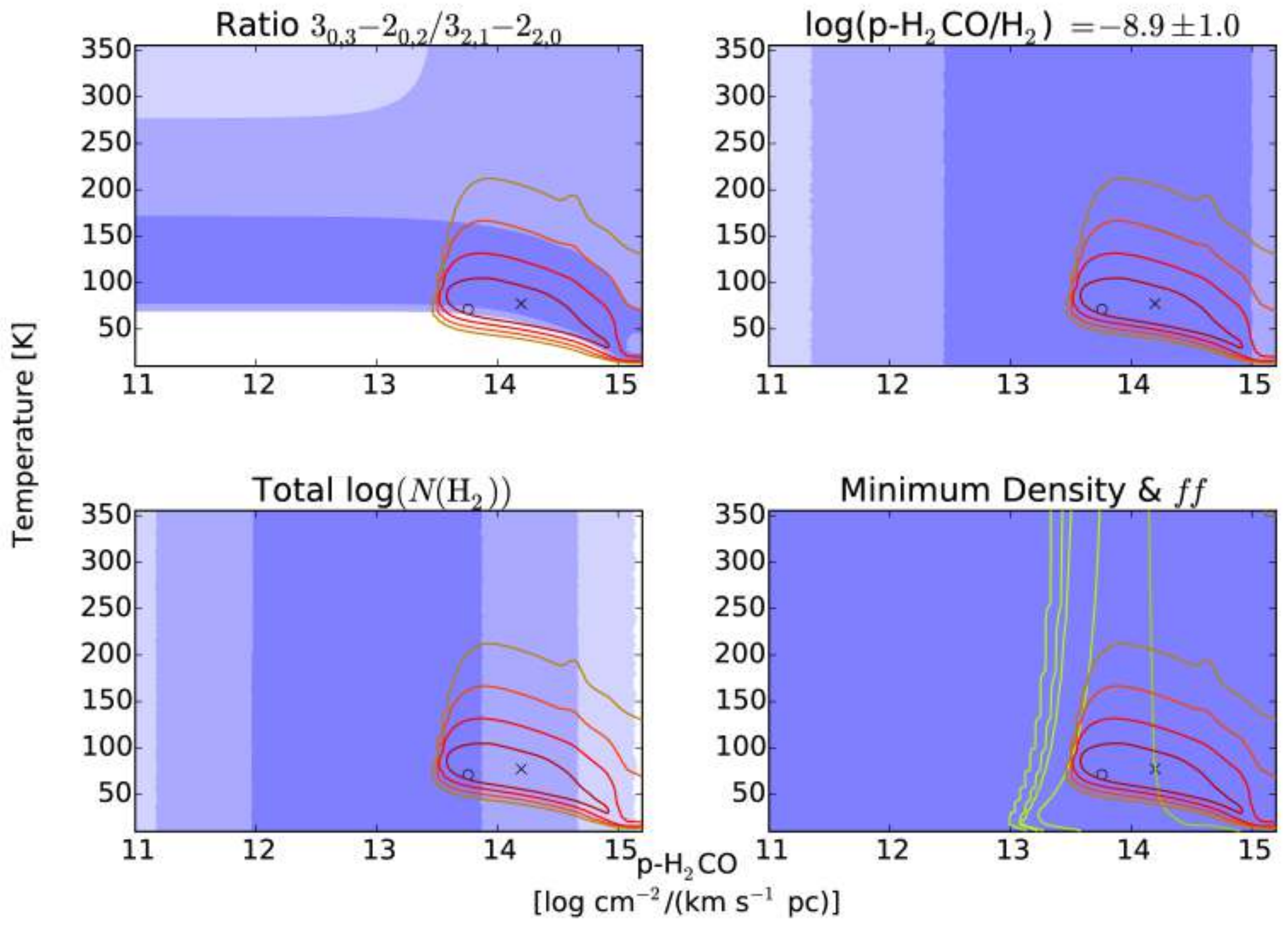}
{The parameter constraints for ``cloud c'' (Figure \ref{fig:cloudcspec})
projected (marginalized) onto the temperature/column density plane.\newline
(top left) Constraints imposed by the measured ratio \Rone are shown in the
background filled colors,
with significance ranges progressing from dark-light from 1-3$\sigma$.  The
red line contours show the joint constraints imposed by including
restrictions on the
total column density, volume density,  filling factor, and abundance, with
1, 2, 3, and 4$\sigma$ contours shown.
The $\times$ marks the expectation value and the $\circ$ marks the maximum
likelihood value.
\newline
(top right) The same colorscheme as before, showing the constraints imposed by
assuming the abundance of \para relative to \hh is as labeled.  
The abundance does not constrain these parameters, but in Figure
\ref{fig:denscolconstraints}, the abundance rules out a substantial region of parameter
space.
\newline
(bottom left) The same colorscheme as before, showing the constraints imposed
by using the measured mean volume density as a lower limit.  In this figure,
the volume density imposes no constraint, but in Figure \ref{fig:denscolconstraints}
and \ref{fig:denstemconstraints} it is important.
\newline
(bottom right) The same colorscheme as before, showing the constraints imposed
by the measured total column density of \hh.  The green contours show
constraints imposed by the lower limit on the area filling factor.}
{fig:coltemconstraints}{0.5}{7in}

\Figure{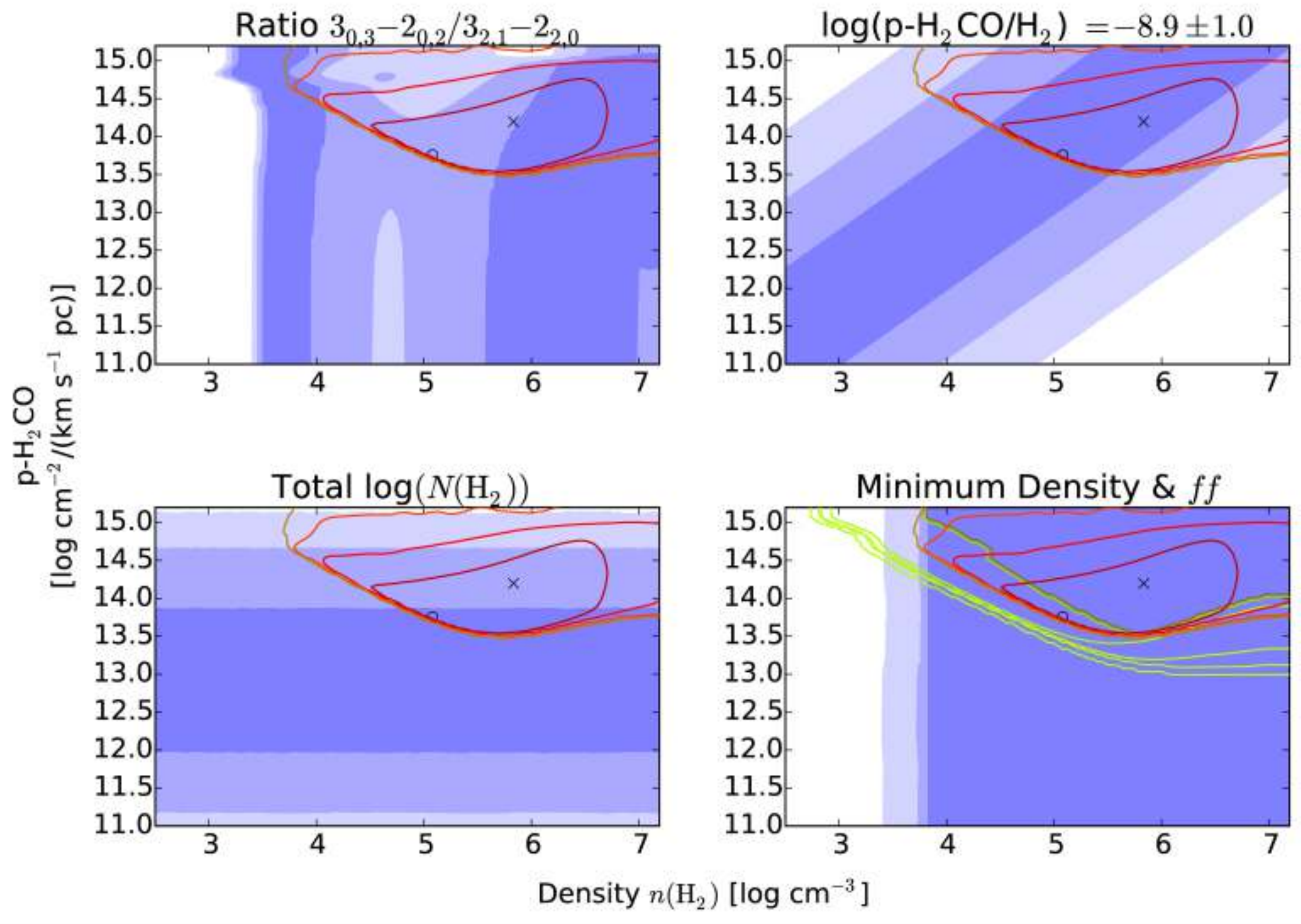}
{The constraints in density-column parameter space.
See Figure \ref{fig:coltemconstraints} for details.}
{fig:denscolconstraints}{0.5}{7in}

\Figure{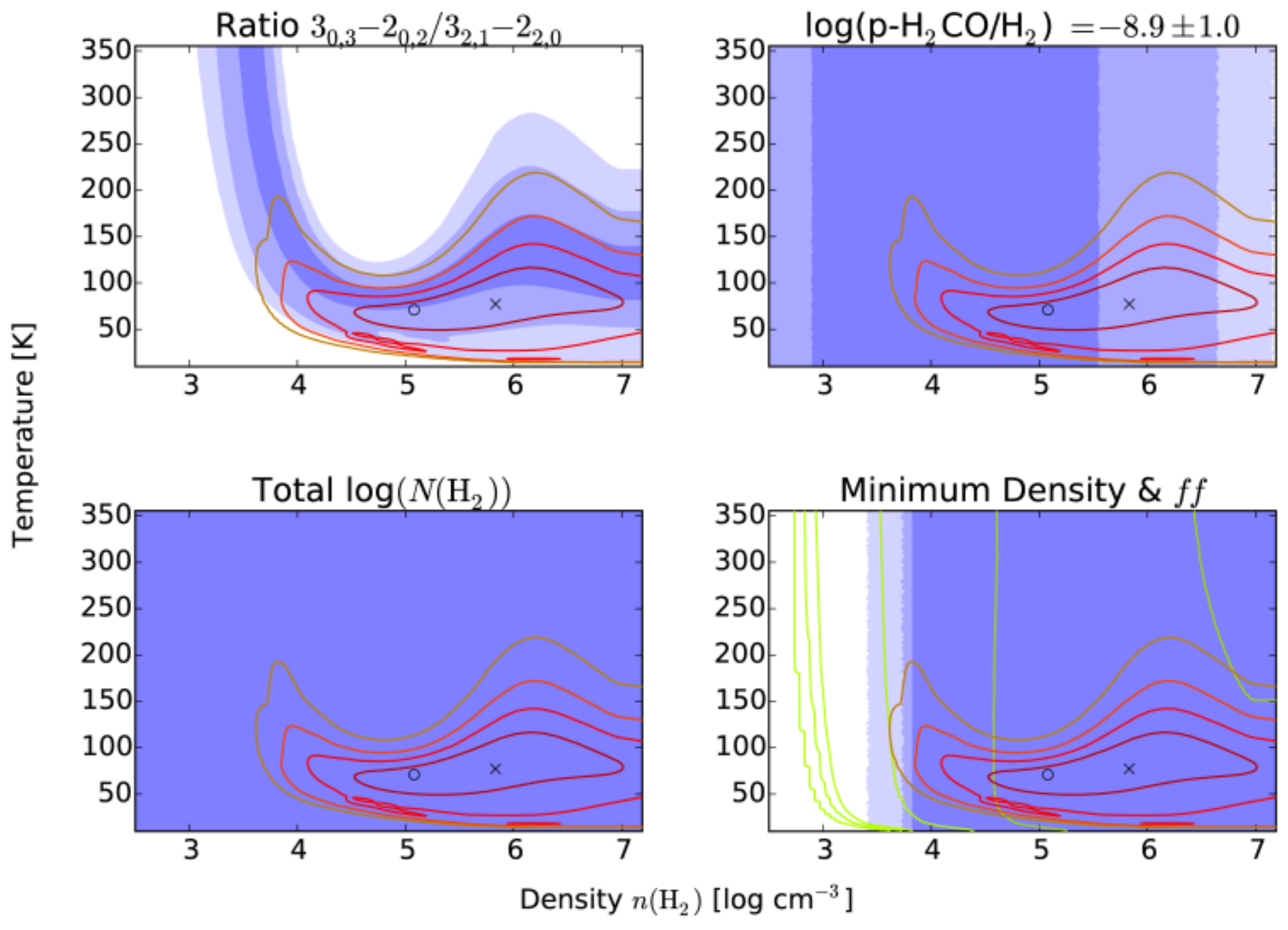}
{The constraints in density-temperature parameter space.
See Figure \ref{fig:coltemconstraints} for details.}
{fig:denstemconstraints}{0.5}{7in}

\Figure{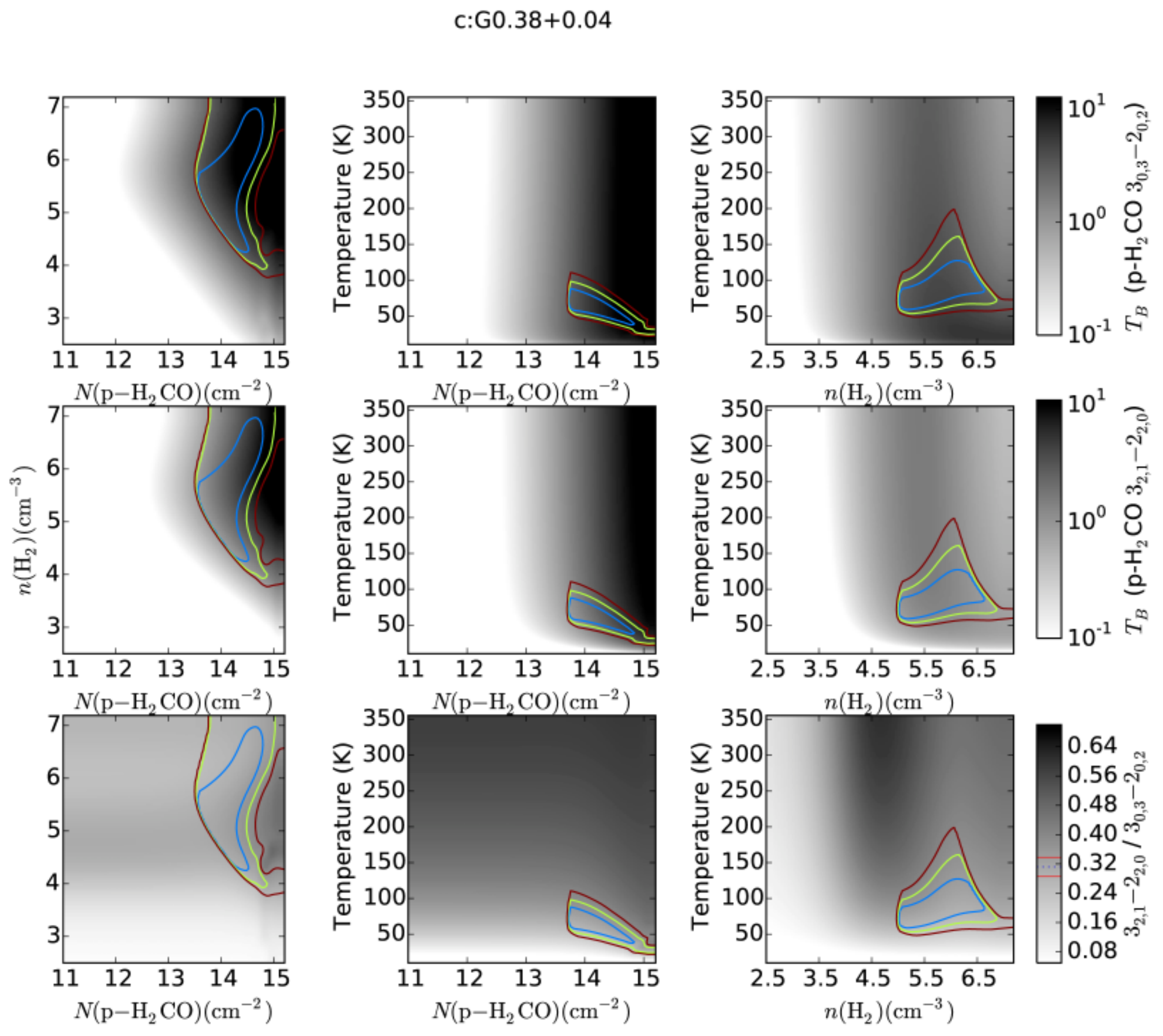}
{The line brightness of \para \threeohthree (top row) and \para \threetwoone
(middle row) and \Rone (bottom row) in the three different projections of parameter space.  The
grayscale images correspond to a slice through the parameter spaces at the
location of the best-fit parameter.  The colored contours show the allowed
marginalized regions in each parameter space as described in the
Figure \ref{fig:coltemconstraints} caption.}
{fig:parsonbrightness}{0.5}{7in}

\Figure{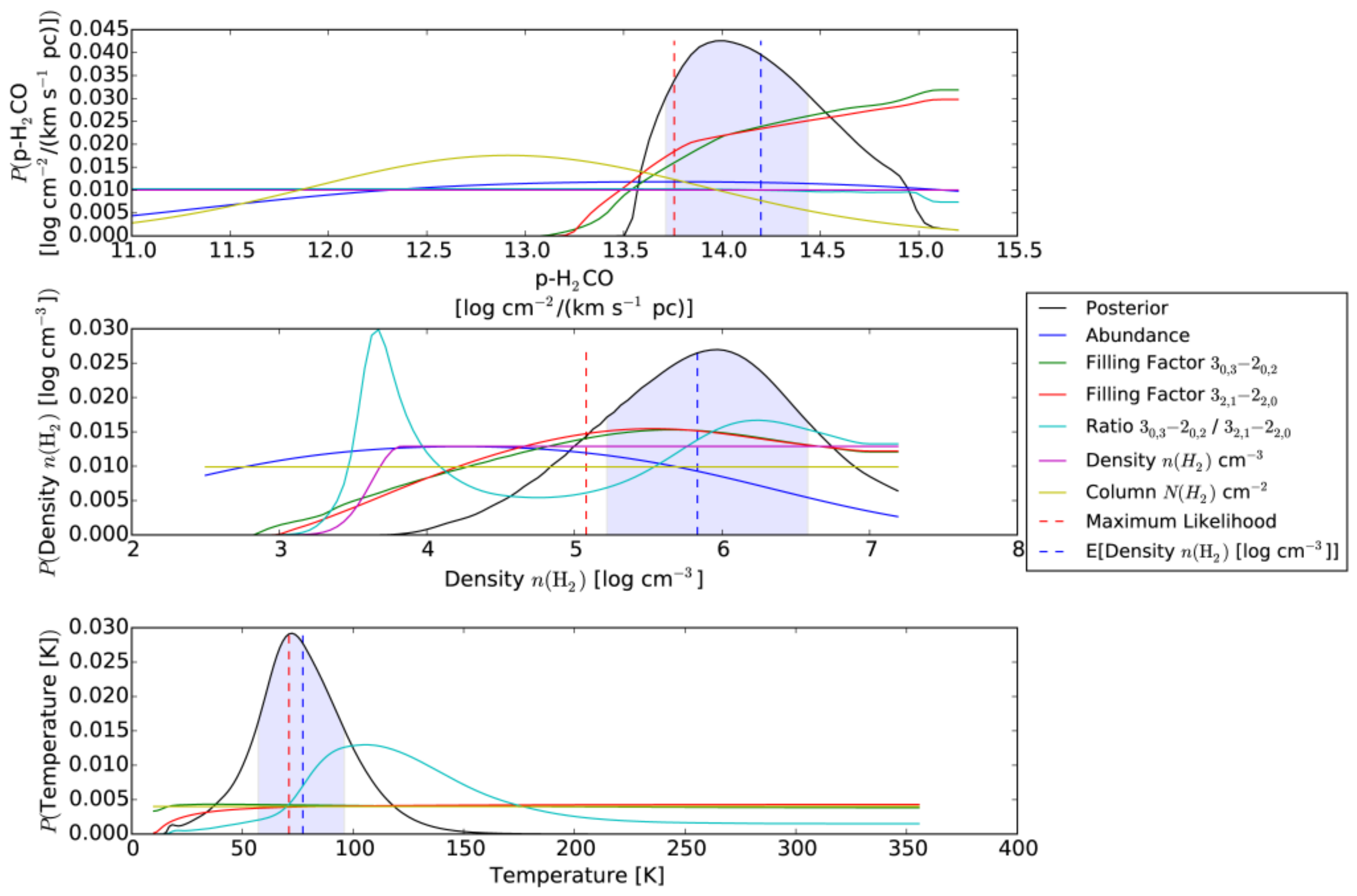}
{The one-dimensional probability distributions for each parameter.  The legend
describes the lines.  The blue shaded area shows the highest 68\% probability
region.  The maximum-likelihood and expectation value are generally
close for temperature but can be very different for column and density.}
{fig:oneddistributions}{0.55}{7in}

\section{Smoothed maps}
\label{appendix:smoothed}
In many regions, signal was detected over large areas though the peak
signal-to-noise ratio remained low.  In this Appendix, we display smoothed
versions of many of the main figures in the paper to provide a record of
features that were weakly detected.  The smoothed maps may also give a sense
of how the temperature varies with spatial scale.

The cubes were smoothed in the spatial dimensions with a
$\sigma_{FWHM}=33.84\arcsec$ Gaussian to achieve a resolution 45\arcsec and
$\sigma_{v, FWHM} = 3$ \kms in the spectral direction.  The smoothed cubes were
also downsampled by a factor of 2 in the spectral direction (the smoothing and
downsampling are independent).  
The same signal extraction analysis described in Section \ref{sec:signal} was used with a
threshold $T_A > 3\sigma$.  

Figure \ref{fig:peaksn} shows the peak signal-to-noise in both the unsmoothed
and the smoothed maps to indicate where the temperature measurements can be
relied upon.  These figures show the variable-opacity mask used in most of the
other figures in this paper, with black corresponding to fully transparent and
white fully opaque.

\RotFigureTwoAA
{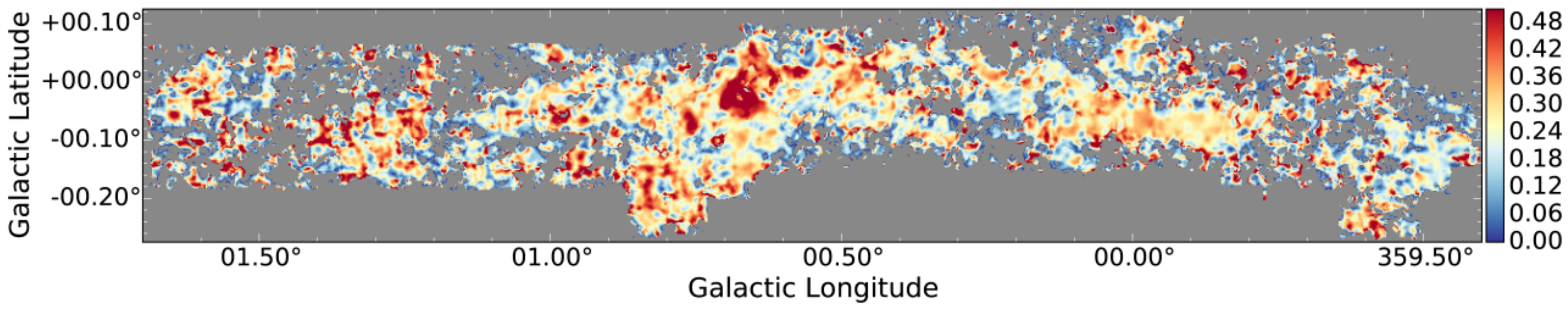}
{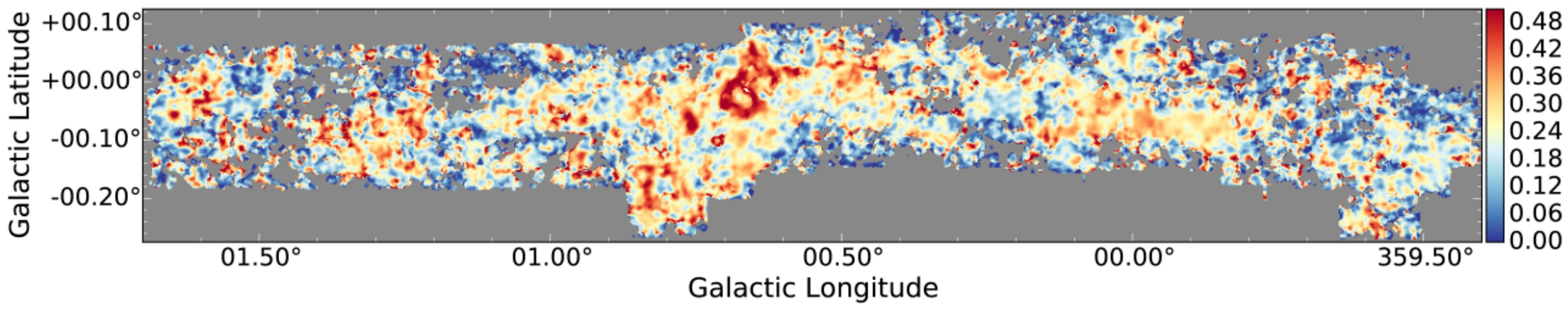}
{Same as Figure \ref{fig:ratiomaps}, but smoothed with a 34\arcsec FWHM Gaussian.
}
{fig:ratiomapssm}{1}{9.5in}

\RotFigureThreeAA
{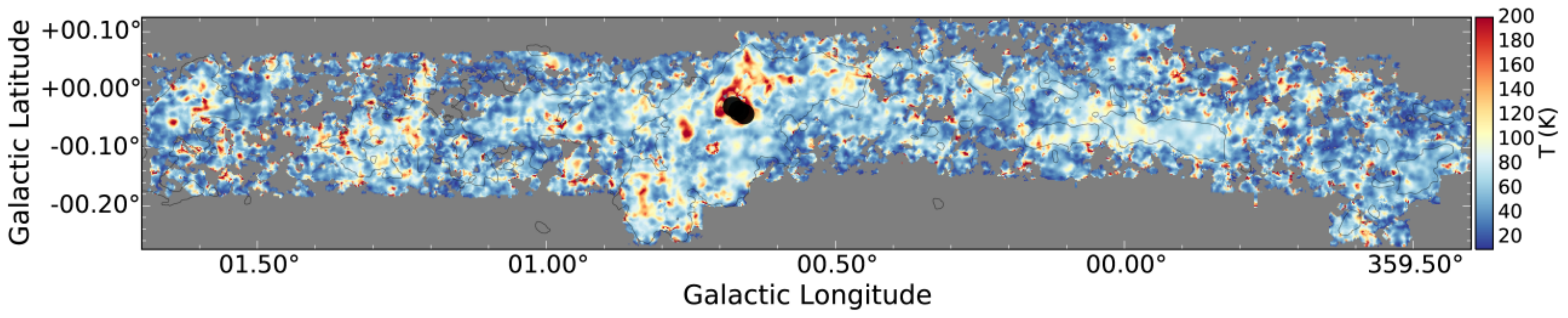}
{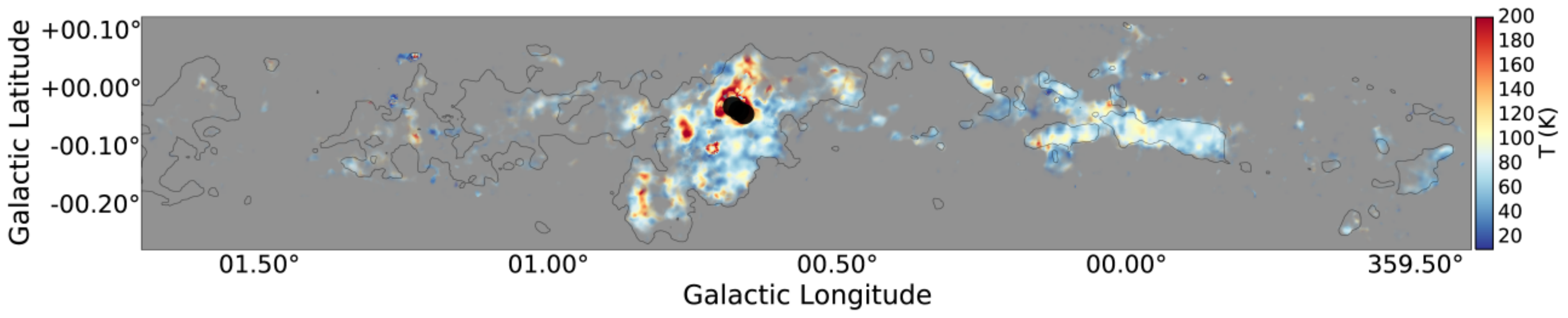}
{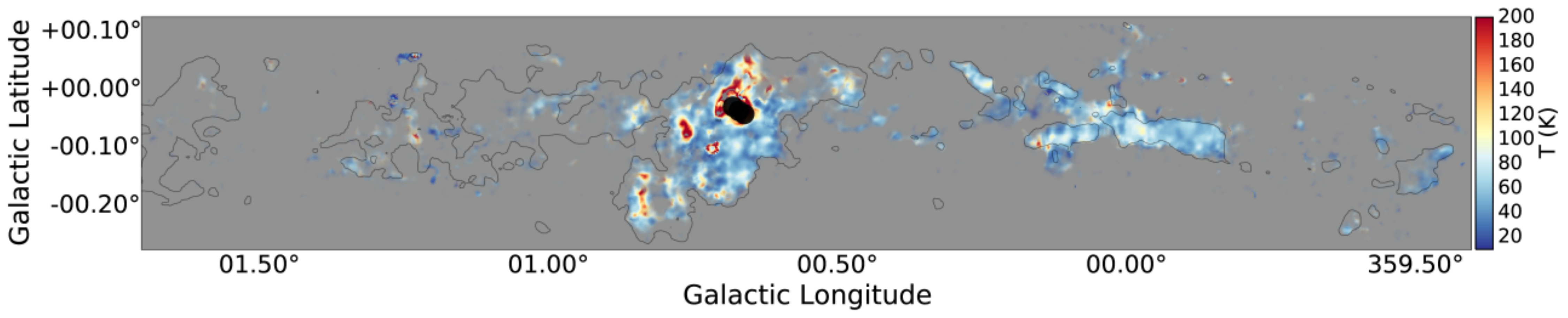}
{Temperature maps using the smoothed data assuming fixed $X_{\para}=1.2\ee{-9}$
and $n(\hh)$.  The maps in (b) and (c) are masked by signal-to-noise as in
Figure \ref{fig:ratiomaps}.  The thin black contours are at a column
$N(\hh)=5\ee{22}$ \persc from the Herschel SED fit maps.
(a) $n(\hh) = 10^4$ \percc, no mask
(b) $n(\hh) = 10^4$ \percc
(c) $n(\hh) = 10^5$ \percc.
Figure \ref{fig:temmapnosmooth_abund} shows the same figures, but with varying
abundance and fixed $n(\hh)$.
}
{fig:temmap}{1}{9.5in}

\RotFigureTwoAA
{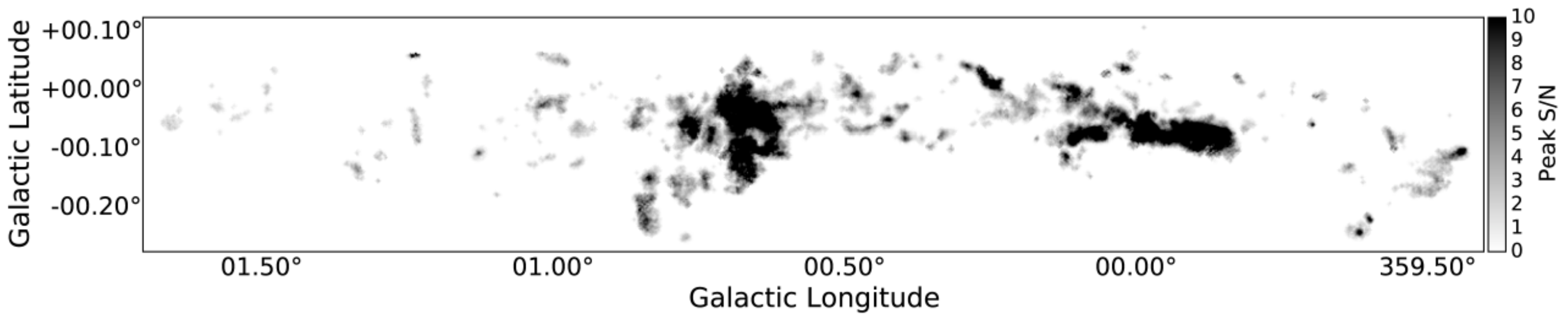}
{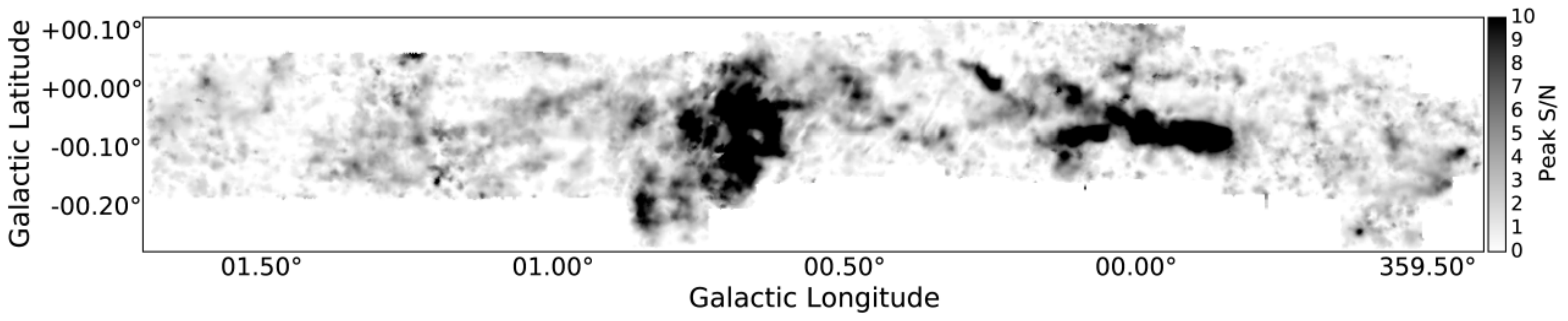}
{Map of the peak signal-to-noise in the \para \threeohthree line with no
smoothing (top) and with 33.84\arcsec smoothing (bottom).  
These maps give an indication of the reliability of the temperatures extracted
in Figure \ref{fig:temmap}.  The colorbars are intentionally saturated at
S/N$>10$ since above this threshold, the temperatures are reliable as long as
they are in the $T_G<150$ K regime.
}
{fig:peaksn}{1}{9.5in}

\clearpage
\subsection{Dendrogram cubes}

The dendrogram-extracted catalog was used to build a temperature PPV data cube
for visualization purposes.  Each voxel was assigned a value corresponding to
the temperature in the smallest dendrogram structure it was included in.  For
example, a voxel corresponding to a leaf has the temperature measured for that
leaf, while a voxel that is not in a leaf but is part of a structure would be
assigned the temperature of the \emph{smallest} structure in which it is
included.  The rest of the
cube was assigned NAN values.  This cube is similar to a smoothed cube, but
smoothed over connected structures rather than with a symmetric beam.  The cube
is interesting for visualization purposes, but should be treated with some
skepticism: structures that are connected in PPV space in the \threeohthree
line do not generally have the same, constant temperature, but they are forced
by construction to have the same temperature in this cube.  Despite this
caveat, comparison of the dendrogram figures (Figure \ref{fig:dendrotemmap}) to
the directly averaged figures (Figures \ref{fig:temmap} and
\ref{fig:temmapnosmooth}) shows that the dendrogram objects are effectively
averages over the included area in most cases.  Since there is less noise in
these figures, it is easier to identify interesting features in them by eye.

\clearpage

\RotFigureThreeAA
{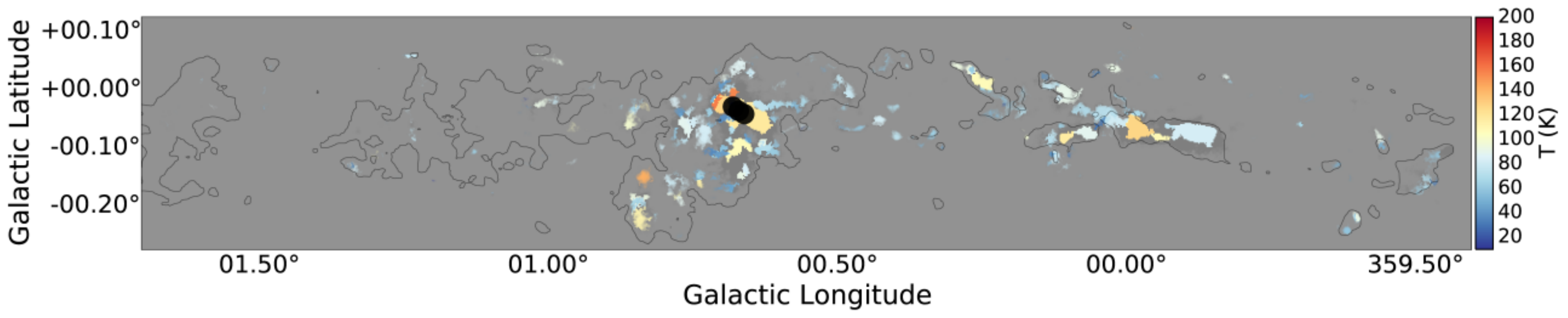}
{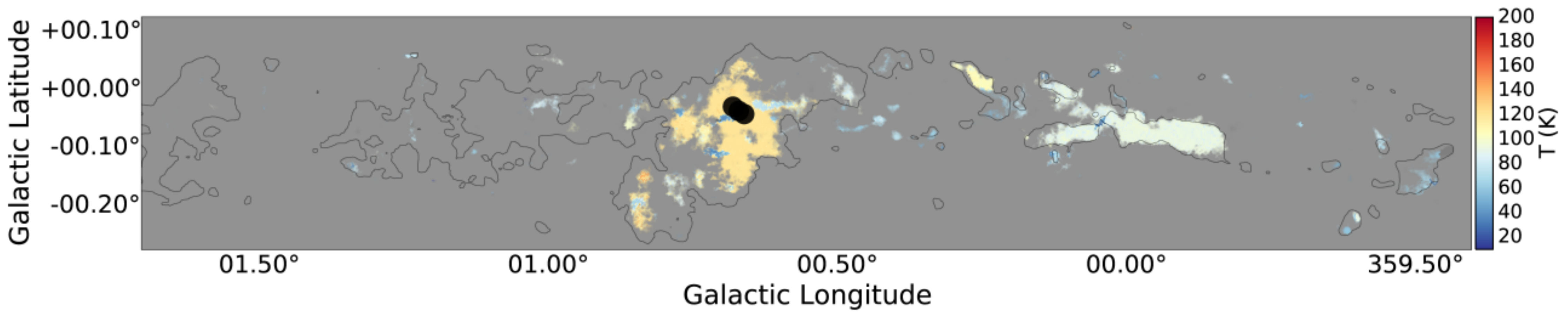}
{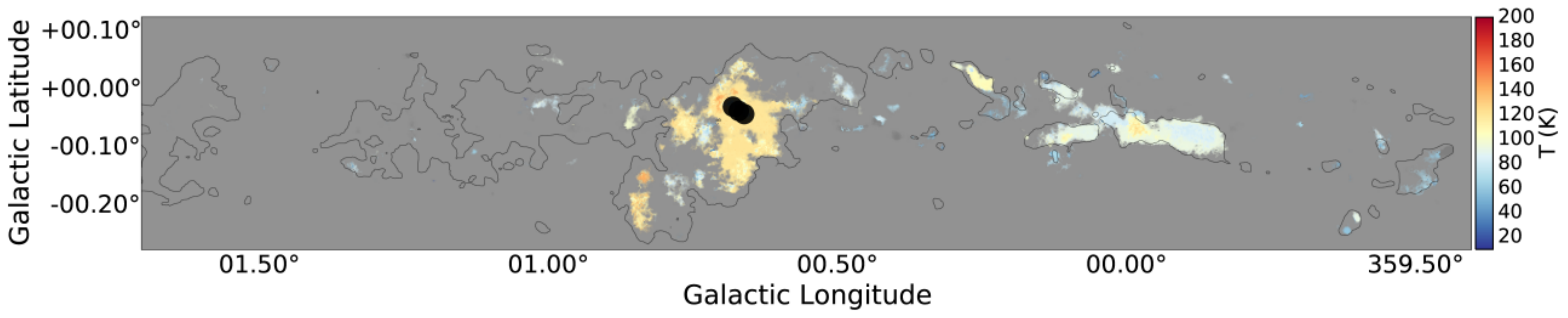}
{Maps generated by building a data cube in which each voxel has been replaced
with the average temperature from the smallest associated dendrogram-extracted
clump.  These maps can be thought of as adaptively-smoothed maps, where the
smoothing kernel is matched to the source size.
(a) The mean temperature along each line of sight through the dendrogram-extracted
cube, where only the leaf nodes have been included
\newline
(b) The same as (a), but including all ancestor nodes in addition to the leaf
nodes. 
\newline
(c) The same as (b), but weighted by the \threeohthree brightness.
\newline
In all three panels, regions of lower signal-to-noise, and therefore less
reliable
temperature, are grayed out with a filter that gets more opaque toward lower
signal-to-noise.  The thin contours are from the Herschel HiGal
dust SED fit at a level $N(\hh)=5\ee{22}$ \persc and are included to provide a
visual reference for comparison between the temperature maps.
}
{fig:dendrotemmap}{1}{9.5in}

\RotFigureThreeAA
{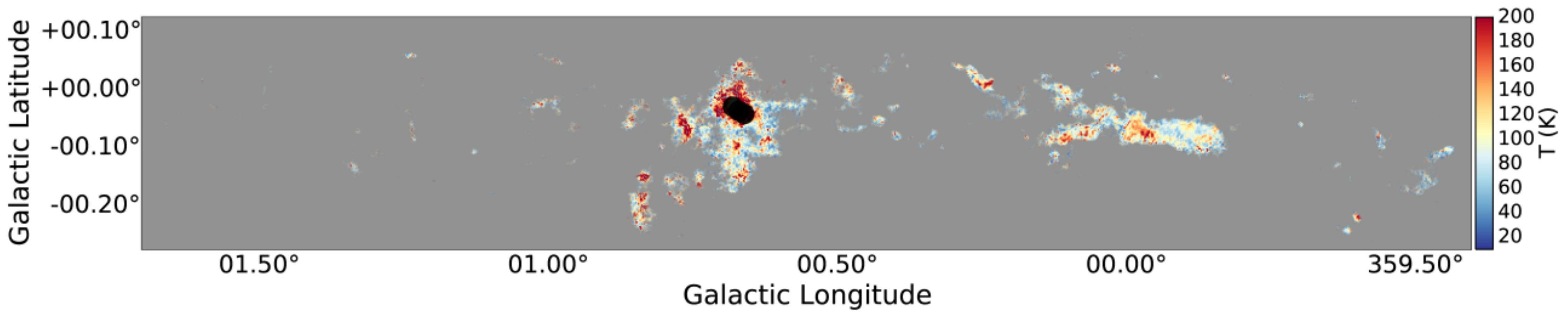}
{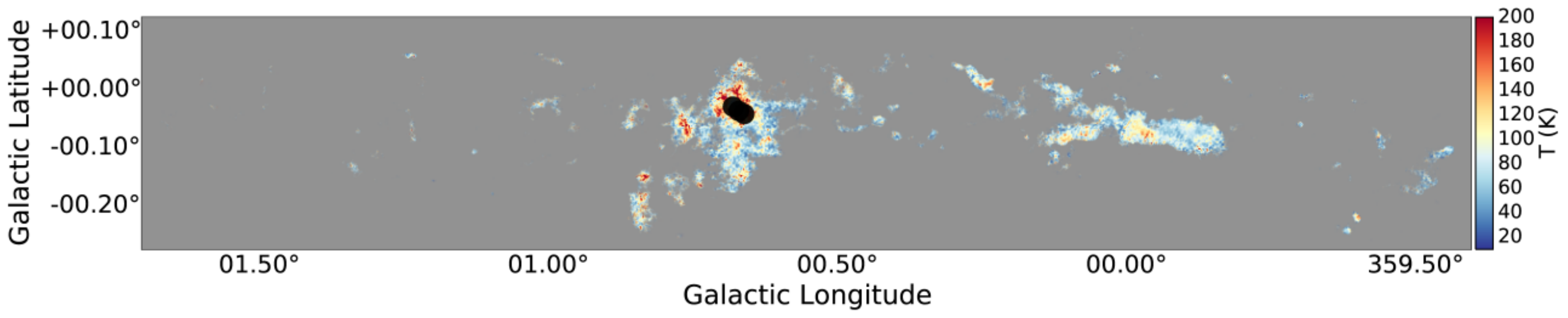}
{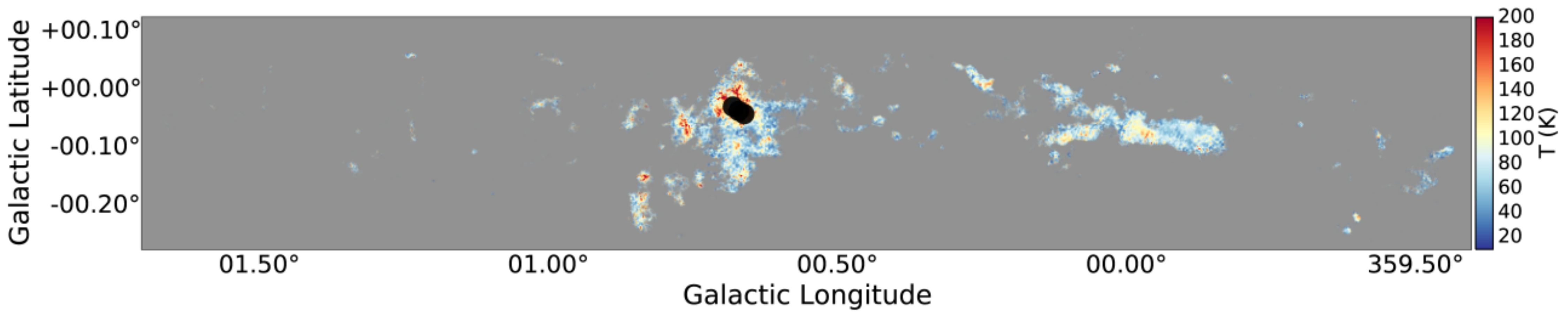}
{Temperature maps assuming fixed $X(\para)$ and $n(\hh)=1\ee{4}$ \percc.  The
maps are masked by signal-to-noise as in Figure \ref{fig:ratiomaps}.  The thin
black contours are at a column $N(\hh)=5\ee{22}$ \persc from the Herschel SED
fit maps.
(a) $X(\para) = 1\ee{-8}$
(b) $X(\para) = 1.2\ee{-9}$
(c) $X(\para) = 1\ee{-10}$
}
{fig:temmapnosmooth_abund}{1}{9.5in}

\section{Temperature balance plots without data}
\label{sec:despoticplots}
Figure \ref{fig:temvsfwhm_regions} shows theoretical curves with all data
points superposed.  For clarity, we reproduce that figure with no data plotted
here in Figure \ref{fig:tvssigma_despotic}.

The legend shows the various modifications to model parameters used.  The
fiducial model is indicated by a dashed black line with $\zeta_{CR} =
1\ee{-17}$ \pers, $n=10^4$ \percc, $L=5$ pc, velocity gradient $dv/dr=5$ \kms
pc$^{-1}$, dust temperature $T_D=25$ K, dust radiation temperature $T_{D,rad} =
10$ K, and interstellar radiation field $G_0$.  The green solid curve shows a
model that follows both a size-linewidth relation $L=5 \sigma_5^{0.7}$ pc and
is isobaric with $n=10^{4.25} \sigma_5^2$ \percc, where $\sigma_5$ is the
velocity dispersion in units of 5 \kms.  The green dashed curve shows the same
with only the size-linewidth relation included.  The green dotted curve shows a
high cosmic-ray ionization rate with no turbulent heating and is meant to show
the absolute floor temperature guaranteed by such a high CRIR.

\Figure{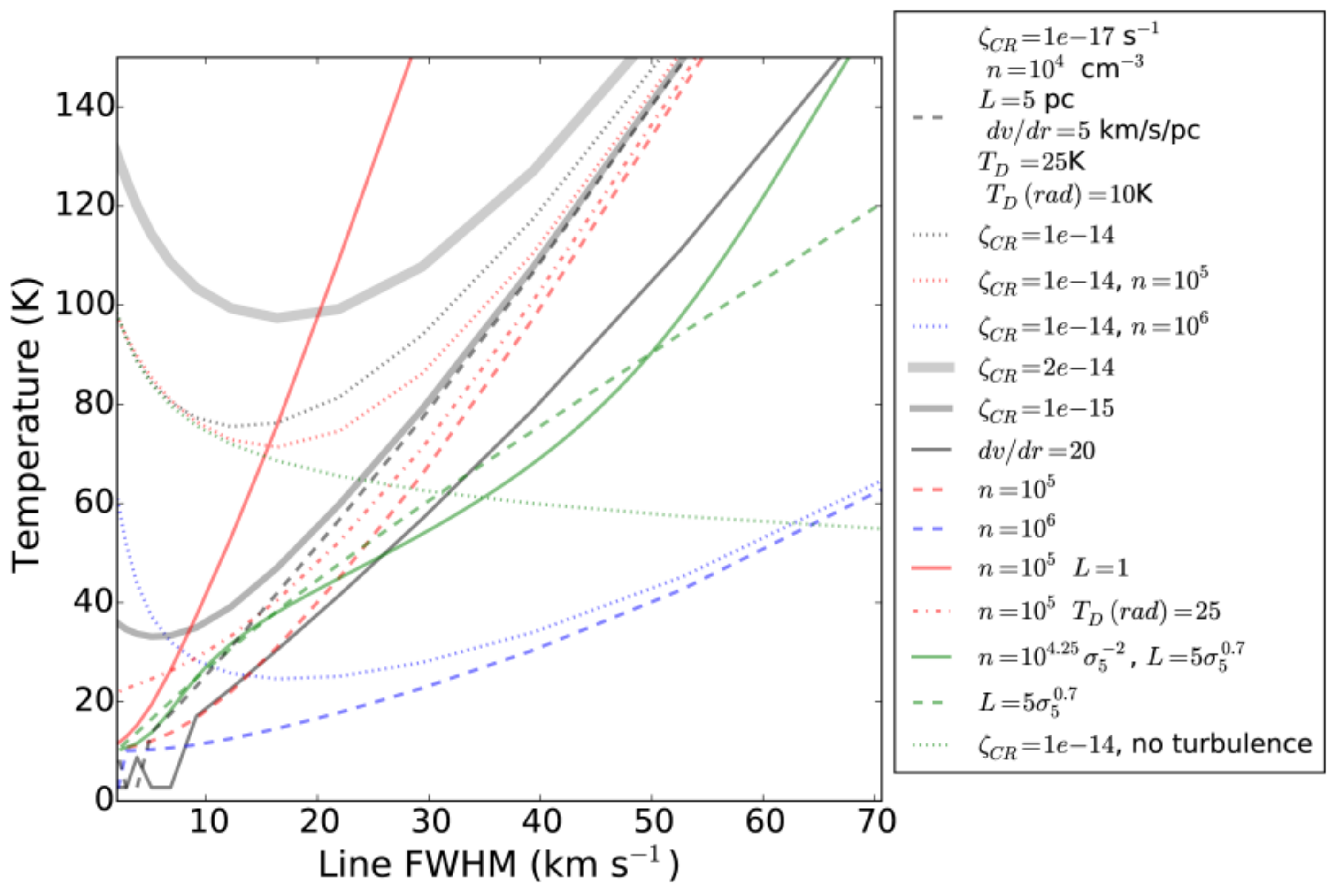}
{Plots of the DESPOTIC-derived equilibrium temperature as a function of line
width with no data overlaid.  The curves are identical to Figure
\ref{fig:temvsfwhm_regions}.}
{fig:tvssigma_despotic}{0.5}{7in}

\section{Source Table}
We include an excerpt from the source tables here.  The full tables are
available in digital form from
\url{https://raw.githubusercontent.com/adamginsburg/APEX_CMZ_H2CO/v1.1/tables/PPV_H2CO_Temperature_orbit.ipac}
and
\url{https://raw.githubusercontent.com/adamginsburg/APEX_CMZ_H2CO/v1.1/tables/fitted_line_parameters_Chi2Constraints_orbit.ipac}.
Both tables are also available on the dataverse page.
A complete description of the columns is available at
\url{https://github.com/adamginsburg/APEX_CMZ_H2CO/blob/v1.1/tables/README.rst}

\begin{table*}[htp]
\caption{\formaldehyde Line Parameters and Fit Properties}
\begin{tabular}{cccccccc}
\label{tab:handregions}
Source Name & $R_1$ & $\sigma_v$ & $v_{lsr}$ & $T_B(H_2CO)$ & log($n(H_2)$) & $T_{gas}$ & $T_{gas, turb}$ \\
$\mathrm{}$ &  & $\mathrm{km\,s^{-1}}$ & $\mathrm{km\,s^{-1}}$ & $\mathrm{K}$ & $\mathrm{}$ & $\mathrm{K}$ & $\mathrm{K}$ \\
\hline
Menten1 & $0.35\pm0.03$ & $4.93\pm0.15$ & $55.85\pm0.15$ & $1.65\pm0.05$ & 22.8 & $83.55^{+16.65}_{-44.41}$ & 36.8 \\
Menten7 & $0.76\pm0.10$ & $10.34\pm0.97$ & $54.41\pm0.98$ & $0.35\pm0.04$ & 22.4 & $350.00^{+81.88}_{-5.55}$ & 92.6 \\
dn:G0.41+0.05box & $0.29\pm0.02$ & $11.19\pm0.27$ & $19.59\pm0.27$ & $1.55\pm0.04$ & 23.1 & $66.90^{+11.10}_{-18.04}$ & 102 \\
20 kms & $0.32\pm0.00$ & $8.15\pm0.02$ & $17.18\pm0.02$ & $12.39\pm0.03$ & 23.3 & $62.73^{+12.49}_{-20.82}$ & 34.7 \\
CoolSpot & $0.22\pm0.02$ & $7.26\pm0.39$ & $27.60\pm0.56$ & $3.79\pm0.13$ & 22.9 & $48.86^{+11.10}_{-6.94}$ & 140 \\
G1.34-0.13 & $0.32\pm0.02$ & $5.19\pm0.16$ & $32.27\pm0.15$ & $1.68\pm0.05$ & 22.7 & $75.22^{+12.49}_{-29.14}$ & 41.3 \\
G0.43-0.05 & $0.29\pm0.01$ & $10.25\pm0.13$ & $89.93\pm0.14$ & $2.84\pm0.03$ & 22.5 & $66.90^{+9.71}_{-15.27}$ & 87 \\
G359.49-0.14 & $0.25\pm0.02$ & $9.82\pm0.20$ & $-56.04\pm0.20$ & $1.57\pm0.03$ & 22.8 & $55.80^{+8.33}_{-9.71}$ & 58.1 \\
G0.76-0.08 & $0.76\pm0.05$ & $12.49\pm0.35$ & $19.83\pm0.26$ & $1.62\pm0.10$ & 23.1 & $350.00^{+38.86}_{-5.55}$ & 131 \\
G0.24-0.05box & $0.20\pm0.02$ & $7.50\pm0.18$ & $79.72\pm0.18$ & $1.31\pm0.03$ & 22.7 & $46.08^{+6.94}_{-6.94}$ & 56.7 \\
G0.85-0.04box & $0.31\pm0.01$ & $17.99\pm0.26$ & $4.03\pm0.27$ & $1.16\pm0.02$ & 22.9 & $73.84^{+8.33}_{-23.59}$ & 129 \\
G359.8\_box & $0.23\pm0.01$ & $10.53\pm0.25$ & $11.55\pm0.17$ & $0.72\pm0.02$ & 22.5 & $53.02^{+5.55}_{-8.33}$ & 47.5 \\
G0.83-0.19box & $0.36\pm0.01$ & $10.11\pm0.14$ & $43.97\pm0.14$ & $1.76\pm0.03$ & 22.9 & $89.10^{+9.71}_{-43.02}$ & 50.3 \\
Map\_002 Off 2 & $0.18\pm0.04$ & $11.23\pm1.33$ & $6.48\pm1.54$ & $0.22\pm0.02$ & 22.8 & $46.08^{+15.27}_{-12.49}$ & 66.6 \\
Map\_007 Off 2 & $0.22\pm0.01$ & $26.86\pm0.22$ & $45.78\pm0.27$ & $1.30\pm0.01$ & 23.2 & $51.63^{+8.33}_{-6.94}$ & 142 \\
Map\_012 Off 2 & $0.31\pm0.06$ & $9.83\pm1.17$ & $7.63\pm1.34$ & $0.24\pm0.02$ & 22.5 & $79.39^{+27.76}_{-51.35}$ & 59.2 \\
Map\_023 Off 3 & $0.08\pm0.04$ & $19.67\pm1.30$ & $33.04\pm1.25$ & $0.28\pm0.01$ & 22.5 & $28.04^{+9.71}_{-5.55}$ & 114 \\
Map\_027 Off 2 & $0.15\pm0.03$ & $10.86\pm0.45$ & $81.72\pm0.43$ & $0.52\pm0.02$ & 22.8 & $37.76^{+8.33}_{-5.55}$ & 51.3 \\
Map\_055 Off 3 & $0.27\pm0.02$ & $8.36\pm0.27$ & $10.94\pm0.20$ & $0.81\pm0.03$ & 22.8 & $62.73^{+9.71}_{-15.27}$ & 47 \\
Map\_116 Off 1 & $0.33\pm0.03$ & $13.42\pm0.53$ & $64.31\pm0.55$ & $0.41\pm0.02$ & 22.5 & $80.78^{+16.65}_{-37.47}$ & 79.3 \\
Map\_122 Off 1 & $0.09\pm0.05$ & $22.05\pm1.61$ & $10.11\pm1.55$ & $0.19\pm0.01$ & 22.7 & $30.82^{+12.49}_{-6.94}$ & 128 \\
\hline
\end{tabular}
\par

\end{table*}
\begin{table*}[htp]
\caption{\formaldehyde Parameters and Fit Properties for dendrogram-selected clumps}
\begin{tabular}{cccccccc}
\label{tab:dendroregions}
Source ID & $R_1$ & $\sigma_v$ & $v_{lsr}$ & Max $T_B(3_{0,3})$ & log($n(H_2)$) & $T_{gas}$ & $T_{gas, turb}$ \\
$\mathrm{}$ &  & $\mathrm{km\,s^{-1}}$ & $\mathrm{km\,s^{-1}}$ & $\mathrm{K}$ & $\mathrm{}$ & $\mathrm{K}$ & $\mathrm{K}$ \\
\hline
0 & $0.127\pm0.022$ & 0.00211 & -111 & 0.513 & 22.3 & $34.98^{+5.55}_{-4.16}$ & 35 \\
10 & $0.206\pm0.010$ & 0.0033 & -57.7 & 0.525 & 22.8 & $50.24^{+6.94}_{-4.16}$ & 56 \\
20 & $0.055\pm0.026$ & 0.00251 & -38 & 0.474 & 22.7 & $23.88^{+6.94}_{-1.39}$ & 59 \\
30 & $0.245\pm0.006$ & 0.00744 & -15 & 0.683 & 22.7 & $59.96^{+6.94}_{-6.94}$ & 118 \\
40 & $0.322\pm0.000$ & 0.019 & 27.7 & 2.3 & 23.1 & $79.39^{+8.33}_{-26.37}$ & 122 \\
50 & $0.332\pm0.002$ & 0.0141 & 28.2 & 0.99 & 23.1 & $86.33^{+9.71}_{-23.59}$ & 147 \\
60 & $0.317\pm0.001$ & 0.00886 & 14.7 & 2.3 & 23.3 & $78.00^{+11.10}_{-20.82}$ & 88 \\
70 & $0.382\pm0.001$ & 0.0172 & 47.4 & 3.97 & 23.3 & $98.82^{+11.10}_{-68.00}$ & 76 \\
80 & $0.420\pm0.002$ & 0.0116 & 25 & 0.604 & 23 & $121.02^{+8.33}_{-116.57}$ & 155 \\
90 & $0.382\pm0.001$ & 0.0172 & 47.4 & 3.97 & 23.3 & $105.76^{+18.04}_{-63.84}$ & 78 \\
100 & $0.235\pm0.014$ & 0.00394 & 11.4 & 0.411 & 23 & $54.41^{+5.55}_{-8.33}$ & 118 \\
110 & $0.460\pm0.001$ & 0.016 & 49 & 3.97 & 23.6 & $140.45^{+2.78}_{-215.10}$ & 104 \\
120 & $0.289\pm0.003$ & 0.00591 & 31.3 & 0.883 & 23.1 & $69.67^{+5.55}_{-16.65}$ & 47 \\
130 & $0.222\pm0.005$ & 0.00512 & 28.5 & 0.82 & 22.6 & $53.02^{+5.55}_{-6.94}$ & 54 \\
140 & $0.089\pm0.018$ & 0.00371 & 28.1 & 0.527 & 22.6 & $29.43^{+5.55}_{-1.39}$ & 107 \\
150 & $0.257\pm0.014$ & 0.0051 & 31.8 & 0.469 & 22.7 & $64.12^{+9.71}_{-6.94}$ & 128 \\
160 & $0.311\pm0.009$ & 0.00281 & 29.2 & 0.816 & 23.1 & $72.45^{+5.55}_{-23.59}$ & 145 \\
170 & $0.353\pm0.003$ & 0.00576 & 45 & 0.819 & 23 & $91.88^{+6.94}_{-37.47}$ & 56 \\
180 & $0.318\pm0.003$ & 0.00261 & 29.9 & 1.16 & 23.2 & $75.22^{+4.16}_{-26.37}$ & 88 \\
190 & $0.311\pm0.001$ & 0.00734 & 50.7 & 1.72 & 22.8 & $73.84^{+4.16}_{-24.98}$ & 37 \\
200 & $0.321\pm0.001$ & 0.00706 & 50.6 & 1.72 & 22.8 & $76.61^{+2.78}_{-29.14}$ & 38 \\
210 & $0.351\pm0.003$ & 0.0061 & 49.7 & 1.27 & 23.1 & $86.33^{+4.16}_{-40.24}$ & 138 \\
220 & $0.161\pm0.022$ & 0.00306 & 46.4 & 0.416 & 23 & $40.53^{+6.94}_{-4.16}$ & 82 \\
230 & $0.215\pm0.015$ & 0.00261 & 47.4 & 0.405 & 22.9 & $53.02^{+8.33}_{-4.16}$ & 57 \\
240 & $0.265\pm0.009$ & 0.00225 & 49.4 & 0.626 & 22.6 & $61.35^{+4.16}_{-12.49}$ & 75 \\
250 & $0.313\pm0.021$ & 0.00435 & 58 & 0.418 & 23.1 & $75.22^{+11.10}_{-22.20}$ & 93 \\
260 & $0.190\pm0.015$ & 0.00473 & 59.6 & 0.433 & 22.7 & $48.86^{+8.33}_{-2.78}$ & 98 \\
270 & $0.261\pm0.007$ & 0.0042 & 65.2 & 0.827 & 22.7 & $61.35^{+4.16}_{-11.10}$ & 74 \\
280 & $0.653\pm0.007$ & 0.00284 & 70.7 & 1 & 24 & $350.00^{+1.39}_{-5.55}$ & 85 \\
290 & $0.220\pm0.009$ & 0.0044 & 79.9 & 0.373 & 23 & $55.80^{+9.71}_{-2.78}$ & 80 \\
300 & $0.242\pm0.020$ & 0.00357 & 80.2 & 0.359 & 23.1 & $55.80^{+8.33}_{-9.71}$ & 93 \\
310 & $0.246\pm0.012$ & 0.00331 & 82.7 & 0.446 & 22.9 & $62.73^{+11.10}_{-4.16}$ & 57 \\
320 & $0.251\pm0.023$ & 0.00295 & 85.9 & 0.456 & 23.1 & $57.18^{+8.33}_{-12.49}$ & 69 \\
330 & $0.140\pm0.028$ & 0.00247 & 90 & 0.371 & 22.9 & $37.76^{+8.33}_{-4.16}$ & 61 \\
340 & $0.054\pm0.028$ & 0.00345 & 97.1 & 0.532 & 22.6 & $22.49^{+6.94}_{-2.78}$ & 70 \\
350 & $0.164\pm0.014$ & 0.00334 & 117 & 0.504 & 22.7 & $43.31^{+6.94}_{-1.39}$ & 59 \\
\hline
\end{tabular}
\par

\end{table*}


\begin{thebibliography}{114}
\expandafter\ifx\csname natexlab\endcsname\relax\def\natexlab#1{#1}\fi

\bibitem[{{Aguirre} {et~al.}(2011){Aguirre}, {Ginsburg}, {Dunham}, {Drosback},
  {Bally}, {Battersby}, {Bradley}, {Cyganowski}, {Dowell}, {Evans}, {Glenn},
  {Harvey}, {Rosolowsky}, {Stringfellow}, {Walawender}, \&
  {Williams}}]{Aguirre2011a}
{Aguirre}, J.~E. {et~al.} 2011, \apjs, 192, 4

\bibitem[{{Ao} {et~al.}(2013){Ao}, {Henkel}, {Menten}, {Requena-Torres},
  {Stanke}, {Mauersberger}, {Aalto}, {M{\"u}hle}, \& {Mangum}}]{Ao2013a}
{Ao}, Y. {et~al.} 2013, \aap, 550, A135

\bibitem[{{Astropy Collaboration} {et~al.}(2013){Astropy Collaboration},
  {Robitaille}, {Tollerud}, {Greenfield}, {Droettboom}, {Bray}, {Aldcroft},
  {Davis}, {Ginsburg}, {Price-Whelan}, {Kerzendorf}, {Conley}, {Crighton},
  {Barbary}, {Muna}, {Ferguson}, {Grollier}, {Parikh}, {Nair}, {Unther},
  {Deil}, {Woillez}, {Conseil}, {Kramer}, {Turner}, {Singer}, {Fox}, {Weaver},
  {Zabalza}, {Edwards}, {Azalee Bostroem}, {Burke}, {Casey}, {Crawford},
  {Dencheva}, {Ely}, {Jenness}, {Labrie}, {Lim}, {Pierfederici}, {Pontzen},
  {Ptak}, {Refsdal}, {Servillat}, \& {Streicher}}]{Astropy-Collaboration2013a}
{Astropy Collaboration} {et~al.} 2013, \aap, 558, A33

\bibitem[{{Bally} {et~al.}(2010){Bally}, {Aguirre}, {Battersby}, {Bradley},
  {Cyganowski}, {Dowell}, {Drosback}, {Dunham}, {Evans}, {Ginsburg}, {Glenn},
  {Harvey}, {Mills}, {Merello}, {Rosolowsky}, {Schlingman}, {Shirley},
  {Stringfellow}, {Walawender}, \& {Williams}}]{Bally2010a}
{Bally}, J. {et~al.} 2010, \apj, 721, 137

\bibitem[{{Battersby} {et~al.}(2014){Battersby}, {Bally}, {Dunham}, {Ginsburg},
  {Longmore}, \& {Darling}}]{Battersby2014a}
{Battersby}, C., {Bally}, J., {Dunham}, M., {Ginsburg}, A., {Longmore}, S., \&
  {Darling}, J. 2014, \apj, 786, 116

\bibitem[{{Battersby} {et~al.}(2011){Battersby}, {Bally}, {Ginsburg},
  {Bernard}, {Brunt}, {Fuller}, {Martin}, {Molinari}, {Mottram}, {Peretto},
  {Testi}, \& {Thompson}}]{Battersby2011a}
{Battersby}, C. {et~al.} 2011, \aap, 535, A128

\bibitem[{{Bayet} {et~al.}(2011){Bayet}, {Yates}, \& {Viti}}]{Bayet2011a}
{Bayet}, E., {Yates}, J., \& {Viti}, S. 2011, \apj, 728, 114

\bibitem[{{Belloche} {et~al.}(2013){Belloche}, {M{\"u}ller}, {Menten},
  {Schilke}, \& {Comito}}]{Belloche2013a}
{Belloche}, A., {M{\"u}ller}, H.~S.~P., {Menten}, K.~M., {Schilke}, P., \&
  {Comito}, C. 2013, ArXiv e-prints

\bibitem[{{Bonnell} {et~al.}(2006){Bonnell}, {Clarke}, \&
  {Bate}}]{Bonnell2006b}
{Bonnell}, I.~A., {Clarke}, C.~J., \& {Bate}, M.~R. 2006, \mnras, 368, 1296

\bibitem[{{Carey} {et~al.}(1998){Carey}, {Clark}, {Egan}, {Price}, {Shipman},
  \& {Kuchar}}]{Carey1998a}
{Carey}, S.~J., {Clark}, F.~O., {Egan}, M.~P., {Price}, S.~D., {Shipman},
  R.~F., \& {Kuchar}, T.~A. 1998, \apj, 508, 721

\bibitem[{{Clark} {et~al.}(2013){Clark}, {Glover}, {Ragan}, {Shetty}, \&
  {Klessen}}]{Clark2013a}
{Clark}, P.~C., {Glover}, S.~C.~O., {Ragan}, S.~E., {Shetty}, R., \& {Klessen},
  R.~S. 2013, \apjl, 768, L34

\bibitem[{{Dale} {et~al.}(2012){Dale}, {Ercolano}, \& {Bonnell}}]{Dale2012a}
{Dale}, J.~E., {Ercolano}, B., \& {Bonnell}, I.~A. 2012, \mnras, 427, 2852

\bibitem[{{Dame}(2011)}]{Dame2011b}
{Dame}, T.~M. 2011, ArXiv e-prints

\bibitem[{{de Vicente} {et~al.}(1997){de Vicente}, {Martin-Pintado}, \&
  {Wilson}}]{de-Vicente1997a}
{de Vicente}, P., {Martin-Pintado}, J., \& {Wilson}, T.~L. 1997, \aap, 320, 957

\bibitem[{{Dunham} {et~al.}(2010){Dunham}, {Rosolowsky}, {Evans}, {Cyganowski},
  {Aguirre}, {Bally}, {Battersby}, {Bradley}, {Dowell}, {Drosback}, {Ginsburg},
  {Glenn}, {Harvey}, {Merello}, {Schlingman}, {Shirley}, {Stringfellow},
  {Walawender}, \& {Williams}}]{Dunham2010a}
{Dunham}, M.~K. {et~al.} 2010, \apj, 717, 1157

\bibitem[{{Etxaluze} {et~al.}(2013){Etxaluze}, {Goicoechea}, {Cernicharo},
  {Polehampton}, {Noriega-Crespo}, {Molinari}, {Swinyard}, {Wu}, \&
  {Bally}}]{Etxaluze2013a}
{Etxaluze}, M. {et~al.} 2013, \aap, 556, A137

\bibitem[{{Falgarone} \& {Puget}(1995)}]{Falgarone1995a}
{Falgarone}, E. \& {Puget}, J.-L. 1995, \aap, 293, 840

\bibitem[{{Federrath} \& {Klessen}(2012)}]{Federrath2012a}
{Federrath}, C. \& {Klessen}, R.~S. 2012, \apj, 761, 156

\bibitem[{{Flower} {et~al.}(2000){Flower}, {Le Bourlot}, {Pineau des
  For{\^e}ts}, \& {Roueff}}]{Flower2000a}
{Flower}, D.~R., {Le Bourlot}, J., {Pineau des For{\^e}ts}, G., \& {Roueff}, E.
  2000, \mnras, 314, 753

\bibitem[{{Ghez} {et~al.}(2008){Ghez}, {Salim}, {Weinberg}, {Lu}, {Do}, {Dunn},
  {Matthews}, {Morris}, {Yelda}, {Becklin}, {Kremenek}, {Milosavljevic}, \&
  {Naiman}}]{Ghez2008a}
{Ghez}, A.~M. {et~al.} 2008, \apj, 689, 1044

\bibitem[{{Gillessen} {et~al.}(2009){Gillessen}, {Eisenhauer}, {Fritz},
  {Bartko}, {Dodds-Eden}, {Pfuhl}, {Ott}, \& {Genzel}}]{Gillessen2009b}
{Gillessen}, S., {Eisenhauer}, F., {Fritz}, T.~K., {Bartko}, H., {Dodds-Eden},
  K., {Pfuhl}, O., {Ott}, T., \& {Genzel}, R. 2009, \apjl, 707, L114

\bibitem[{{Gillessen} {et~al.}(2013){Gillessen}, {Eisenhauer}, {Fritz},
  {Pfuhl}, {Ott}, \& {Genzel}}]{Gillessen2013b}
{Gillessen}, S., {Eisenhauer}, F., {Fritz}, T.~K., {Pfuhl}, O., {Ott}, T., \&
  {Genzel}, R. 2013, in IAU Symposium, Vol. 289, IAU Symposium, ed. R.~{de
  Grijs}, 29--35

\bibitem[{{Ginsburg} {et~al.}(2013){Ginsburg}, {Glenn}, {Rosolowsky},
  {Ellsworth-Bowers}, {Battersby}, {Dunham}, {Merello}, {Shirley}, {Bally},
  {Evans}, {Stringfellow}, \& {Aguirre}}]{Ginsburg2013a}
{Ginsburg}, A. {et~al.} 2013, \apjs, 208, 14

\bibitem[{{Ginsburg} \& {Mirocha}(2011)}]{Ginsburg2011c}
{Ginsburg}, A. \& {Mirocha}, J. 2011, {PySpecKit: Python Spectroscopic
  Toolkit}, Astrophysics Source Code Library

\bibitem[{{Goicoechea} {et~al.}(2013){Goicoechea}, {Etxaluze}, {Cernicharo},
  {Gerin}, {Neufeld}, {Contursi}, {Bell}, {De Luca}, {Encrenaz}, {Indriolo},
  {Lis}, {Polehampton}, \& {Sonnentrucker}}]{Goicoechea2013a}
{Goicoechea}, J.~R. {et~al.} 2013, \apjl, 769, L13

\bibitem[{{Goldsmith}(2001)}]{Goldsmith2001a}
{Goldsmith}, P.~F. 2001, \apj, 557, 736

\bibitem[{{Goto} {et~al.}(2014){Goto}, {Geballe}, {Indriolo}, {Yusef-Zadeh},
  {Usuda}, {Henning}, \& {Oka}}]{Goto2014a}
{Goto}, M., {Geballe}, T.~R., {Indriolo}, N., {Yusef-Zadeh}, F., {Usuda}, T.,
  {Henning}, T., \& {Oka}, T. 2014, \apj, 786, 96

\bibitem[{{Goto} {et~al.}(2013){Goto}, {Indriolo}, {Geballe}, \&
  {Usuda}}]{Goto2013a}
{Goto}, M., {Indriolo}, N., {Geballe}, T.~R., \& {Usuda}, T. 2013, Journal of
  Physical Chemistry A, 117, 9919

\bibitem[{{Goto} {et~al.}(2011){Goto}, {Usuda}, {Geballe}, {Indriolo},
  {McCall}, {Henning}, \& {Oka}}]{Goto2011a}
{Goto}, M., {Usuda}, T., {Geballe}, T.~R., {Indriolo}, N., {McCall}, B.~J.,
  {Henning}, T., \& {Oka}, T. 2011, \pasj, 63, L13

\bibitem[{{Green}(1991)}]{Green1991a}
{Green}, S. 1991, \apjs, 76, 979

\bibitem[{{G{\"u}sten} {et~al.}(2006){G{\"u}sten}, {Booth}, {Cesarsky},
  {Menten}, {Agurto}, {Anciaux}, {Azagra}, {Belitsky}, {Belloche}, {Bergman},
  {De Breuck}, {Comito}, {Dumke}, {Duran}, {Esch}, {Fluxa}, {Greve}, {Hafok},
  {H{\"a}upl}, {Helldner}, {Henseler}, {Heyminck}, {Johansson}, {Kasemann},
  {Klein}, {Korn}, {Kreysa}, {Kurz}, {Lapkin}, {Leurini}, {Lis}, {Lundgren},
  {Mac-Auliffe}, {Martinez}, {Melnick}, {Morris}, {Muders}, {Nyman}, {Olberg},
  {Olivares}, {Pantaleev}, {Patel}, {Pausch}, {Philipp}, {Philipps},
  {Sridharan}, {Polehampton}, {Reveret}, {Risacher}, {Roa}, {Sauer}, {Schilke},
  {Santana}, {Schneider}, {Sepulveda}, {Siringo}, {Spyromilio}, {Stenvers},
  {van der Tak}, {Torres}, {Vanzi}, {Vassilev}, {Weiss}, {Willmeroth},
  {Wunsch}, \& {Wyrowski}}]{Gusten2006b}
{G{\"u}sten}, R. {et~al.} 2006, in Society of Photo-Optical Instrumentation
  Engineers (SPIE) Conference Series, Vol. 6267, Society of Photo-Optical
  Instrumentation Engineers (SPIE) Conference Series, 14

\bibitem[{{G{\"u}sten} {et~al.}(1981){G{\"u}sten}, {Walmsley}, \&
  {Pauls}}]{Guesten1981a}
{G{\"u}sten}, R., {Walmsley}, C.~M., \& {Pauls}, T. 1981, \aap, 103, 197

\bibitem[{{Heiderman} {et~al.}(2010){Heiderman}, {Evans}, {Allen}, {Huard}, \&
  {Heyer}}]{Heiderman2010a}
{Heiderman}, A., {Evans}, II, N.~J., {Allen}, L.~E., {Huard}, T., \& {Heyer},
  M. 2010, \apj, 723, 1019

\bibitem[{{Hennebelle} \& {Chabrier}(2011)}]{Hennebelle2011a}
{Hennebelle}, P. \& {Chabrier}, G. 2011, \apjl, 743, L29

\bibitem[{{Hennebelle} \& {Chabrier}(2013)}]{Hennebelle2013a}
---. 2013, \apj, 770, 150

\bibitem[{{Hollenbach} \& {Tielens}(1999)}]{Hollenbach1999a}
{Hollenbach}, D.~J. \& {Tielens}, A.~G.~G.~M. 1999, Reviews of Modern Physics,
  71, 173

\bibitem[{Hopkins(2013)}]{Hopkins2013a}
Hopkins, P.~F. 2013, Monthly Notices of the Royal Astronomical Society, 430,
  1653

\bibitem[{{H{\"u}ttemeister} {et~al.}(1993){H{\"u}ttemeister}, {Wilson},
  {Bania}, \& {Martin-Pintado}}]{Huettemeister1993a}
{H{\"u}ttemeister}, S., {Wilson}, T.~L., {Bania}, T.~M., \& {Martin-Pintado},
  J. 1993, \aap, 280, 255

\bibitem[{{Immer} {et~al.}(2012){Immer}, {Menten}, {Schuller}, \&
  {Lis}}]{Immer2012a}
{Immer}, K., {Menten}, K.~M., {Schuller}, F., \& {Lis}, D.~C. 2012, \aap, 548,
  A120

\bibitem[{{Jappsen} {et~al.}(2005){Jappsen}, {Klessen}, {Larson}, {Li}, \& {Mac
  Low}}]{Jappsen2005a}
{Jappsen}, A.-K., {Klessen}, R.~S., {Larson}, R.~B., {Li}, Y., \& {Mac Low},
  M.-M. 2005, \aap, 435, 611

\bibitem[{{Johnston} {et~al.}(2014){Johnston}, {Beuther}, {Linz}, {Schmiedeke},
  {Ragan}, \& {Henning}}]{Johnston2014a}
{Johnston}, K.~G., {Beuther}, H., {Linz}, H., {Schmiedeke}, A., {Ragan}, S.~E.,
  \& {Henning}, T. 2014

\bibitem[{{Juvela} {et~al.}(2012){Juvela}, {Pelkonen}, {White}, {K{\"o}nyves},
  {Kirk}, \& {Andr{\'e}}}]{Juvela2012a}
{Juvela}, M., {Pelkonen}, V.-M., {White}, G.~J., {K{\"o}nyves}, V., {Kirk}, J.,
  \& {Andr{\'e}}, P. 2012, \aap, 544, A14

\bibitem[{{Kamenetzky} {et~al.}(2012){Kamenetzky}, {Glenn}, {Rangwala},
  {Maloney}, {Bradford}, {Wilson}, {Bendo}, {Baes}, {Boselli}, {Cooray},
  {Isaak}, {Lebouteiller}, {Madden}, {Panuzzo}, {Schirm}, {Spinoglio}, \&
  {Wu}}]{Kamenetzky2012a}
{Kamenetzky}, J. {et~al.} 2012, \apj, 753, 70

\bibitem[{{Kamenetzky} {et~al.}(2014){Kamenetzky}, {Rangwala}, {Glenn},
  {Maloney}, \& {Conley}}]{Kamenetzky2014a}
{Kamenetzky}, J., {Rangwala}, N., {Glenn}, J., {Maloney}, P.~R., \& {Conley},
  A. 2014

\bibitem[{{Kauffmann} {et~al.}(2013){Kauffmann}, {Pillai}, \&
  {Zhang}}]{Kauffmann2013a}
{Kauffmann}, J., {Pillai}, T., \& {Zhang}, Q. 2013, \apjl, 765, L35

\bibitem[{{Kennicutt} \& {Evans}(2012)}]{Kennicutt2012a}
{Kennicutt}, R.~C. \& {Evans}, N.~J. 2012, \araa, 50, 531

\bibitem[{{Kennicutt}(1998)}]{Kennicutt1998a}
{Kennicutt}, Jr., R.~C. 1998, \apj, 498, 541

\bibitem[{{Klein} {et~al.}(2012){Klein}, {Hochg{\"u}rtel}, {Kr{\"a}mer},
  {Bell}, {Meyer}, \& {G{\"u}sten}}]{Klein2012a}
{Klein}, B., {Hochg{\"u}rtel}, S., {Kr{\"a}mer}, I., {Bell}, A., {Meyer}, K.,
  \& {G{\"u}sten}, R. 2012, \aap, 542, L3

\bibitem[{{Kruijssen} {et~al.}(2015){Kruijssen}, {Dale}, \&
  {Longmore}}]{Kruijssen2015a}
{Kruijssen}, J.~M.~D., {Dale}, J.~E., \& {Longmore}, S.~N. 2015, \mnras, 447,
  1059

\bibitem[{{Kruijssen} \& {Longmore}(2013)}]{Kruijssen2013a}
{Kruijssen}, J.~M.~D. \& {Longmore}, S.~N. 2013, \mnras, 435, 2598

\bibitem[{{Kruijssen} {et~al.}(2014){Kruijssen}, {Longmore}, {Elmegreen},
  {Murray}, {Bally}, {Testi}, \& {Kennicutt}}]{Kruijssen2014c}
{Kruijssen}, J.~M.~D., {Longmore}, S.~N., {Elmegreen}, B.~G., {Murray}, N.,
  {Bally}, J., {Testi}, L., \& {Kennicutt}, R.~C. 2014, \mnras, 440, 3370

\bibitem[{{Krumholz}(2014)}]{Krumholz2014c}
{Krumholz}, M.~R. 2014, \mnras, 437, 1662

\bibitem[{{Krumholz} \& {Kruijssen}(2015)}]{Krumholz2015b}
{Krumholz}, M.~R. \& {Kruijssen}, J.~M.~D. 2015, 1505.07111, submitted to
  \mnras

\bibitem[{{Krumholz} \& {McKee}(2005)}]{Krumholz2005c}
{Krumholz}, M.~R. \& {McKee}, C.~F. 2005, \apj, 630, 250

\bibitem[{{Larson}(2005)}]{Larson2005a}
{Larson}, R.~B. 2005, \mnras, 359, 211

\bibitem[{{Le Bourlot} {et~al.}(1999){Le Bourlot}, {Pineau des For{\^e}ts}, \&
  {Flower}}]{Le-Bourlot1999a}
{Le Bourlot}, J., {Pineau des For{\^e}ts}, G., \& {Flower}, D.~R. 1999, \mnras,
  305, 802

\bibitem[{{Leroy} {et~al.}(2013){Leroy}, {Walter}, {Sandstrom}, {Schruba},
  {Munoz-Mateos}, {Bigiel}, {Bolatto}, {Brinks}, {de Blok}, {Meidt}, {Rix},
  {Rosolowsky}, {Schinnerer}, {Schuster}, \& {Usero}}]{Leroy2013a}
{Leroy}, A.~K. {et~al.} 2013, \aj, 146, 19

\bibitem[{{Lis} {et~al.}(1991){Lis}, {Carlstrom}, \& {Keene}}]{Lis1991a}
{Lis}, D.~C., {Carlstrom}, J.~E., \& {Keene}, J. 1991, \apj, 380, 429

\bibitem[{{Lis} {et~al.}(1999){Lis}, {Li}, {Dowell}, \& {Menten}}]{Lis1999a}
{Lis}, D.~C., {Li}, Y., {Dowell}, C.~D., \& {Menten}, K.~M. 1999, in ESA
  Special Publication, Vol. 427, The Universe as Seen by ISO, ed. P.~{Cox} \&
  M.~{Kessler}, 627

\bibitem[{{Lis} \& {Menten}(1998)}]{Lis1998a}
{Lis}, D.~C. \& {Menten}, K.~M. 1998, \apj, 507, 794

\bibitem[{{Lis} {et~al.}(2014){Lis}, {Schilke}, {Bergin}, {Gerin}, {Black},
  {Comito}, {Luca}, {Godard}, {Higgins}, {Petit}, {Pearson}, {Pellegrini},
  {Phillips}, \& {Yu}}]{Lis2014a}
{Lis}, D.~C. {et~al.} 2014

\bibitem[{{Loenen} {et~al.}(2008){Loenen}, {Spaans}, {Baan}, \&
  {Meijerink}}]{Loenen2008a}
{Loenen}, A.~F., {Spaans}, M., {Baan}, W.~A., \& {Meijerink}, R. 2008, \aap,
  488, L5

\bibitem[{{Longmore} {et~al.}(2013{\natexlab{a}}){Longmore}, {Bally}, {Testi},
  {Purcell}, {Walsh}, {Bressert}, {Pestalozzi}, {Molinari}, {Ott}, {Cortese},
  {Battersby}, {Murray}, {Lee}, {Kruijssen}, {Schisano}, \&
  {Elia}}]{Longmore2013b}
{Longmore}, S.~N. {et~al.} 2013{\natexlab{a}}, \mnras, 429, 987

\bibitem[{{Longmore} {et~al.}(2013{\natexlab{b}}){Longmore}, {Kruijssen},
  {Bally}, {Ott}, {Testi}, {Rathborne}, {Bastian}, {Bressert}, {Molinari},
  {Battersby}, \& {Walsh}}]{Longmore2013a}
---. 2013{\natexlab{b}}, \mnras, 433, L15

\bibitem[{{Longmore} {et~al.}(2012){Longmore}, {Rathborne}, {Bastian}, {Alves},
  {Ascenso}, {Bally}, {Testi}, {Longmore}, {Battersby}, {Bressert}, {Purcell},
  {Walsh}, {Jackson}, {Foster}, {Molinari}, {Meingast}, {Amorim}, {Lima},
  {Marques}, {Moitinho}, {Pinhao}, {Rebordao}, \& {Santos}}]{Longmore2012b}
---. 2012, \apj, 746, 117

\bibitem[{{Mangum} {et~al.}(2013){Mangum}, {Darling}, {Henkel}, \&
  {Menten}}]{Mangum2013a}
{Mangum}, J.~G., {Darling}, J., {Henkel}, C., \& {Menten}, K.~M. 2013, \apj,
  766, 108

\bibitem[{{Mangum} {et~al.}(2007){Mangum}, {Emerson}, \&
  {Greisen}}]{Mangum2007a}
{Mangum}, J.~G., {Emerson}, D.~T., \& {Greisen}, E.~W. 2007, \aap, 474, 679

\bibitem[{{Mangum} \& {Wootten}(1993)}]{Mangum1993a}
{Mangum}, J.~G. \& {Wootten}, A. 1993, \apjs, 89, 123

\bibitem[{{Mauersberger} {et~al.}(1986){Mauersberger}, {Henkel}, {Wilson}, \&
  {Walmsley}}]{Mauersberger1986a}
{Mauersberger}, R., {Henkel}, C., {Wilson}, T.~L., \& {Walmsley}, C.~M. 1986,
  \aap, 162, 199

\bibitem[{{Meijerink} {et~al.}(2011){Meijerink}, {Spaans}, {Loenen}, \& {van
  der Werf}}]{Meijerink2011a}
{Meijerink}, R., {Spaans}, M., {Loenen}, A.~F., \& {van der Werf}, P.~P. 2011,
  \aap, 525, A119

\bibitem[{{Menten} {et~al.}(2009){Menten}, {Wilson}, {Leurini}, \&
  {Schilke}}]{Menten2009a}
{Menten}, K.~M., {Wilson}, R.~W., {Leurini}, S., \& {Schilke}, P. 2009, \apj,
  692, 47

\bibitem[{{Mills} \& {Morris}(2013)}]{Mills2013a}
{Mills}, E.~A.~C. \& {Morris}, M.~R. 2013

\bibitem[{{Molinari} {et~al.}(2011){Molinari}, {Bally}, {Noriega-Crespo},
  {Compi{\`e}gne}, {Bernard}, {Paradis}, {Martin}, {Testi}, {Barlow}, {Moore},
  {Plume}, {Swinyard}, {Zavagno}, {Calzoletti}, {Di Giorgio}, {Elia},
  {Faustini}, {Natoli}, {Pestalozzi}, {Pezzuto}, {Piacentini}, {Polenta},
  {Polychroni}, {Schisano}, {Traficante}, {Veneziani}, {Battersby}, {Burton},
  {Carey}, {Fukui}, {Li}, {Lord}, {Morgan}, {Motte}, {Schuller},
  {Stringfellow}, {Tan}, {Thompson}, {Ward-Thompson}, {White}, \&
  {Umana}}]{Molinari2011a}
{Molinari}, S. {et~al.} 2011, \apjl, 735, L33

\bibitem[{{Molinari} {et~al.}(2010){Molinari}, {Swinyard}, {Bally}, {Barlow},
  {Bernard}, {Martin}, {Moore}, {Noriega-Crespo}, {Plume}, {Testi}, {Zavagno},
  {Abergel}, {Ali}, {Anderson}, {Andr{\'e}}, {Baluteau}, {Battersby},
  {Beltr{\'a}n}, {Benedettini}, {Billot}, {Blommaert}, {Bontemps}, {Boulanger},
  {Brand}, {Brunt}, {Burton}, {Calzoletti}, {Carey}, {Caselli}, {Cesaroni},
  {Cernicharo}, {Chakrabarti}, {Chrysostomou}, {Cohen}, {Compiegne}, {de
  Bernardis}, {de Gasperis}, {di Giorgio}, {Elia}, {Faustini}, {Flagey},
  {Fukui}, {Fuller}, {Ganga}, {Garcia-Lario}, {Glenn}, {Goldsmith}, {Griffin},
  {Hoare}, {Huang}, {Ikhenaode}, {Joblin}, {Joncas}, {Juvela}, {Kirk},
  {Lagache}, {Li}, {Lim}, {Lord}, {Marengo}, {Marshall}, {Masi}, {Massi},
  {Matsuura}, {Minier}, {Miville-Desch{\^e}nes}, {Montier}, {Morgan}, {Motte},
  {Mottram}, {M{\"u}ller}, {Natoli}, {Neves}, {Olmi}, {Paladini}, {Paradis},
  {Parsons}, {Peretto}, {Pestalozzi}, {Pezzuto}, {Piacentini}, {Piazzo},
  {Polychroni}, {Pomar{\`e}s}, {Popescu}, {Reach}, {Ristorcelli}, {Robitaille},
  {Robitaille}, {Rod{\'o}n}, {Roy}, {Royer}, {Russeil}, {Saraceno}, {Sauvage},
  {Schilke}, {Schisano}, {Schneider}, {Schuller}, {Schulz}, {Sibthorpe},
  {Smith}, {Smith}, {Spinoglio}, {Stamatellos}, {Strafella}, {Stringfellow},
  {Sturm}, {Taylor}, {Thompson}, {Traficante}, {Tuffs}, {Umana}, {Valenziano},
  {Vavrek}, {Veneziani}, {Viti}, {Waelkens}, {Ward-Thompson}, {White},
  {Wilcock}, {Wyrowski}, {Yorke}, \& {Zhang}}]{Molinari2010a}
---. 2010, \aap, 518, L100

\bibitem[{{Morris} \& {Serabyn}(1996)}]{Morris1996a}
{Morris}, M. \& {Serabyn}, E. 1996, \araa, 34, 645

\bibitem[{{Morris} {et~al.}(1973){Morris}, {Zuckerman}, {Palmer}, \&
  {Turner}}]{Morris1973a}
{Morris}, M., {Zuckerman}, B., {Palmer}, P., \& {Turner}, B.~E. 1973, \apj,
  186, 501

\bibitem[{{M{\"u}hle} {et~al.}(2007){M{\"u}hle}, {Seaquist}, \&
  {Henkel}}]{Muhle2007a}
{M{\"u}hle}, S., {Seaquist}, E.~R., \& {Henkel}, C. 2007, \apj, 671, 1579

\bibitem[{{Mundy} {et~al.}(1987){Mundy}, {Evans}, {Snell}, \&
  {Goldsmith}}]{Mundy1987a}
{Mundy}, L.~G., {Evans}, II, N.~J., {Snell}, R.~L., \& {Goldsmith}, P.~F. 1987,
  \apj, 318, 392

\bibitem[{{Nelson} \& {Langer}(1999)}]{Nelson1999a}
{Nelson}, R.~P. \& {Langer}, W.~D. 1999, \apj, 524, 923

\bibitem[{{Neufeld} \& {Dalgarno}(1989)}]{Neufeld1989a}
{Neufeld}, D.~A. \& {Dalgarno}, A. 1989, \apj, 340, 869

\bibitem[{{Nummelin} {et~al.}(1998){Nummelin}, {Bergman}, {Hjalmarson},
  {Friberg}, {Irvine}, {Millar}, {Ohishi}, \& {Saito}}]{Nummelin1998a}
{Nummelin}, A., {Bergman}, P., {Hjalmarson}, {\AA}., {Friberg}, P., {Irvine},
  W.~M., {Millar}, T.~J., {Ohishi}, M., \& {Saito}, S. 1998, \apjs, 117, 427

\bibitem[{{Offner} {et~al.}(2013){Offner}, {Clark}, {Hennebelle}, {Bastian},
  {Bate}, {Hopkins}, {Moraux}, \& {Whitworth}}]{Offner2013b}
{Offner}, S.~S.~R., {Clark}, P.~C., {Hennebelle}, P., {Bastian}, N., {Bate},
  M.~R., {Hopkins}, P.~F., {Moraux}, E., \& {Whitworth}, A.~P. 2013, ArXiv
  e-prints

\bibitem[{{Ott} {et~al.}(2014){Ott}, {Weiss}, {Staveley-Smith}, {Henkel}, \&
  {Meier}}]{Ott2014a}
{Ott}, J., {Weiss}, A., {Staveley-Smith}, L., {Henkel}, C., \& {Meier}, D.~S.
  2014

\bibitem[{{Padoan} \& {Nordlund}(2011)}]{Padoan2011b}
{Padoan}, P. \& {Nordlund}, {\AA}. 2011, \apj, 730, 40

\bibitem[{{Papadopoulos}(2010)}]{Papadopoulos2010a}
{Papadopoulos}, P.~P. 2010, \apj, 720, 226

\bibitem[{{Papadopoulos} \& {Thi}(2013)}]{Papadopoulos2013a}
{Papadopoulos}, P.~P. \& {Thi}, W.-F. 2013, in Astrophysics and Space Science
  Proceedings, Vol.~34, Cosmic Rays in Star-Forming Environments, ed. D.~F.
  {Torres} \& O.~{Reimer}, 41

\bibitem[{{Papadopoulos} {et~al.}(2011){Papadopoulos}, {Thi}, {Miniati}, \&
  {Viti}}]{Papadopoulos2011a}
{Papadopoulos}, P.~P., {Thi}, W.-F., {Miniati}, F., \& {Viti}, S. 2011, \mnras,
  414, 1705

\bibitem[{P\'erez \& Granger(2007)}]{Perez2007a}
P\'erez, F. \& Granger, B.~E. 2007, Computing in Science and Engineering, 9, 21

\bibitem[{{Pillai} {et~al.}(2006){Pillai}, {Wyrowski}, {Carey}, \&
  {Menten}}]{Pillai2006a}
{Pillai}, T., {Wyrowski}, F., {Carey}, S.~J., \& {Menten}, K.~M. 2006, \aap,
  450, 569

\bibitem[{{Pineda} {et~al.}(2010){Pineda}, {Goodman}, {Arce}, {Caselli},
  {Foster}, {Myers}, \& {Rosolowsky}}]{Pineda2010a}
{Pineda}, J.~E., {Goodman}, A.~A., {Arce}, H.~G., {Caselli}, P., {Foster},
  J.~B., {Myers}, P.~C., \& {Rosolowsky}, E.~W. 2010, \apjl, 712, L116

\bibitem[{{Rathborne} {et~al.}(2014){Rathborne}, {Longmore}, {Jackson},
  {Kruijssen}, {Alves}, {Bally}, {Bastian}, {Contreras}, {Foster}, {Garay},
  {Testi}, \& {Walsh}}]{Rathborne2014b}
{Rathborne}, J.~M. {et~al.} 2014, \apjl, 795, L25

\bibitem[{{Reid} {et~al.}(2009){Reid}, {Menten}, {Zheng}, {Brunthaler}, \&
  {Xu}}]{Reid2009a}
{Reid}, M.~J., {Menten}, K.~M., {Zheng}, X.~W., {Brunthaler}, A., \& {Xu}, Y.
  2009, \apj, 705, 1548

\bibitem[{{Riquelme} {et~al.}(2013){Riquelme}, {Amo-Baladr{\'o}n},
  {Mart{\'{\i}}n-Pintado}, {Mauersberger}, {Mart{\'{\i}}n}, \&
  {Bronfman}}]{Riquelme2013a}
{Riquelme}, D., {Amo-Baladr{\'o}n}, M.~A., {Mart{\'{\i}}n-Pintado}, J.,
  {Mauersberger}, R., {Mart{\'{\i}}n}, S., \& {Bronfman}, L. 2013, \aap, 549,
  A36

\bibitem[{{Rodriguez-Fernandez} {et~al.}(2006){Rodriguez-Fernandez}, {Combes},
  {Martin-Pintado}, {Wilson}, \& {Apponi}}]{Rodriguez-Fernandez2006b}
{Rodriguez-Fernandez}, N.~J., {Combes}, F., {Martin-Pintado}, J., {Wilson},
  T.~L., \& {Apponi}, A. 2006, \aap, 455, 963

\bibitem[{{Rosolowsky} \& {Blitz}(2005)}]{Rosolowsky2005b}
{Rosolowsky}, E. \& {Blitz}, L. 2005, \apj, 623, 826

\bibitem[{{Rosolowsky} \& {Leroy}(2006)}]{Rosolowsky2006a}
{Rosolowsky}, E. \& {Leroy}, A. 2006, \pasp, 118, 590

\bibitem[{{Rosolowsky} {et~al.}(2008){Rosolowsky}, {Pineda}, {Kauffmann}, \&
  {Goodman}}]{Rosolowsky2008c}
{Rosolowsky}, E.~W., {Pineda}, J.~E., {Kauffmann}, J., \& {Goodman}, A.~A.
  2008, \apj, 679, 1338

\bibitem[{{Sch{\"o}ier} {et~al.}(2005){Sch{\"o}ier}, {van der Tak}, {van
  Dishoeck}, \& {Black}}]{Schoier2005a}
{Sch{\"o}ier}, F.~L., {van der Tak}, F.~F.~S., {van Dishoeck}, E.~F., \&
  {Black}, J.~H. 2005, \aap, 432, 369

\bibitem[{{Shetty} {et~al.}(2012){Shetty}, {Beaumont}, {Burton}, {Kelly}, \&
  {Klessen}}]{Shetty2012a}
{Shetty}, R., {Beaumont}, C.~N., {Burton}, M.~G., {Kelly}, B.~C., \& {Klessen},
  R.~S. 2012, \mnras, 425, 720

\bibitem[{{Shirley}(2015)}]{Shirley2015a}
{Shirley}, Y.~L. 2015

\bibitem[{{Sofue}(1995)}]{Sofue1995a}
{Sofue}, Y. 1995, \pasj, 47, 527

\bibitem[{{Traficante} {et~al.}(2011){Traficante}, {Calzoletti}, {Veneziani},
  {Ali}, {de Gasperis}, {di Giorgio}, {Faustini}, {Ikhenaode}, {Molinari},
  {Natoli}, {Pestalozzi}, {Pezzuto}, {Piacentini}, {Piazzo}, {Polenta}, \&
  {Schisano}}]{Traficante2011a}
{Traficante}, A. {et~al.} 2011, \mnras, 416, 2932

\bibitem[{{van der Tak} {et~al.}(2007){van der Tak}, {Black}, {Sch{\"o}ier},
  {Jansen}, \& {van Dishoeck}}]{van-Der-Tak2007a}
{van der Tak}, F.~F.~S., {Black}, J.~H., {Sch{\"o}ier}, F.~L., {Jansen}, D.~J.,
  \& {van Dishoeck}, E.~F. 2007, \aap, 468, 627

\bibitem[{{van der Werf} {et~al.}(2010){van der Werf}, {Isaak}, {Meijerink},
  {Spaans}, {Rykala}, {Fulton}, {Loenen}, {Walter}, {Wei{\ss}}, {Armus},
  {Fischer}, {Israel}, {Harris}, {Veilleux}, {Henkel}, {Savini}, {Lord},
  {Smith}, {Gonz{\'a}lez-Alfonso}, {Naylor}, {Aalto}, {Charmandaris}, {Dasyra},
  {Evans}, {Gao}, {Greve}, {G{\"u}sten}, {Kramer}, {Mart{\'{\i}}n-Pintado},
  {Mazzarella}, {Papadopoulos}, {Sanders}, {Spinoglio}, {Stacey}, {Vlahakis},
  {Wiedner}, \& {Xilouris}}]{van-der-Werf2010a}
{van der Werf}, P.~P. {et~al.} 2010, \aap, 518, L42

\bibitem[{{Vassilev} {et~al.}(2008){Vassilev}, {Meledin}, {Lapkin}, {Belitsky},
  {Nystr{\"o}m}, {Henke}, {Pavolotsky}, {Monje}, {Risacher}, {Olberg},
  {Strandberg}, {Sundin}, {Fredrixon}, {Ferm}, {Desmaris}, {Dochev},
  {Pantaleev}, {Bergman}, \& {Olofsson}}]{Vassilev2008a}
{Vassilev}, V. {et~al.} 2008, \aap, 490, 1157

\bibitem[{{Walsh} {et~al.}(2011){Walsh}, {Breen}, {Britton}, {Brooks},
  {Burton}, {Cunningham}, {Green}, {Harvey-Smith}, {Hindson}, {Hoare},
  {Indermuehle}, {Jones}, {Lo}, {Longmore}, {Lowe}, {Phillips}, {Purcell},
  {Thompson}, {Urquhart}, {Voronkov}, {White}, \& {Whiting}}]{Walsh2011a}
{Walsh}, A.~J. {et~al.} 2011, \mnras, 416, 1764

\bibitem[{{Wei{\ss}} {et~al.}(2001){Wei{\ss}}, {Neininger}, {Henkel},
  {Stutzki}, \& {Klein}}]{Weis2001a}
{Wei{\ss}}, A., {Neininger}, N., {Henkel}, C., {Stutzki}, J., \& {Klein}, U.
  2001, \apjl, 554, L143

\bibitem[{{Wiesenfeld} \& {Faure}(2013)}]{Wiesenfeld2013a}
{Wiesenfeld}, L. \& {Faure}, A. 2013, \mnras, 432, 2573

\bibitem[{{Wolniewicz} {et~al.}(1998){Wolniewicz}, {Simbotin}, \&
  {Dalgarno}}]{Wolniewicz1998a}
{Wolniewicz}, L., {Simbotin}, I., \& {Dalgarno}, A. 1998, \apjs, 115, 293

\bibitem[{{Wootten} {et~al.}(1978){Wootten}, {Evans}, {Snell}, \& {vanden
  Bout}}]{Wootten1978a}
{Wootten}, A., {Evans}, II, N.~J., {Snell}, R., \& {vanden Bout}, P. 1978,
  \apjl, 225, L143

\bibitem[{{Wrathmall} {et~al.}(2007){Wrathmall}, {Gusdorf}, \&
  {Flower}}]{Wrathmall2007a}
{Wrathmall}, S.~A., {Gusdorf}, A., \& {Flower}, D.~R. 2007, \mnras, 382, 133

\bibitem[{{Yusef-Zadeh} {et~al.}(2009){Yusef-Zadeh}, {Hewitt}, {Arendt},
  {Whitney}, {Rieke}, {Wardle}, {Hinz}, {Stolovy}, {Lang}, {Burton}, \&
  {Ramirez}}]{Yusef-Zadeh2009a}
{Yusef-Zadeh}, F. {et~al.} 2009, \apj, 702, 178

\bibitem[{{Yusef-Zadeh} {et~al.}(2013){Yusef-Zadeh}, {Hewitt}, {Wardle},
  {Tatischeff}, {Roberts}, {Cotton}, {Uchiyama}, {Nobukawa}, {Tsuru}, {Heinke},
  \& {Royster}}]{Yusef-Zadeh2013b}
---. 2013, \apj, 762, 33

\bibitem[{{Yusef-Zadeh} {et~al.}(2010){Yusef-Zadeh}, {Lacy}, {Wardle},
  {Whitney}, {Bushouse}, {Roberts}, \& {Arendt}}]{Yusef-Zadeh2010a}
{Yusef-Zadeh}, F., {Lacy}, J.~H., {Wardle}, M., {Whitney}, B., {Bushouse}, H.,
  {Roberts}, D.~A., \& {Arendt}, R.~G. 2010, \apj, 725, 1429

\end{thebibliography}
\end{document}